\documentclass[twocolumn,twocolappendix]{aastex7}

\usepackage{amsmath}

\shorttitle{The rare shock breakout event EP260321a}
\shortauthors{O'Connor et al.}
\received{June 3, 2026}
\revised{June 25, 2026}
\accepted{June 26, 2026}
\submitjournal{ApJL}

\begin{document}

\title{
EP260321a/SN 2026gzf: The Faintest Shock Breakout Associated with a Broad-Lined Supernova \\ 
}

\correspondingauthor{Brendan O'Connor, Xander J. Hall}
\author[0000-0002-9700-0036]{Brendan O'Connor}
    \altaffiliation{McWilliams Fellow}
    \affiliation{McWilliams Center for Cosmology and Astrophysics, Department of Physics, Carnegie Mellon University, Pittsburgh, PA 15213, USA}
    \email[show]{boconno2@andrew.cmu.edu}  

\author[0000-0002-9364-5419]{Xander J. Hall}
    \affiliation{McWilliams Center for Cosmology and Astrophysics, Department of Physics, Carnegie Mellon University, Pittsburgh, PA 15213, USA}
    \email[show]{xhall@cmu.edu}

\author[0009-0001-0574-2332
]{Malte Busmann}
    \affiliation{University Observatory, Faculty of Physics, Ludwig-Maximilians-Universität München, Scheinerstr. 1, 81679 Munich, Germany}
    \affiliation{Excellence Cluster ORIGINS, Boltzmannstr. 2, 85748 Garching, Germany}
    \email{m.busmann@physik.lmu.de}

\author[0000-0003-3270-7644]{Daniel Gruen}
	\affiliation{University Observatory, Faculty of Physics, Ludwig-Maximilians-Universität München, Scheinerstr. 1, 81679 Munich, Germany}
	\affiliation{Excellence Cluster ORIGINS, Boltzmannstr. 2, 85748 Garching, Germany}
	\email{daniel.gruen@lmu.de}

\author[0009-0004-1671-5454]{Alberto Floris}
    \affiliation{Institute of Astrophysics, FORTH, N.Plastira 100, Vassilika Vouton, 70013 Heraklion, Greece}
    \affiliation{Department of Physics University of Crete, Voutes University Campus, 70013 Heraklion, Greece}
    \affiliation{National Institute for Astrophysics (INAF), Astronomical Observatory of Padova, IT-35122 Padova, Italy}
	\affiliation{University Observatory, Faculty of Physics, Ludwig-Maximilians-Universität München, Scheinerstr. 1, 81679 Munich, Germany}
    \email{afloris@ia.forth.gr}

\author[0000-0002-1270-7666]{Tom\'as Cabrera}
\affiliation{McWilliams Center for Cosmology and Astrophysics, Department of Physics, Carnegie Mellon University, Pittsburgh, PA 15213, USA}
\email{tcabrera@andrew.cmu.edu}

\author[0009-0001-8023-5701]{Ziyuan Zhu}
	\affiliation{University Observatory, Faculty of Physics, Ludwig-Maximilians-Universität München, Scheinerstr. 1, 81679 Munich, Germany}
    \email{ziyuan.zhu@campus.lmu.de}
    
\author[0000-0002-6011-0530]{Antonella Palmese}
\affiliation{McWilliams Center for Cosmology and Astrophysics, Department of Physics, Carnegie Mellon University, Pittsburgh, PA 15213, USA}
\email{palmese@cmu.edu}

\author[0000-0002-0676-3661]{Dylan Green}
    \affiliation{Lawrence Berkeley National Laboratory, 1 Cyclotron Road, Berkeley, CA 94720, USA}
    \email{dylangreen@lbl.gov}

\author[0000-0003-0776-8859]{John Banovetz}
    \affiliation{Lawrence Berkeley National Laboratory, 1 Cyclotron Road, Berkeley, CA 94720, USA}
    \email{jdbanovetz@lbl.gov}

\author[0009-0008-2754-1946]{Julius Gassert}
	\affiliation{University Observatory, Faculty of Physics, Ludwig-Maximilians-Universität München, Scheinerstr. 1, 81679 Munich, Germany}
    \affiliation{McWilliams Center for Cosmology and Astrophysics, Department of Physics, Carnegie Mellon University, Pittsburgh, PA 15213, USA}
    \email{julius.gassert@campus.lmu.de}

\author[0000-0003-2624-0056]{Christopher L. Fryer}
    \affiliation{Center for Nonlinear Studies, Los Alamos National Laboratory, Los Alamos, NM 87545 USA}
    \affiliation{Department of Physics, The George Washington University, Washington, DC 20052, USA}
    \email{fryer@lanl.gov}

\author[0000-0003-4631-1528]{Roberto Ricci}
    \affiliation{Dipartimento di Fisica, Universit\`a di Tor Vergata, Via della Ricerca Scientifica, 1, 00133 Rome, Italy}
    \affiliation{INAF-Istituto di Radioastronomia, Via Gobetti 101, I-40129 Bologna, Italy}
    \email{ricci@ira.inaf.it}

\author[0000-0002-1869-7817]{Eleonora Troja}
    \affiliation{Dipartimento di Fisica, Universit\`a di Tor Vergata, Via della Ricerca Scientifica, 1, 00133 Rome, Italy}
    \email{eleonora.troja@uniroma2.it}

\author[0009-0006-1510-4648]{Surya Shivaprasad}
    \affiliation{University Observatory, Faculty of Physics, Ludwig-Maximilians-Universität München, Scheinerstr. 1, 81679 Munich, Germany}
    \affiliation{Excellence Cluster ORIGINS, Boltzmannstr. 2, 85748 Garching, Germany}
    \email{surya.shivaprasad@physik.lmu.de}

\author[0000-0003-2307-0629]{Gregory R. Zeimann}
    \affiliation{University of Texas, Hobby–Eberly Telescope, McDonald Observatory, TX 79734, USA}
    \email{grzeimann@gmail.com}


\author[0000-0003-3433-2698]{Ariel J. Amsellem}
    \affiliation{McWilliams Center for Cosmology and Astrophysics, Department of Physics, Carnegie Mellon University, Pittsburgh, PA 15213, USA}
    \email{aamselle@andrew.cmu.edu}

\author[0000-0003-4162-6619]{Stephen Bailey}
    \affiliation{Lawrence Berkeley National Laboratory, 1 Cyclotron Road, Berkeley, CA 94720, USA}
    \email{stephenbailey@lbl.gov}

\author[0000-0001-5537-4710]{Segev BenZvi}
    \affiliation{Department of Physics \& Astronomy, University of Rochester, 206 Bausch and Lomb Hall, P.O. Box 270171, Rochester, NY 14627-0171, USA}
    \email{sbenzvi@ur.rochester.edu}

\author[0000-0001-6849-1270]{Simone Dichiara}
    \affiliation{Department of Astronomy and Astrophysics, The Pennsylvania State University, 525 Davey Lab, University Park, PA 16802, USA}
    \email{sbd5667@psu.edu}

\author[0000-0002-8680-8718]{Hendrik van Eerten}
\affiliation{Department of Physics, University of Bath, Building 3 West, Bath BA2 7AY, United Kingdom}
\email{hjve20@bath.ac.uk }

\author[0000-0002-8548-482X]{Jeremy Hare}
    \affiliation{Astrophysics Science Division, NASA Goddard Space Flight Center, 8800 Greenbelt Rd, Greenbelt, MD 20771, USA}
    \affiliation{Center for Research and Exploration in Space Science and Technology, NASA/GSFC, Greenbelt, Maryland 20771, USA}
    \affiliation{The Catholic University of America, 620 Michigan Ave., N.E. Washington, DC 20064, USA}
    \email{harej10@gmail.com}

\author[0000-0001-7201-1938]{Lei Hu}
    \affiliation{McWilliams Center for Cosmology and Astrophysics, Department of Physics, Carnegie Mellon University, Pittsburgh, PA 15213, USA}
    \email{leihu@andrew.cmu.edu}

\author[]{Christopher M. Irwin}
    \affiliation{Astronomical Institute, Tohoku University, Sendai, Miyagi 980-8578, Japan}
    \email{irwincm@g.ecc.u-tokyo.ac.jp}

\author[0009-0000-4830-1484]{Keerthi Kunnumkai}
    \affiliation{McWilliams Center for Cosmology and Astrophysics, Department of Physics, Carnegie Mellon University, Pittsburgh, PA 15213, USA}
    \email{kkunnumk@andrew.cmu.edu}

\author[0000-0001-7179-7406]{Konstantin Malanchev}
    \affiliation{McWilliams Center for Cosmology and Astrophysics, Department of Physics, Carnegie Mellon University, Pittsburgh, PA 15213, USA}
    \email{malanchev@cmu.edu}

\author[0009-0001-5050-6232]{Mitra Maleki}
    \affiliation{University Observatory, Faculty of Physics, Ludwig-Maximilians-Universität München, Scheinerstr. 1, 81679 Munich, Germany}
    \email{Mitra.Maleki@campus.lmu.de}

\author[0000-0002-1103-7082]{Michael J. Moss}
    \affiliation{NASA Postdoctoral Program Fellow, NASA Goddard Space Flight Center, Greenbelt, MD 20771, USA}
    \email{mikejmoss3@gmail.com}

\author{Adam D. Myers}
    \affiliation{Department of Physics \& Astronomy, University  of Wyoming, 1000 E. University, Dept.\textasciitilde{}3905, Laramie, WY 82071, USA}
    \email{amyers14@uwyo.edu}

\author[0000-0003-1386-7861]{Dheeraj Pasham}
	\affiliation{Eureka Scientific, 2452 Delmer Street Suite 100, Oakland, CA 94602-3017, USA}
    \affiliation{Department of Physics, The George Washington University, Washington, DC 20052, USA}
    \email{p.dheerajreddy@gmail.com}

\author[0009-0006-6697-8548]{Christoph Ries}
    \affiliation{University Observatory, Faculty of Physics, Ludwig-Maximilians-Universität München, Scheinerstr. 1, 81679 Munich, Germany}
    \email{cries@usm.lmu.de}

\author[0000-0001-9068-7157]{Geoffrey Ryan}
\affiliation{Perimeter Institute for Theoretical Physics, Waterloo, Ontario N2L 2Y5, Canada}
\email{gryan@perimeterinstitute.ca}

\author[0000-0002-5042-5088]{David Schlegel}
    \affiliation{Lawrence Berkeley National Laboratory, 1 Cyclotron Road, Berkeley, CA 94720, USA}
    \email{djschlegel@lbl.gov}

\author[0009-0003-1323-9774]{Michael Schmidt}
    \affiliation{University Observatory, Faculty of Physics, Ludwig-Maximilians-Universität München, Scheinerstr. 1, 81679 Munich, Germany}
    \email{mschmidt@usm.lmu.de}

\author{Silona Wilke}
    \affiliation{University Observatory, Faculty of Physics, Ludwig-Maximilians-Universität München, Scheinerstr. 1, 81679 Munich, Germany}
    \email{Silona.Wilke@lmu.de}

\author[0000-0003-0691-6688]{Yu-Han Yang}
    \affiliation{Dipartimento di Fisica, Universit\`a di Tor Vergata, Via della Ricerca Scientifica, 1, 00133 Rome, Italy}
    \email{yyang@roma2.infn.it}

\begin{abstract}

The explosion of a star is first marked by the shock wave breaking out of the stellar surface, producing a burst of ultraviolet and X-ray radiation. These events are observationally rare, despite likely accompanying the majority of supernovae. Here, we report on our multi-wavelength observing campaign of the closest Einstein Probe fast X-ray transient EP260321a at $z=0.0344$. The thermal ($kT=130$ eV) X-ray emission with peak luminosity $1.0\times10^{45}$ erg s$^{-1}$ points to a shock breakout origin. We demonstrate that EP260321a is accompanied by a broad-lined Type Ic supernova, SN 2026gzf. The supernova properties, including its spectral evolution, lightcurve evolution, and expansion velocities, are all typical of the energetic stripped-envelope supernovae associated with gamma-ray bursts. However, deep X-ray upper limits obtained with the \textit{Chandra X-ray Observatory} do not detect an X-ray afterglow, and instead exclude the afterglow of known gamma-ray bursts or fast X-ray transients. If the stellar explosion launched a successful relativistic jet, we require that it had both a low Lorentz factor $\Gamma_0$\,$<$\,$30$ and a kinetic energy $E_\textrm{kin}$\,$<$\,$10^{49}$ erg for a stellar wind density of $A_*$\,$\gtrsim$\,$1$. We propose that EP260321a originated from a mildly relativistic, weak outflow that was choked by the progenitor star. This scenario is capable of naturally explaining its low X-ray luminosity and lack of prompt gamma-ray emission. EP260321a bridges the gap between SN 2008D and low-luminosity GRBs, suggesting a greater diversity in the physical parameters of stripped stars as they undergo terminal collapse.

\end{abstract}



\section{Introduction} 
\label{sec:intro}





Supernova shock breakout marks the first electromagnetic signal produced when the explosion reaches the surface of the progenitor star or its surrounding circumstellar material. This brief flash encodes the radius, density structure, and immediate environment (e.g., mass loss) of the star at the moment of core collapse, and, therefore, provides the most direct probe of the final stage of massive star evolution \citep{Colgate1974,Weaver1976,Falk1978,KleinChevalier1978,Piro2010,NakarSari2010,NakarSari2012,Sapir2011,Katz2012-1,Katz2012-2,Svirski2014-1,Svirski2014-2,WaxmanKatz2017,FaranSari2019,LevinsonNakar2020,Fryer2020,Irwin2021,Goldberg2022,Irwin2025-1,Irwin2025-2,Fryer2026}. Although shock breakout is a generic prediction of supernova theory, such that all supernovae should exhibit breakout emission, direct detections are rare because the emission is short lived and peaks in the extreme ultraviolet or soft X-ray band \citep{Campana060218,Soderberg2008,Gezari2008,Schawinski2008}. The X-ray transient XRF 080109 associated with SN 2008D remains the clearest example of shock breakout to date  \citep{Soderberg2008,Mazzali2008,ChevalierFransson2008,Modjaz2009,Malesani2009}. Similar physics has also been invoked to explain the thermal high-energy components of nearby low-luminosity gamma-ray bursts, including GRB 060218/SN 2006aj, GRB 100316D/SN 2010bh, and GRB 171205A/SN 2017iuk \citep{Soderberg2006grb060218,Campana060218,Wang2007,Waxman2007,Li2007,NakarSari2012,Nakar2015,Irwin2016,DElia2018,Izzo2019,Irwin2025-2}. These events strongly suggest that shock breakout emission spans a continuum between ordinary supernovae and relativistic explosions, depending on the progenitor's radius \citep[e.g.,][]{WaxmanKatz2017}, circumstellar environment and late stage mass loss \citep[e.g.,][]{Kasen2010,Bayless2015,Lovegrove2017,Smith2017,Goldberg2022,Niblett2025}, outflow collimation \citep[e.g.,][]{Nakar2015,Irwin2016}, Lorentz factor \citep[e.g.,][]{Fryer2026}, and degree to which the jet successfully breaks out (or fails to break out) of the star \citep[e.g.,][]{Nakar2015,Fryer2026}.

The soft X-ray band provides an optimal discovery channel to identify shock breakout emission from stripped-envelope supernovae \citep[e.g.,][]{Alp2020,Bayless2022,Fryer2026}. These sources naturally display themselves as Fast X-ray Transients (FXTs). For example, SN 2008D was first serendipitously identified as an X-ray transient by the \textit{Swift} X-ray Telescope \citep{Burrows2005} before its supernova nature was recognized \citep{Soderberg2008,Mazzali2008}, demonstrating that untargeted X-ray monitoring can reveal core-collapse explosions at phases inaccessible to optical surveys \citep[see, e.g.,][]{Bayless2022}. The launch of the \textit{Einstein Probe} \citep[EP;][]{EP2015,EP2022,Yuan2025} has expanded this discovery space on a much larger scale. Instead of relying on serendipitous discovery with narrow field-of-view X-ray instruments (e.g., the \textit{Neil Gehrels Swift Observatory}, \citealt{Gehrels2004,Soderberg2008}; the \textit{Chandra X-ray Observatory}, \citealt{Jonker2013,Glennie2015,Bauer2017,Alp2020,Quirola2022,Quirola2023}), the 3600 deg$^2$ view of EP has produced a rapid increase in the rate of fast X-ray transient discoveries \citep{Yuan2025,OConnor2025z}. Nearby EP transients associated with Type Ic-BL supernovae offer a particularly important opportunity to probe the diversity of stellar death, the continuum of shock breakout emission from stripped stars, and their link to the collapsar progenitors of long-duration GRBs \citep[see, e.g.,][]{Woosley1993,MacFadyen1999,Woosley2006,Hjorth2012sn,Cano2017}. 

Several EP events have already been interpreted as shock breakout candidates \citep[e.g.,][]{Sun2024,Hamidani2025,Li2025,Liang2026-tns-ep}. In particular, EP250108a and EP250827b have been proposed as relativistic shock breakouts associated with nearby Type Ic-BL supernovae \citep{Li2025,Rastinejad2025EP,Eyles-Ferris2025EP,Srinivasaragavan2025EP0108a,Srinivasaragavan2026}. These detections suggest that EP is finally revealing the predicted population of soft high-energy transients produced by shock breakout, mildly relativistic outflows, failed jets or cocoons, or dirty fireballs with lower bulk Lorentz factors than classical GRBs \citep[e.g.,][]{Rhoads2003}. Therefore, EP is not only improving the census of fast X-ray transients in general, but also providing the first sizeable sample of events at the boundary between ordinary stripped-envelope supernovae \citep[e.g.,][]{Clocchiatti1996ApJ...462..462C,Filippenko1997ARA&A..35..309F,Prentice2017}, low-luminosity GRBs \citep[e.g.,][]{Campana060218,Soderberg2006grb060218}, and jet driven explosions \citep[e.g.,][]{Fryer2025}.

EP260321a is a recent fast X-ray transient detected by the Einstein Probe on 2026 March 21 \citep{2026GCN.44075....1H,Yuan2026}. Initial reports of optical follow-up failed to identify a counterpart with the exception of an archival source with blue color that displayed historical variability \citep{2026GCN.44070....1L,Chen2026}. Optical spectroscopy carried out with the Very Large Telescope showed that the archival point source was not a star, but instead co-located at a consistent redshift $z$\,$=$\,$0.0344$ with the coincident galaxy \citep{2026GCN.44082....1T}. Notably this redshift is comparable to the distance of the low-luminosity GRBs 060218 \citep[$z$\,$=$\,$0.0331$;][]{Campana2006,Pian2006,Soderberg2006grb060218} and 171205A \citep[$z$\,$=$\,$0.0368$;][]{DElia2018,Izzo2019}, which are among the most nearby GRBs.

At this redshift ($z$\,$=$\,$0.0344$), EP260321a displayed a peak X-ray luminosity of $L_{\rm X}$\,$=$\,$(1.0\pm0.3)\times 10^{45}$ erg s$^{-1}$, which combined with its soft, thermal spectrum with $kT$\,$=$\,$132^{+30}_{-21}$ eV pointed to a supernova shock breakout origin \citep{2026GCN.44075....1H,Yuan2026}. In this scenario, the archival optical variability of the source can be explained by stellar outbursts as it approached its death \citep{2026GCN.44082....1T,Chen2026}. Additional photometric and spectroscopic follow-up revealed the clear presence of excess emission at $<$\,$1$ d \citep{2026GCN.44087....1L,Chen2026,MartinCarrillo2026}, possibly attributed to the combination of shock cooling and circumstellar interaction \citep{Chen2026,MartinCarrillo2026,Yuan2026}. This was followed by a clearly rising supernova lightcurve \citep{2026GCN.44089....1S,2026GCN.44091....1D} and the onset of supernova features in the optical spectra 
\citep{2026GCN.44092....1X,2026GCN.44105....1C,2026GCN.44107....1R}, solidifying the shock breakout interpretation \citep{Yuan2026}.

Here we present optical, near-infrared, and X-ray follow-up observations of EP260321a/SN 2026gzf over the first 60 days. We present a high cadence multi-wavelength lightcurve spanning the $ugrizJK_s$ filters starting less than 10 hours after discovery and a series of 12 optical spectra starting $3$ days after discovery. We compare the lightcurves and spectra to other Ic-BL supernovae and derive the properties of the supernova ejecta. We further present X-ray and radio observations obtained with the \textit{Chandra X-ray Observatory} and Very Large Array that constrain the presence of a relativistic outflow from EP260321a. For additional studies of EP260321a/SN 2026gzf that occurred both independently and simultaneously, see \citet{Yuan2026,Rastinejad2026,MartinCarrillo2026,Chen2026}.

Throughout the manuscript we adopt a standard $\Lambda$CDM cosmology \citep{Planck2020} with $H_0$\,$=$\,$67.4$ km s$^{-1}$ Mpc$^{-1}$, $\Omega_\textrm{m}$\,$=$\,$0.315$, and $\Omega_\Lambda$\,$=$\,$0.685$. At the transient's redshift $z$\,$=$\,$0.034328\pm0.000006$ (see \S \ref{sec:hetenv}), this corresponds to a luminosity distance of 158 Mpc. We adopt a rounded value of $z$\,$=$\,$0.0344$ throughout the manuscript. All upper limits are reported at the $3\sigma$ level.

\section{Observations}
\label{sec:obs}


\subsection{Einstein Probe Discovery}

EP260321a was first detected by the Einstein Probe on 2026-03-21 at 12:30:18 UTC \citep{2026GCN.44075....1H,Yuan2026}, which we hereafter adopt as $T_0$. The source was observed by the Wide-field X-ray Telescope (WXT) until $T_0+2200$ s and Follow-up X-ray Telescope (FXT) from $T_0+1065$ to $2359$ s \citep{Yuan2026}. 
The transient peaked at a $0.3$\,$-$\,$10$ keV X-ray luminosity of $L_{\rm X}$\,$=$\,$(1.0\pm0.3)\times 10^{45}$ erg s$^{-1}$ at 690 s after the trigger \citep{Yuan2026}. The spectrum of the source was best fit with an absorbed blackbody with hydrogen column density $N_H$\,$=$\,$8.4\times10^{20}$ cm$^{-2}$ and a temperature ranging from $kT$\,$=$\,$132^{+30}_{-21}$ eV measured over the first 500 s by WXT to $kT$\,$=$\,$107.9^{+3.2}_{-3.0}$ over the last 500 s of the FXT observation \citep[see Figure 1 and Table 2 of][]{Yuan2026}. 
The emission has been interpreted as supernova shock breakout \citep{2026GCN.44075....1H,Yuan2026}. We point the reader to \citet{Yuan2026} for further details on the observations and analysis of EP260321a. All values presented here are taken from \citet{Yuan2026}.

\subsection{Fraunhofer Telescope Wendelstein (FTW)}
\label{sec:FTW}

We observed the optical and near-infrared (OIR) counterpart of EP260321a with the Three Channel Imager (3KK; \citealt{2016SPIE.9908E..44L}) mounted on the 2.1-m Fraunhofer Telescope at Wendelstein Observatory (FTW; \citealt{2014SPIE.9145E..2DH}). Observations began on 2026-03-21 at 22:29:50 UT, corresponding to 0.41 d (9.84 hr) after the trigger. The 3KK camera is capable of obtaining images in three filters simultaneously: two optical and one near-infrared (NIR). We carried out daily observations in the $griz$ bands as weather allowed. NIR observations ($JK_s$) were initially obtained simultaneously, but were not available after 2026-03-28 due to an issue with the cooling system. The cooling system was repaired and NIR observations began again on 2026-04-22. The complete log of observations is tabulated in Table \ref{tab:photometry}.

The FTW data were reduced using a custom pipeline \citep{2002A&A...381.1095G} to perform bias and dark subtraction, flat-fielding, and cosmic ray rejection. For more details, specifically with regard to the near-infrared data reduction, see \citet{Busmann2025}. For optical images, we performed difference imaging with the Saccadic Fast Fourier Transform (\texttt{SFFT}) software\footnote{\url{https://github.com/thomasvrussell/sfft}} \citep{Hu2022} using archival PS1 \citep{Chambers2016}, SDSS \citep[e.g.,][]{sdss-dr7}, and Legacy Survey images as templates. For the near-infrared imaging, we performed difference imaging with respect to the UKIRT Infrared Deep Sky Survey \citep[UKIDSS;][]{Lawrence2007}. We performed aperture photometry on the difference images using \texttt{Photutils} \citep{Bradley2024} with AB magnitude zeropoints calibrated to the PS1 \citep{Chambers2016} and 2MASS \citep{Skrutskie2006} catalogs for optical and near-infrared data, respectively. The near-infrared photometry was converted from the Vega to the AB magnitude system using standard offsets. The photometry is reported in Table \ref{tab:photometry}.

Despite the archival variability \citep{Chen2026} of the existing blue point source (Figure \ref{fig:decamarch}), we find that this variability is low level and does not have an impact on our lightcurve. We refer the reader to \citet{Chen2026} for a more in depth discussion of the source's variability and the link to SN 2026gzf. Here, we tested the possible impact of flux variations of the archival source in a few ways. First, we performed difference imaging versus multiple surveys (PS1, SDSS, and LS) using templates taken across multiple times, yielding negligible photometric deviations across all bands. Secondly, we performed an analysis of an archival time series and found no significant variability that would impact our photometry (see Figure \ref{fig:decamarch} in Appendix \ref{sec:decamarch}).

\begin{figure}
    \centering
    \includegraphics[width=\linewidth]{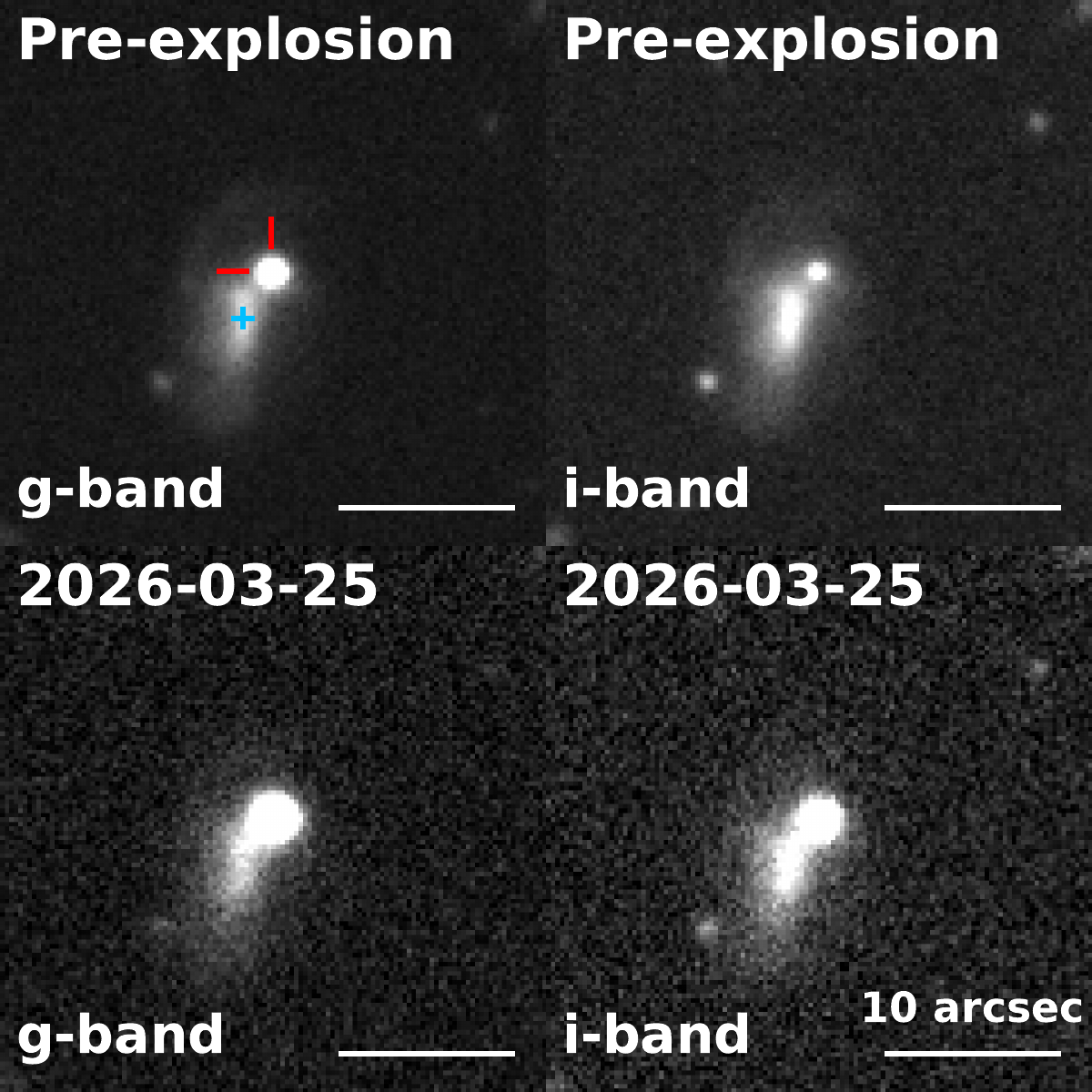}
    \caption{Finding chart of EP260321a/SN 2026gzf using DECam imaging in the $g$ and $i$ filters. Archival pre-explosion DECam images from 2013 (13 years before discovery) are shown in the top panels, while the bottom panels show imaging obtained on 2026-03-25 ($T_0+3.6$ d). A blue point source ($g-i$\,$\approx$\,$-1.3$ mag) is visible at the location of transient in archival imaging (see Appendix \ref{sec:decamarch}). The transient explosion site (red crosshairs) is offset by $3.4\arcsec$ from the center of the host galaxy (blue cross), corresponding to $2.5$ kpc at $z$\,$=$\,$0.0344$. 
    }
    \label{fig:FC}
\end{figure}

\subsection{Dark Energy Camera}

\subsubsection{Targeted Follow-up}

We observed EP260321a with the Dark Energy Camera (DECam) instrument mounted on the Blanco 4\,m Telescope at the Cerro Tololo Inter-American Observatory (CTIO) using the $g$, $r$, and $z$ bands (PI: Palmese; Table \ref{tab:photometry}) starting on 2026-03-25, corresponding to 3.6 d after discovery. The images were astrometrically calibrated against Gaia DR3 \citep{Gaia2021, 2021A&A...649A...2L, gaiaEDR3}. We use \texttt{SFFT} \citep{Hu2022} for image subtraction against archival DECam images and photometrically calibrate using the PS1 \citep[PS1;][]{Chambers2016}. See \citet{Cabrera2024,Hu2025,Hu2026} for further details on the DECam analysis pipeline. We show DECam imaging of the field as a finding chart in Figure \ref{fig:FC}. 

\subsubsection{Archival Image Analysis}

We analyzed all archival $griz$ images previously obtained by DECam. A blue compact source is detected at the location of the transient in all pre-explosion images, see Figure \ref{fig:FC}. These observations span the time range of 2013-06-21 to 2018-05-07, corresponding to 12.8 to 7.9 years prior to the explosion. In total there are 25, 17, 13, and 8 images in the $g$, $r$, $i$, and $z$ bands, respectively.  We performed point spread function (PSF)  photometry of these images in each band to build archival lightcurves. This is displayed in Figure \ref{fig:decamarch} in Appendix \ref{sec:decamarch}, where we have further discussion of the origin of the pre-explosion source at the location of SN 2026gzf.

\subsection{Dark Energy Spectroscopic Instrument}

The Dark Energy Spectroscopic Instrument (DESI) includes 5,000 independently positionable science fibers \citep{schlafly_survey_2023,poppett_overview_2024}. The spectrograph covers $3600-9824$ \AA\ at a spectral resolution $R \sim 2000 - 5500$ \citep{desi_collaboration_desi_2016, miller_optical_2024, desi_collaboration_data_2025, desi_collaboration_desi_2025}. The data were reduced and flux calibrated with the DESI spectroscopic data pipeline \citep{guy_spectroscopic_2023}. DESI spectra are automatically corrected for Galactic extinction, which in this case is $E(B-V)=0.02$ mag \citep{Schlafly2011}.

DESI observed EP260321a on 2026-04-08, 2026-04-23, and 2026-05-14 (see Table \ref{tab:spectra}) as part of a spare fiber program to observe Rubin transients \citep{Hall-desi-rubin}. For other examples of DESI spare fiber targets and observing strategy, see \citet{Myers:2022azg,DESI:2023ytc,HallDESIulz,HallDTS}. The DESI spectra are publicly available through the DESI Transients Survey Zenodo \citep[][]{hall_2026_19653826} as well as the Transient Name Server\footnote{\url{https://www.wis-tns.org/object/2026gzf}} and the Weizmann Interactive Supernova Data Repository \citep[WISeREP;][]{wisrep}.

\subsection{Southern African Large Telescope}

We obtained a spectroscopic sequence of EP260321a with the 10-m class Southern African Large Telescope \citep[SALT;][]{Buckley2006} starting on 2026-03-24. In total we obtained eight spectra between the end of March and the end of May 2026 (Table \ref{tab:spectra}). The initial spectrum was obtained through program 2025-2-SCI-016 (PI: X. Hall). Subsequent observations were acquired through a series of SALT Director's Discretionary Time proposals (2025-2-DDT-004, PI: B. O'Connor; 2026-1-DDT-003, PI: B. O'Connor). These additional spectra were obtained on 2026-03-27, 2026-04-01, 2026-04-05, and 2026-04-13 using the pg0700 grating at a camera station of 22.75 deg covering between $3592-7479$ \AA\, with a resolution of 735 at the central wavelength of 5580 \AA\, and on 2026-05-02, 2026-05-09, and 2026-05-14 using the pg0700 grating at a camera station of 29.5 deg covering between $5250-9043$ \AA\, with a resolution of 953 at the central wavelength of 7202 \AA. Each spectrum was obtained as a single exposure of 2100 s with 2$\times$2 binning using a 1.5\arcsec slit. The data were reduced using the RSS Long-slit spectra processing and extraction app \texttt{rsslsspectra}\footnote{\url{https://astronomers.salt.ac.za/wp-content/uploads/sites/71/2024/09/rsslsspectra.pdf}}. The complete log of spectral observations is tabulated in Table \ref{tab:spectra}. The spectra are available through WISeREP.

\subsection{Hobby-Eberly Telescope}

We observed EP260321a with the 11-m Hobby-Eberly Telescope (HET; \citealt{1998SPIE.3352...34R, 2021AJ....162..298H}) at McDonald Observatory under program M26-1-005 (PI: D. Gruen) on 2026-03-25. The observations were scheduled using the HET queue-scheduling system \citep{2007PASP..119..556S}. We used the low-resolution integral-field spectrograph (LRS2; \citealt{Chonis2014,Chonis2016}) to obtain spectra in the blue channel LRS2-B. 
The raw LRS2 data were initially processed with \texttt{Panacea}\footnote{\url{https://github.com/grzeimann/Panacea}}, which performs bias subtraction, dark subtraction, fiber tracing, fiber wavelength evaluation, fiber extraction, fiber-to-fiber normalization, source detection, source extraction, and flux calibration for each channel. The absolute flux calibration was performed using a spectrophotometric standard observed shortly after our observation.
We extracted the flux-calibrated one-dimensional spectrum centered on the transient using the \texttt{LRS2Multi}\footnote{\url{https://github.com/grzeimann/LRS2Multi}} package. We present an analysis of the spatially resolved environment in \S \ref{sec:hetenv}.

\subsection{Publicly Available Photometry}

We queried publicly available photometry of EP260321a using the Babamul alert broker \citep{Babamul}. The object ID is 314003014107006318\footnote{\url{https://babamul.caltech.edu/objects/LSST/314003014107006318}}. These include data obtained by the Zwicky Transient Facility (ZTF; \citealt{bellm_zwicky_2018,graham_zwicky_2019,masci_zwicky_2018}) and Vera C. Rubin Observatory \citep{Ivezic2019}. As EP260321a falls in the COSMOS Deep Drilling Field, the Rubin Legacy Survey of Space and Time (LSST) commissioning alerts are currently publicly available. While Rubin exhibits some possible pre-explosion detections, there are possible issues with the difference imaging at that stage \citep{2026GCN.44084....1A}. Therefore, we focus on Rubin data obtained after the discovery of EP260321a and do not include any pre-explosion Rubin photometry. ZTF data were obtained in the $gri$ filters and Rubin data in the $ugriz$ filters. A log of photometry from ZTF and Rubin is presented in Table \ref{tab:rubin-ztf-phot}.

\subsection{Chandra X-ray Observatory}

We observed the position of EP260321a with the \textit{Chandra X-ray Observatory} (CXO) under program 27400339 (PI: B. O'Connor). Observations were carried out with ACIS-S starting on 2026-04-05 at 22:00 UT ($T_0$+15.4 d after the EP trigger) for a total of 19.81 ks. A second epoch was obtained starting on 2026-04-29 at 11:33:52 UT ($T_0$+39.0 d) for a total of 59.1 ks. The \textit{Chandra} data were retrieved from the 
\textit{Chandra} Data Archive (CDA)\footnote{\url{https://cda.harvard.edu/chaser/}}. We re-processed the data using the \texttt{CIAO v4.17.0} data reduction package with \texttt{CALDB v4.11.6}. The data were filtered to the $0.5$\,$-$\,$7$ keV energy range. No source is detected at the location of EP260321a in either epoch, with a total of 0 photons within a $1.5\arcsec$ radius at either phase. Adopting an annular background region with inner radius 30\arcsec\, and outer radius of 60\arcsec, we find an expected number of 0.14 and 0.46 background counts within the $1.5\arcsec$ radius circular source aperture in the first and second epochs, respectively. We adopt a typical GRB afterglow spectral shape with $p$\,$=$\,$2.2$ which leads to a photon index of $\Gamma$\,$=$\,$1.6$ for emission between the peak frequency and the cooling frequency \citep{Granot2002}. This is a typical value identified by particle acceleration simulations \citep{Sironi2015}. We adopt a Galactic hydrogen column density $N_\textrm{H}$\,$=$\,$2.5\times10^{20}$ cm$^{-2}$ \citep{Willingale2013}. This absorbed powerlaw spectral shape yields $3\sigma$ upper limits of $<5.2\times10^{-15}$ and $<2.4\times10^{-15}$ erg cm$^{-2}$ s$^{-1}$, respectively, on the $0.3$\,$-$\,$10$ keV unabsorbed flux, which was derived using the \texttt{aplimits} tool available within \texttt{CIAO} \citep{Kashyap2010}. We note that these inferences are not particularly sensitive to the choice in photon index within the reasonably expected range for typical GRB afterglows. For a softer photon index $\Gamma$\,$=$\,$3$, as observed for GRBs 060218 and 100316D \citep{Soderberg2006grb060218,Starling2011,Margutti2013-100316D}, the derived $3\sigma$ limits are $<9.6\times10^{-15}$ and $<4.6\times10^{-15}$ erg cm$^{-2}$ s$^{-1}$. This corresponds to an X-ray luminosity upper limit of roughly $<(1.3-2.9)\times10^{40}$ erg s$^{-1}$ at $z$\,$=$\,$0.0344$.

\subsection{Very Large Array}
\label{sec:vla}

The Karl G. Jansky Very Large Array (VLA)  observed EP260321a on 2026 May 19 at a mid-time of 23:15 UT (Program: SN078192; PI: E. Troja), corresponding to 59.5 days after the EP trigger. The target was observed for 33 minutes (400 s on source) in A-array configuration using C-band at the center frequency of 6 GHz with a bandwidth of 4 GHz. The primary calibrator was 3C286 and the phase calibrator was J1024-0052. The data were flagged and calibrated in \texttt{CASA v6.6.1} using the VLA continuum calibration pipeline \citep{CASA2022}. The calibrated dataset was imaged using the \texttt{CASA} task \texttt{tclean} with the weighting Briggs parameter set to 0.5 and 5000 clean iterations. No detection was found at the optical transient position down to a $3\sigma$ flux density upper limit of 42 $\mu$Jy/beam.

\section{Results}
\label{sec:results}

\subsection{Supernova Lightcurve Analysis}

\subsubsection{Empirical Modeling}

\begin{figure}
    \centering
    \includegraphics[width=\linewidth]{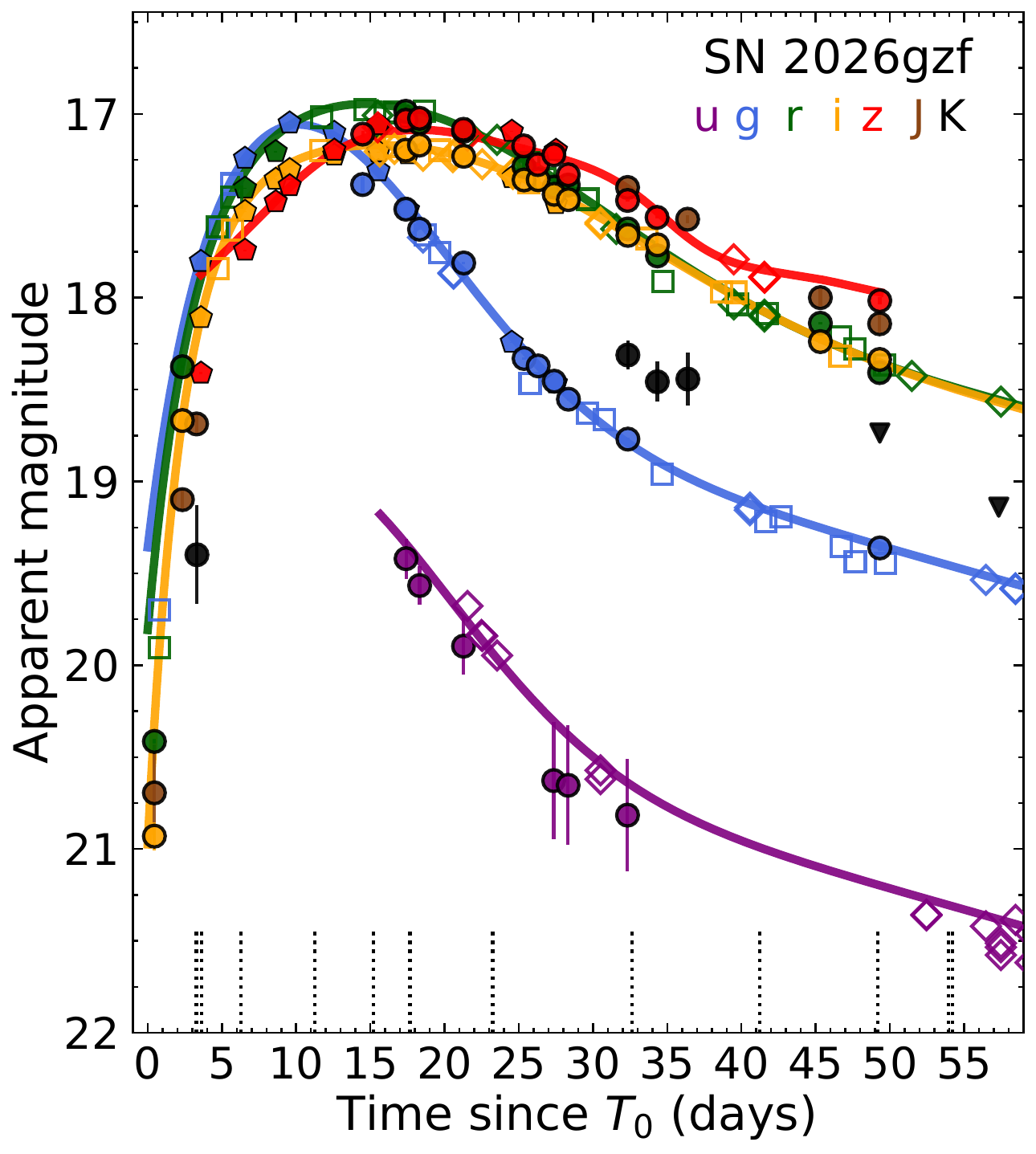}
    \caption{Multi-band lightcurve ($ugrizJK$) of SN 2026gzf from FTW (filled circles), DECam (filled pentagons), ZTF (empty squares), and Rubin (empty diamonds). Solid lines are empirical fits using Equation \ref{eqn:taddia} from \citet{Taddia2018}. The vertical dashed lines at the bottom mark the timing of our spectroscopic observations. 
    }
    \label{fig:lightcurve}
\end{figure}

We performed an empirical fit to the lightcurves of SN 2026gzf in the $griz$ bands following the approach outlined by \citet{Taddia2018,Taddia2019} for modeling stripped-envelope supernovae. The empirical model is given by 
\begin{equation}
\label{eqn:taddia}
m(t) =
\frac{
y_0 + m(t-t_0)
+ g_0 \exp\left[-\frac{(t-t_0)^2}{2\sigma_0^2}\right]
}{
1-\exp\left[\frac{\tau-t}{\theta}\right]
},
\end{equation}
where $y_0$ and $m$ describe the late-time linear decline, $g_0$,
$t_0$, and $\sigma_0$ describe the Gaussian-like peak, and $\tau$ and $\theta$
control the exponential rise. The fit was performed using \texttt{emcee} \citep{emcee} to each band individually. We used these fits to derive the peak time of the lightcurve $t_p$ and the light-curve decline rate parameter $\Delta m_{15}$ in each band, where $\Delta m_{15}$ is defined as the change in magnitude in a given filter over 15 days post-peak. The fit to each band is shown in Figure \ref{fig:lightcurve}.  

\begin{figure}
    \centering
    \includegraphics[width=\linewidth]{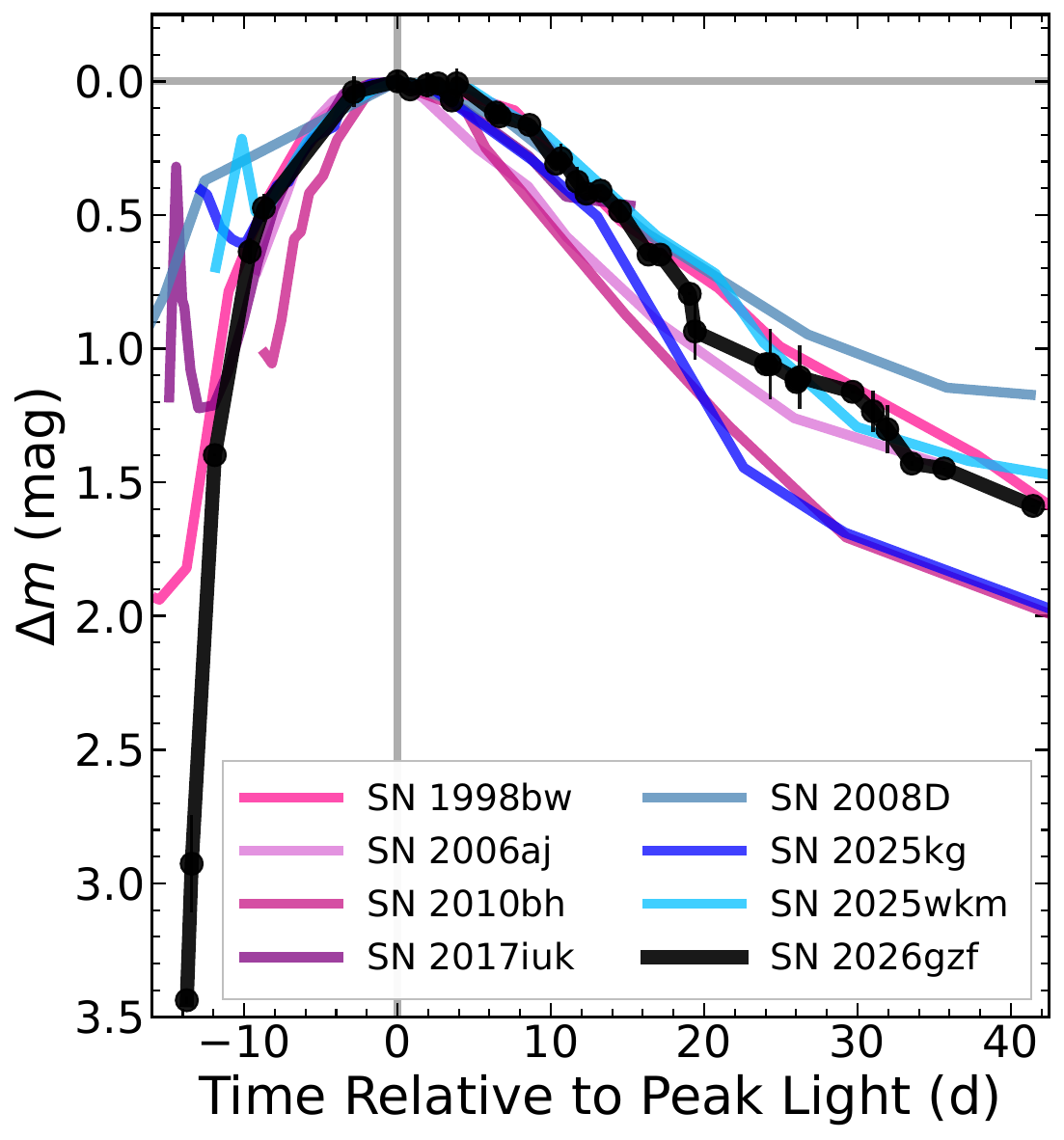}
    \caption{Lightcurve shape relative to peak brightness for SN 2026gzf (black) versus GRB-SNe (SNe 1998bw, 2006aj, 2010bh, and 2017iuk; \citealt{Galama1998,Campana2006,Sollerman2006,Starling2011,DElia2018,Izzo2019}) and FXT-SNe (SNe 2008D, 2025kg and 2025wkm; \citealt{Soderberg2008,Eyles-Ferris2025EP,Rastinejad2025EP,Srinivasaragavan2025EP0108a,Srinivasaragavan2026}). The data are presented in either the $r$ or $R$ filters and time has been converted to the rest-frame.}
    \label{fig:lightcurveshapecompare}
\end{figure}

\begin{figure*}
    \centering
    \includegraphics[width=\linewidth]{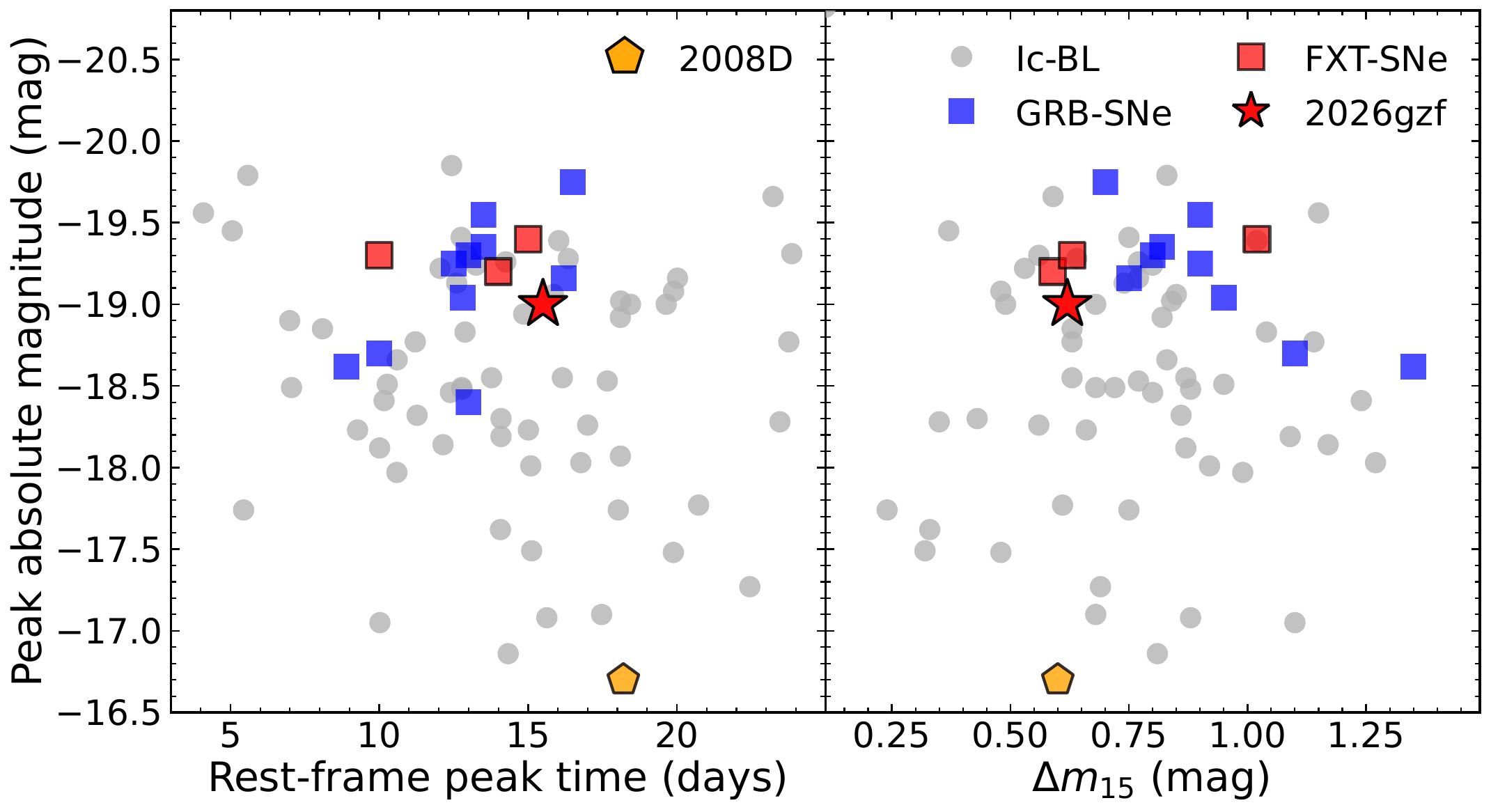}
    \caption{A comparison of EP260321a/SN 2026gzf to other Ic-BL SNe both associated with GRBs and identified independently through optical surveys. The left panel shows the peak absolute magnitude versus the rest-frame peak time, and the right panel shows the absolute magnitude versus $\Delta m_{15}$ which represents the fade rate over 15 days relative to the peak time. GRB-SNe are shown using $V$-band, while other Ic-BL, including SN 2026gzf, are shown using $r$-band. The data have been taken from \citep{Cano2013,Cano2014,Cano2017,Taddia2018,Srinivasaragavan2024,Rastinejad2025EP,Srinivasaragavan2026,Cotter2026}.}
    \label{fig:compare}
\end{figure*}

From this fit, for the $r$-band, we derive $t_{p,r}$\,$=$\,$15.0\pm0.5$ d and 
$\Delta m_{15,r}$\,$=$\,$0.52^{+0.04}_{-0.02}$ mag, see also Figure \ref{fig:lightcurveshapecompare}. We compare this to other Ic-BL supernovae in Figure \ref{fig:compare}, which is discussed further in \S \ref{sec:grbsncompare}. 
In the other bands we derive $t_{p,g}=10.1\pm0.3$ d and
$\Delta m_{15,g}=1.19^{+0.03}_{-0.04}$ mag, $t_{p,i}=15.5\pm0.6$ d and
$\Delta m_{15,i}=0.40^{+0.03}_{-0.04}$ mag, and $t_{p,z}=18.3\pm0.7$ d and
$\Delta m_{15,z}=0.38^{+0.04}_{-0.05}$ mag. 

In Figure \ref{fig:lightcurveshapecompare} we compare the $r$-band lightcurve of SN 2026gzf to Ic-BL SNe associated with GRBs (GRB-SNe; \citealt{Hjorth2012sn,Cano2017}) and Ic-BL SNe associated with fast X-ray transients detected by EP (hereafter, FXT-SNe). We note that SN 2026gzf is itself an FXT-SN. Overall, the lightcurve shape of SN 2026gzf has good agreement with these Ic-BL SNe and with the Ib SN 2008D, despite SN 2008D being $\sim$\,$2$ mag fainter at peak \citep[after correcting for intrinsic dust extinction;][]{Soderberg2008}, see Figure \ref{fig:compare}. 
We discuss this comparison to other SNe with high-energy associations further in \S \ref{sec:grbsncompare}. 

\subsubsection{Bolometric Lightcurve}

We used simultaneous multi-band imaging (Figure \ref{fig:lightcurve}) obtained by FTW and DECam to build a bolometric lightcurve of SN 2026gzf. In general DECam obtained data in the $griz$ filters and FTW in the $grizJK_s$ filters, with $u$-band obtained at later phases by FTW and Rubin/LSST. By modeling the individual spectral energy distributions (SEDs) at each epoch with a blackbody, we can gauge the evolution of the total luminosity as well as the radius $R_\textrm{BB}$ and temperature $T_\textrm{BB}$. We note that the lack of pre-peak ultraviolet data can impact the temperature inferences at early times, which should be treated with caution. 

\begin{figure}
    \centering
    \includegraphics[width=\linewidth]{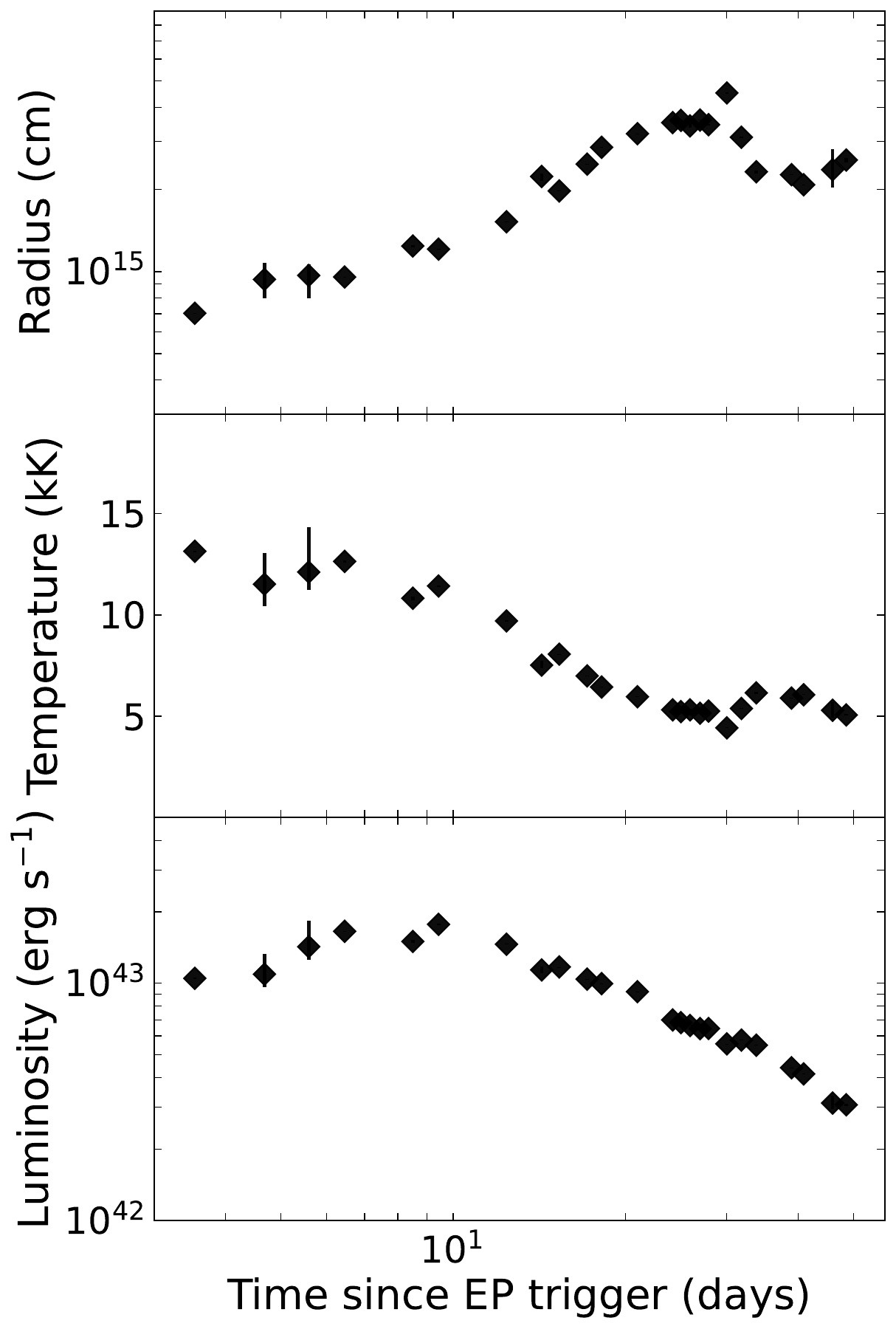}
    \caption{Evolution of the inferred photospheric radius, temperature and bolometric luminosity of SN 2026gzf. Given the lack of UV data, the inferences at $<$\,$1$ d are less secure and are excluded.
    }
    \label{fig:bolometric}
\end{figure}

We use the Hybrid Analytic Flux FittEr for Transients (\texttt{HAFFET}\footnote{\url{https://haffet.readthedocs.io/en/latest/}}; \citealt{HAFFET}) to carry out this analysis. This method was previously used in \citet{Corsi2023,Anand2024,Srinivasaragavan2024}. The bolometric lightcurve, radius, and temperature are shown in Figure \ref{fig:bolometric}. The bolometric lightcurve is consistent, though at the luminous end, with inferences for other Ic-BL SNe both associated with GRBs and without any GRB association \citep{Cano2013,Taddia2018,Prentice2016,Prentice2019,Taddia2019,Srinivasaragavan2024}. The radii and temperature are also consistent with other Ic-BL SNe \citep[e.g.,][]{Taddia2019,Srinivasaragavan2024}. As observed by \citet{Taddia2019,Srinivasaragavan2024}, the radius increases until around 10 days after maximum $r$-band light, which occurs at $\sim$\,$15$ d after EP trigger, and then begins to decrease as the photosphere begins to recede \citep{Dessart2012}. This behavior was also noted by \citet{MartinCarrillo2026}.

We likewise used \texttt{HAFFET} to carry out a semi-analytic Arnett \citep{Arnett1982} model fit to the bolometric lightcurve to derive the nickel mass $M_\textrm{Ni}$ and diffusion timescale $\tau_\textrm{m}$. We adopt a photospheric velocity of $20,000$ km s$^{-1}$ at maximum light. From the bolometric lightcurve in Figure \ref{fig:bolometric}, we infer a $M_\textrm{Ni}$\,$=$\,$0.45\pm0.02$\,$M_\odot$, $\tau_\textrm{m}$\,$=$\,$7.7^{+0.8}_{-0.5}$ d, total ejecta mass $M_\textrm{ej}$\,$=$\,$2.0^{+0.4}_{-0.3}$\,$M_\odot$, and kinetic energy $E_\textrm{k}$\,$=$\,$(9.9^{+2.1}_{-1.3})\times10^{51}$ erg. These values are in good agreement with the average properties inferred from GRB associated Ic-BL SNe by \citet{Cano2017} with typical values of $M_\textrm{Ni}$\,$=$\,$0.4\pm0.2$\,$M_\odot$ and $E_\textrm{k}$\,$=$\,$(2.5\pm1.8)\times10^{52}$ erg. Our inferred properties are consistent with the analyses of \citet{Yuan2026,Chen2026,MartinCarrillo2026,Rastinejad2026} which broadly infer $M_\textrm{Ni}$\,$\approx$\,$0.3-0.5$\,$M_\odot$.

\begin{figure*}
    \centering
    \includegraphics[width=\linewidth]{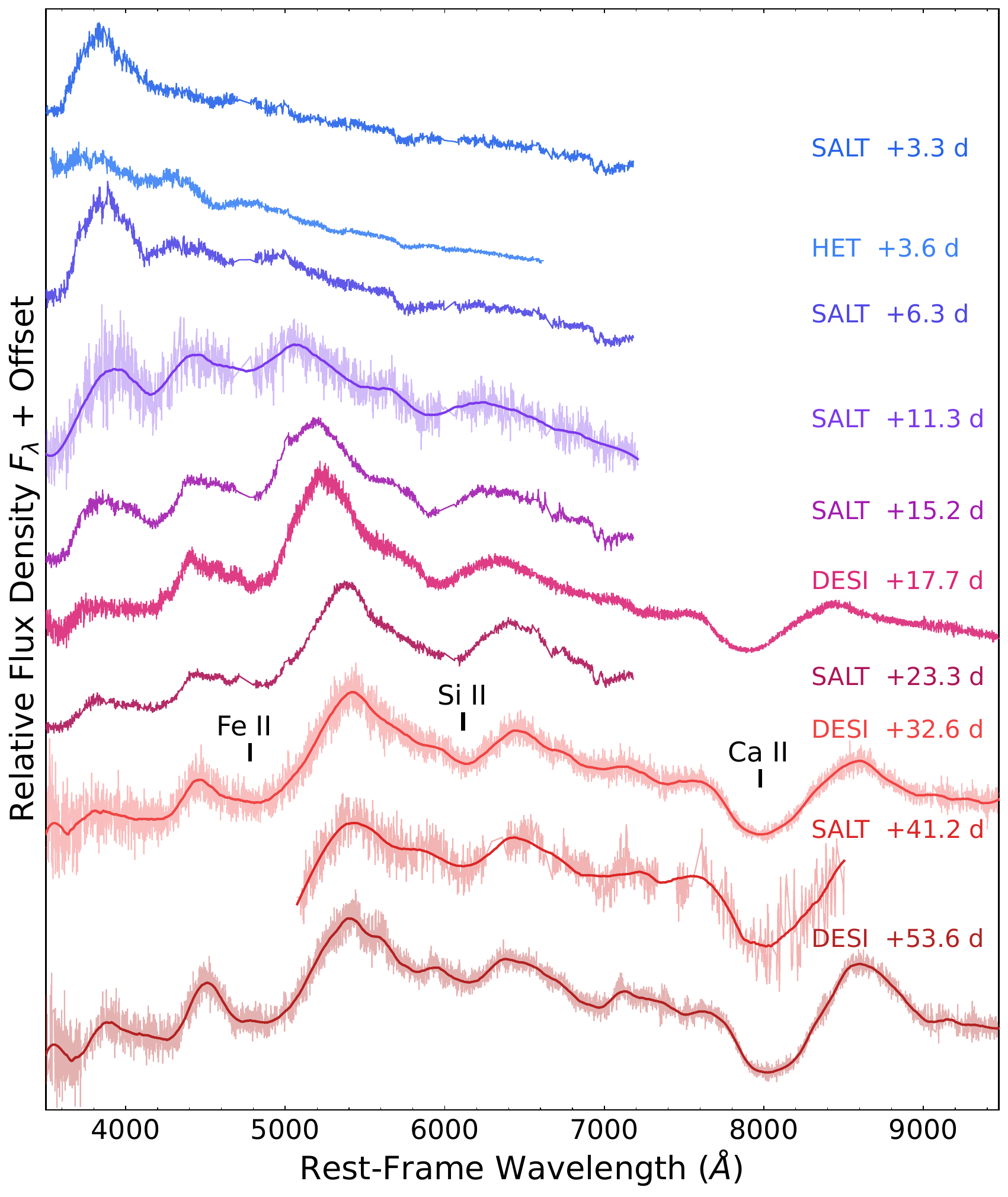}
    \caption{Spectral sequence of EP260321a/SN 2026gzf obtained with SALT, HET, and DESI between 3.3 and 53.6 d after the EP trigger. Blueshifted absorption features of Fe \textsc{ii}, Si \textsc{ii}, and Ca \textsc{ii} that are used for our analysis in \S \ref{sec:spec} and \ref{sec:speccompare} are marked for clarity. Emission lines have been clipped from the spectra for clarity. Some spectra are smoothed with a Savitzky-Golay filter (thick lines) for visualization purposes, and the unsmoothed spectra (thin lines) are also shown for completeness. The spectral shape of the SALT data obtained prior to 10 days is uncertain below 4200 \AA.
    }
    \label{fig:specsequence}
\end{figure*}

\begin{figure}
    \centering
    \includegraphics[width=\linewidth]{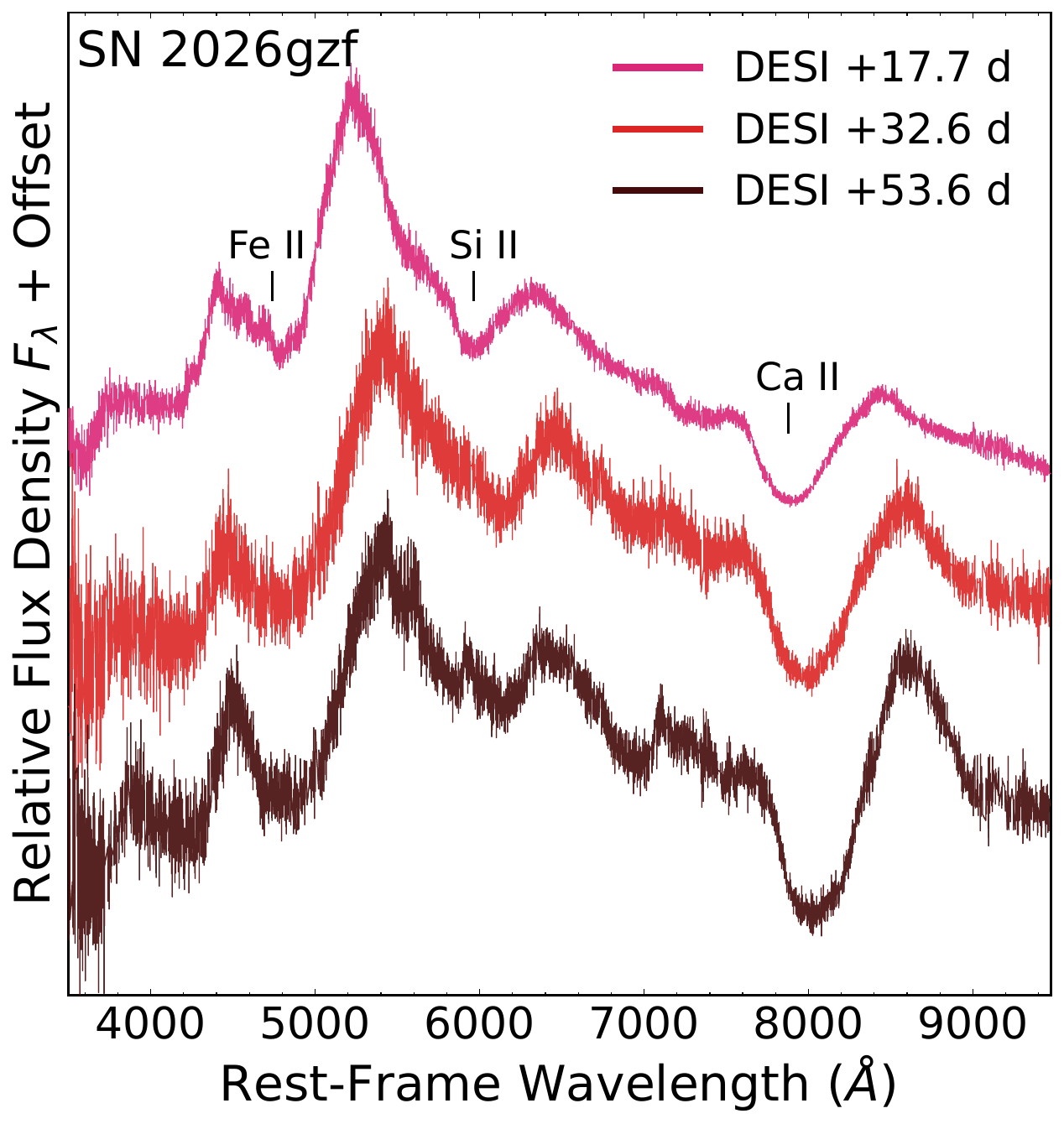}
    \caption{Evolution of the DESI spectra ($R \sim 2000 - 5500$) of SN 2026gzf obtained at 17.7, 32.6, and 53.6 d. Blueshifted absorption features of Fe \textsc{ii}, Si \textsc{ii}, and Ca \textsc{ii} that are used for our analysis in \S \ref{sec:spec} and \ref{sec:speccompare} are marked for clarity.
    Nebular emission lines have been clipped from each spectrum for visualization purposes. Spectra have not been smoothed and are in their native binning. 
    }
    \label{fig:speccompare-desi}
\end{figure}

\begin{figure}
    \centering
    \includegraphics[width=\linewidth]{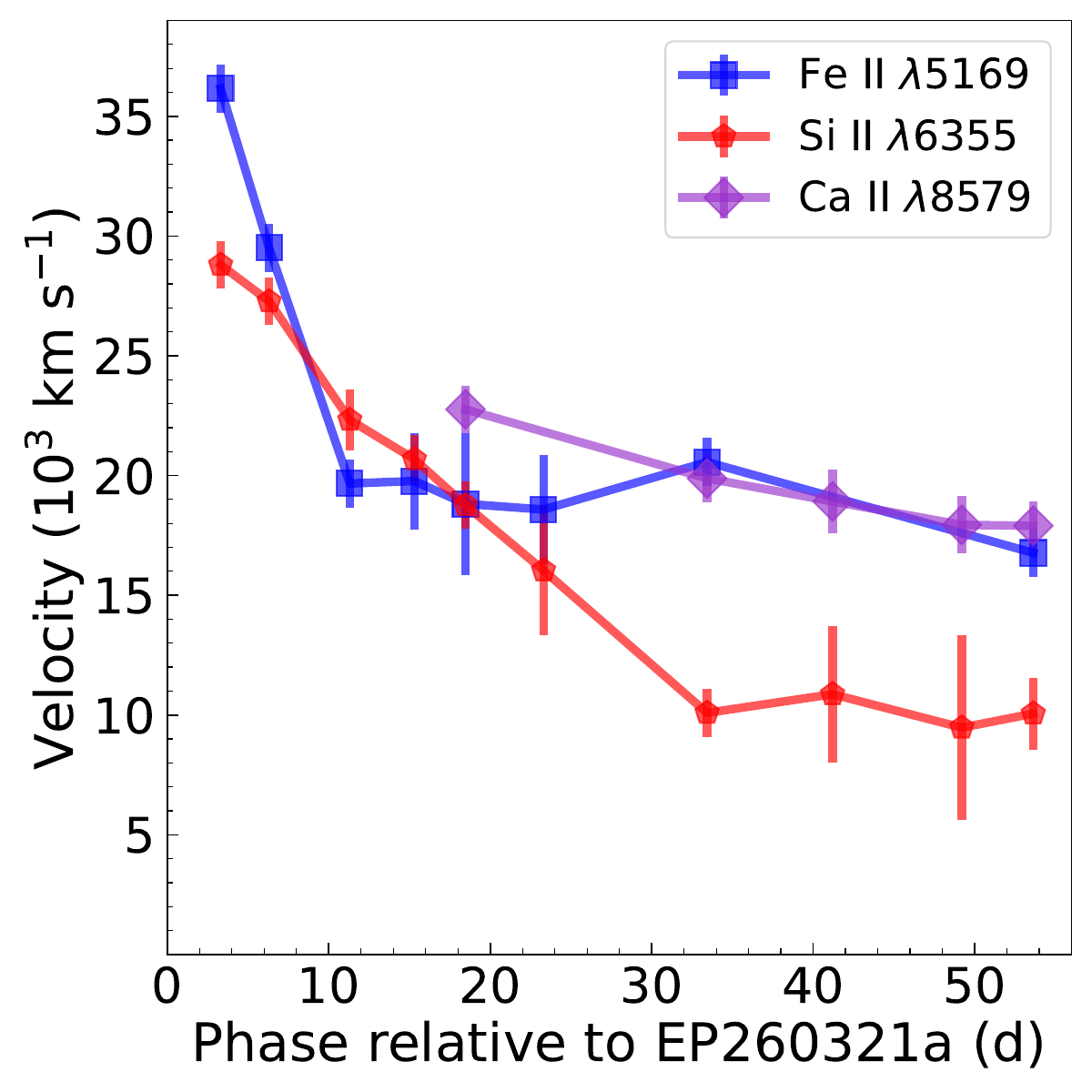}
    \caption{Evolution of the expansion velocity of different absorption features identified in the spectral sequence of SN 2026gzf (Figure \ref{fig:specsequence}).}
    \label{fig:vel}
\end{figure}

\subsection{Supernova Spectral Analysis}

\subsubsection{Spectroscopic Evolution and Velocity Estimates}
\label{sec:spec}

A series of optical spectra (Table \ref{tab:spectra}) were obtained with SALT, HET, and DESI starting $3$ d after discovery and extending to $55$ d. The  sequence of spectra is shown in Figure \ref{fig:specsequence}. The spectra show a clear evolution and display the broad absorption features expected in Ic-BL supernovae. This is consistent with the initial classification of SN 2026gzf as a broad-lined Ic supernova  \citep{2026GCN.44092....1X,2026GCN.44105....1C,2026GCN.44107....1R}. From nebular emission lines within the host galaxy (no absorption lines are detected), we derive a redshift of $z$\,$=$\,$0.0343588 \pm 0.0000015$ using the HET IFU dataset (see \S \ref{sec:hetenv}). The redshift is consistent to the 3rd decimal place across the entire host galaxy, including at the transient location. This is consistent with the redshift ($z$\,$=$\,$0.034328\pm0.000006$) 	 
of the host galaxy\footnote{We note that the DESI fiber placement in this observation is offset from the center of the host galaxy and located in the ``wings'' of its light profile due to an issue with the Legacy Survey source detection (Tractor catalog) that falsely identifies a source at that location.} inferred from archival DESI spectra (Target ID: 39627799320858502\footnote{\url{https://www.legacysurvey.org/viewer/desi-spectrum/dr1/targetid39627799320858502}}) available in Data Release 1\footnote{\url{https://vizier.cds.unistra.fr/viz-bin/VizieR?-source=V/161}} \citep{DESI-DR1}. In what follows, we refer to measured wavelengths in the rest-frame of SN 2026gzf. 

Our first spectra (Figure \ref{fig:specsequence}) were obtained at $\sim$\,$3$ d, roughly $\sim$\,12 days before peak optical light (Figure \ref{fig:lightcurve}). The early spectra show a  blue slope with absorption features at $\sim$\,$4600$ \AA\, and $5900$ \AA. We note that at early times ($<$\,$10$ d) the spectral shape of the SALT data are unreliable below 4200 \AA\, due to a lack of spectrophotometric standards. The HET data obtained at $3.6$ d has a trustworthy spectral shape over the full range as a spectrophotometric standard was observed near simultaneously, and we base our analysis of the $\sim$\,$3$ d spectra on this. In general, this has no impact on our conclusions or inferences. 

By 11.3 d, these early absorption features become clearer and evolve to be significantly deeper at $\sim$\,$4700$ \AA\, and $\sim$\,$6000$ \AA. While the 11.3 d spectrum is noisy due to being obtained in cloudy conditions, these features are clearly observed in all subsequent higher signal-to-noise (SNR) spectra, supporting their existence at 11.3 d. We associate these features around $4700$ \AA\, and $6000$ \AA\, with blueshifted Fe\,{\sc ii} $\lambda\lambda4924,5018,5169$ multiplet 42 (hereafter, Fe\,{\sc ii} $\lambda5169$) and Si\,{\sc ii} $\lambda6355$ absorption, respectively. We utilize these rest-frame wavelengths to derive the photospheric expansion velocity associated with these absorption features as a function of time following the methods outlined by \citet{Finneran2025}. For the Fe\,{\sc ii} $\lambda5169$ feature, we infer a high velocity of $\sim$\,$35,000$ km s$^{-1}$ at 3.6 d decreasing to $\sim$\,$20,000$ km s$^{-1}$ by 11.3 d and remaining relatively constant out to $55$ d. We infer similarly high velocities by associating the absorption feature at $\sim$\,$6000$ \AA\, with Si\,{\sc ii} $\lambda6355$. We observe this feature to continuously decline with time from $\sim$\,$29,000$ km s$^{-1}$ at 3.6 d to $\sim$\,$10,000$ km s$^{-1}$ at 32.6 d. We show the velocity evolution in Figure \ref{fig:vel}.

In DESI spectra obtained at 17.7, 32.6, and 53.6 d (Figures \ref{fig:specsequence} and \ref{fig:speccompare-desi}), spanning from shortly after optical maximum to several weeks after peak, we see  an additional absorption trough at $\sim$\,$8000$ \AA. We interpret this as the Ca\,{\sc ii} triplet consisting of lines at 8498 \AA, 8542 \AA, and 8662 \AA. Here, we adopt the $gf$-weighted centroid of 8579 \AA. We note a mean wavelength of 8567 \AA\, is also sometimes adopted in the literature, but this is less physically motivated than using the oscillator strength weighted value we adopt here. This Ca\,{\sc ii} feature shows a clear P Cygni profile, which becomes increasingly clear at later times, see the 53.6 d DESI spectrum in Figure \ref{fig:speccompare-desi}. 

As our earlier SALT spectra do not cover past $7200$ \AA\, we cannot probe the earlier velocity evolution of this feature, but we can constrain the velocity to be less than $45,000$ km s$^{-1}$ between 3.3 and 15.3 d. Based on the DESI spectra, we infer Ca\,{\sc ii} velocities of $\sim$\,$20,000$ km s$^{-1}$ at 17.7 d. The velocity stays relatively constant out to 55 d (Figure \ref{fig:vel}). In some Ic-BL this feature can become blended with O\,{\sc i} $\lambda7774$ or Mg\,{\sc ii} $\lambda\lambda7877,7896$ ($gf$-weighted value of $\lambda7889$). However, the observed feature lies redward of these lines and they are unlikely to contribute sufficiently. Instead, assuming the same blueshifted expansion velocity as inferred for Ca\,{\sc ii}, we find that O\,{\sc i} $\lambda7774$ and Mg\,{\sc ii} $\lambda\lambda7877,7896$ may be producing a lower significance broad absorption feature near $7200$\,$-$\,$7400$ \AA\,, as seen in the DESI spectra (Figure \ref{fig:specsequence}). 

We also note that in our final DESI spectrum at 53.6 d, additional spectral features, possibly in emission, appear to be developing at $\sim$\,$5900$ \AA\, and $\sim$\,$7100$ \AA, see Figure \ref{fig:speccompare-desi}. The latter feature has a broader visible rise in the continuum over $\sim$\,$7100$ to $7400$ \AA. The exact identification of these features is uncertain \citep[see, e.g.,][]{Mazzali2001,Maeda2006} and additional spectra are required to confirm their significance and evolution. 


\subsubsection{Comparison to Other Ic-BL Spectra and Velocities}
\label{sec:speccompare}

\begin{figure}
    \centering
    \includegraphics[width=\linewidth]{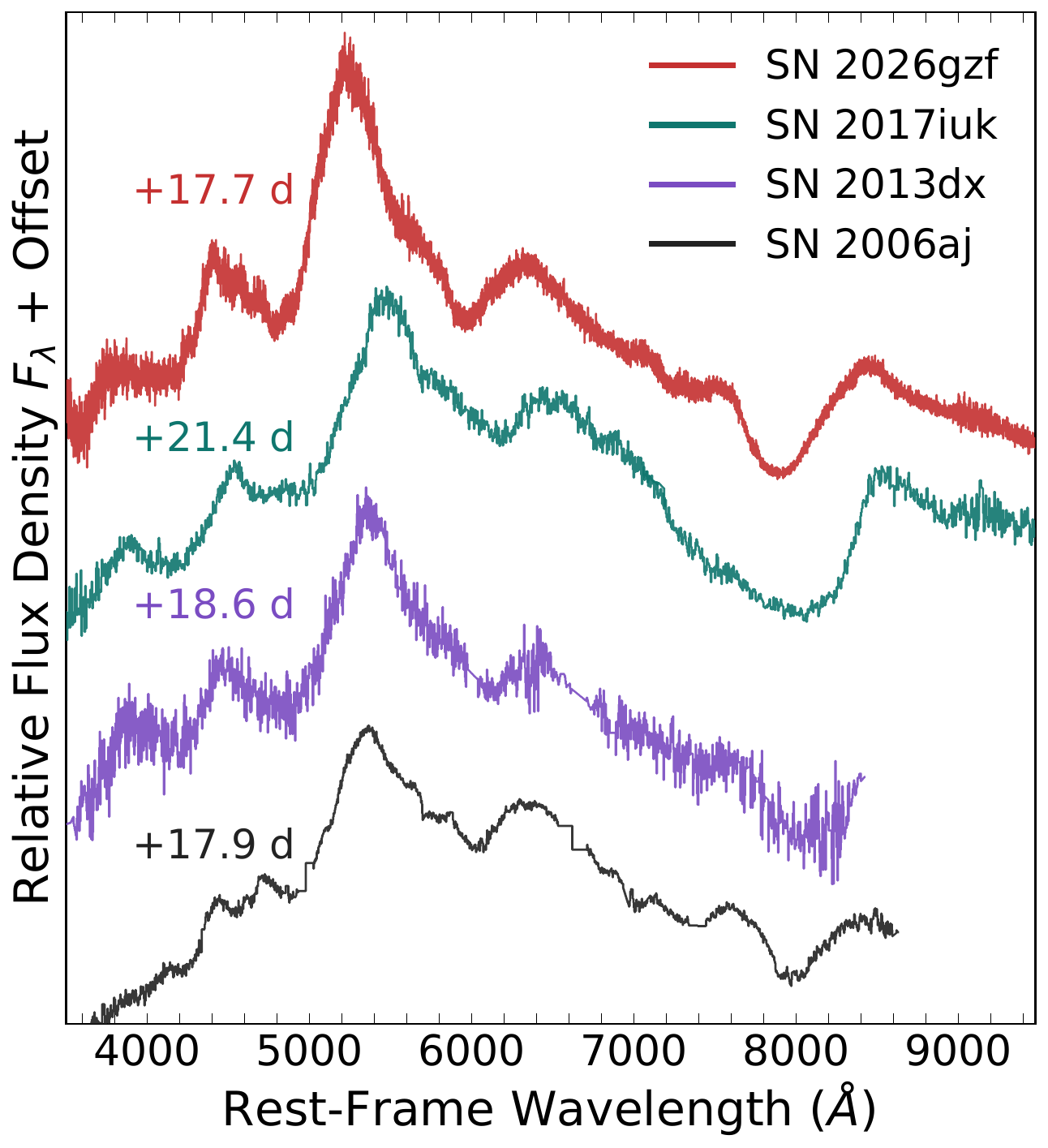}
    \caption{Comparison of the DESI spectra of SN 2026gzf obtained at 17.7 d after discovery versus other Ic-BL supernovae at a similar phase. Nebular emission lines have been clipped from each spectrum for visualization purposes. Spectra have not been smoothed and are in their native binning.}
    \label{fig:speccompare}
\end{figure}

In Figure \ref{fig:speccompare}, we compare the DESI spectra obtained at 17.7 d (near peak optical light) to other Type Ic-BL supernovae. We find a good match to high velocity GRB-SN such as GRB 060218/SN 2006aj \citep{Campana060218,Pian2006,Soderberg2006grb060218,Sollerman2006,Ferrero2006,Mazzali2006}, GRB 130702A/SN 2013dx \citep{DElia2015,Toy2016,Volnova2017,Mazzali2021}, and GRB 171205A/SN 2017iuk \citep{DElia2018,Wang2018sn,Izzo2019} at similar phases. While the exact velocity of each SN differs, similar features clearly exist in each spectrum and provide a definitive classification of SN 2026gzf as a broad-lined Ic supernova.

\begin{figure*}
    \centering
    \includegraphics[width=\linewidth]{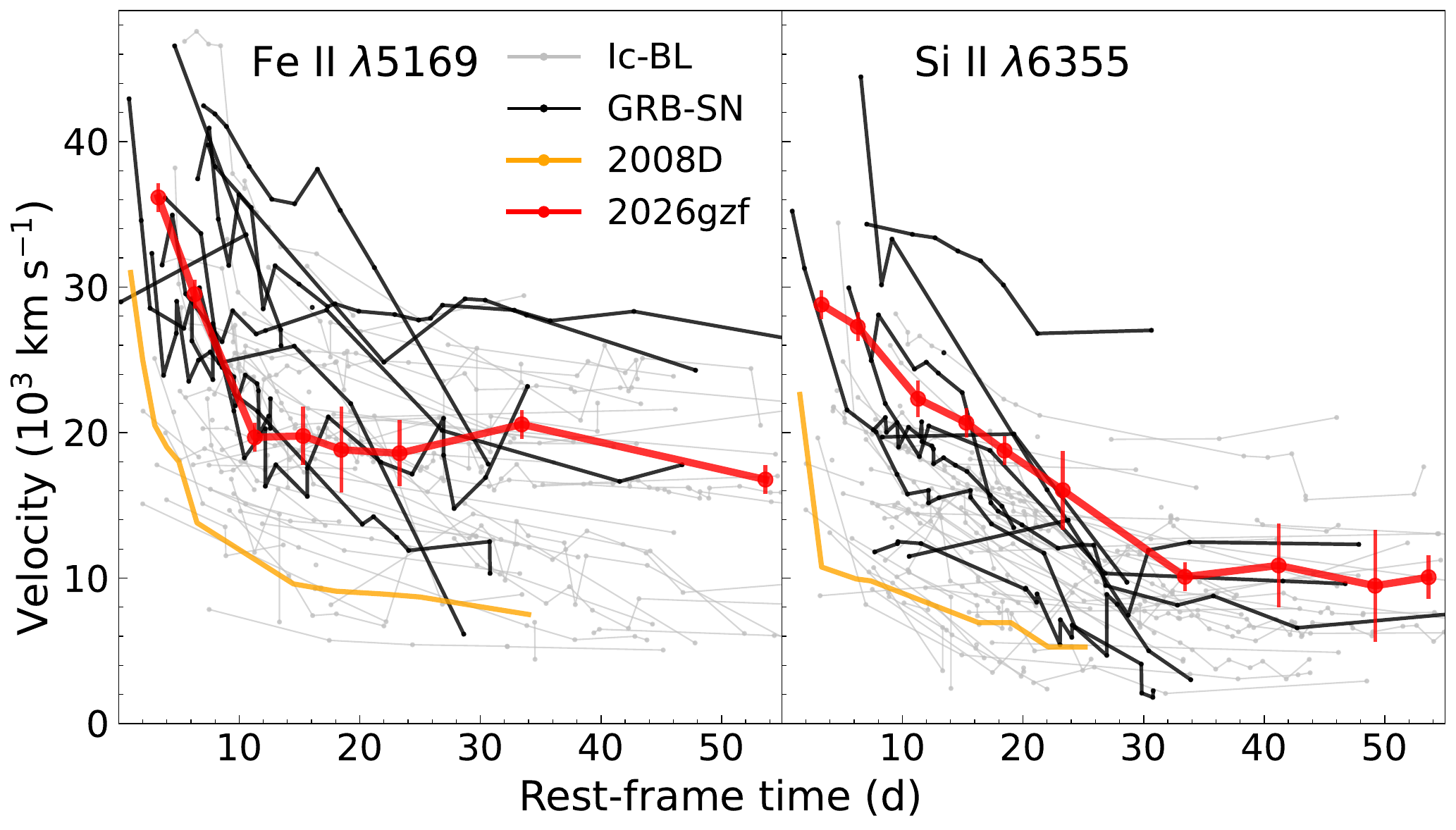}
    \caption{Velocity evolution of Ic-BL SNe (gray) and Ic-BL GRB-SNe (black) from \citet{Finneran2025} to SN 2026gzf (red). The left panel shows the evolution of Fe\,{\sc ii} $\lambda5169$ and the right panel shows Si\,{\sc ii} $\lambda6355$. Velocities for SN 2008D are reproduced from \citet{Mazzali2008,Malesani2009}.
    }
    \label{fig:velcompare}
\end{figure*}

We next compare the inferred Fe\,{\sc ii} and Si\,{\sc ii} velocities and their evolution to a sample of GRB-SNe and non-GRB associated Type Ic-BL SNe from \citet{Finneran2025}. In Figure \ref{fig:velcompare} we show the velocity versus time for both features. There is good agreement between our inferences and other Ic-BL supernovae of both classes. The initially high velocities ($>$\,$30,000$ km s$^{-1}$) for both features have better agreement with GRB-SN, which tend to have higher velocities than non-GRB Ic-BL SNe \citep[see][]{Modjaz2016,Finneran2025}. While SN 2026gzf is not associated with a GRB, and shows instead a soft thermal blackbody prompt emission spectrum \citep{2026GCN.44075....1H}, the existence of high velocities and a luminous peak absolute magnitude (Figure \ref{fig:compare}) may indicate that a central-engine helped drive more energy into the supernova ejecta than in non-GRB/non-FXT associated, optically selected Ic-BL SNe. We discuss this further below. 

From their compiled sample, \citet{Finneran2025} derive median velocities of $\sim$\,$21,000$ km s$^{-1}$ from Fe\,{\sc ii} and $\sim$\,$17,000$ km s$^{-1}$ Si\,{\sc ii} at 15 d. This is in good agreement with our results for SN 2026gzf (Figure \ref{fig:velcompare}). It is not uncommon to infer larger velocities (by up to even $\Delta v$\,$\approx$\,$10,000$\,$-$\,$20,000$ km s$^{-1}$; see Fig. 25 of  \citealt{Finneran2025}) from Fe\,{\sc ii} than from Si\,{\sc ii} and the velocity evolution can have a different temporal dependence, as seen also for SN 2026gzf.

\subsection{Spatially Resolved Analysis of the Environment}
\label{sec:hetenv}

We extracted spectra from individual HET LRS2 spaxels at selected positions across the IFU field in order to investigate both the transient environment and the properties of the host galaxy. The data cube was resampled to a spaxel scale of $0.4\arcsec \times 0.4\arcsec$, corresponding to a projected physical scale of approximately $280$ pc $\times$ $280$ pc at the redshift of the host. We note that the transient is located $\sim$\,$3.4\arcsec$ from the center of the host galaxy (Figure \ref{fig:FC}), which corresponds to roughly $2.5$ kpc.

We measured the redshift in multiple locations employing the brightest detected emission lines across the galaxy, in particular H$\beta$, H$\alpha$ and the [O \textsc{iii}]$\lambda\lambda4959,5007$ doublet, which are consistently detected across the IFU field. From these measurements, we obtain a weighted mean redshift of $z = 0.0343588 \pm 0.0000015$, with the weights determined from the signal-to-noise ratio (S/N) of each extracted spectrum. The measured redshift is consistent, within the uncertainties, across the full spatial extent of the galaxy. 

\begin{figure}
    \centering
    \includegraphics[width=\linewidth]{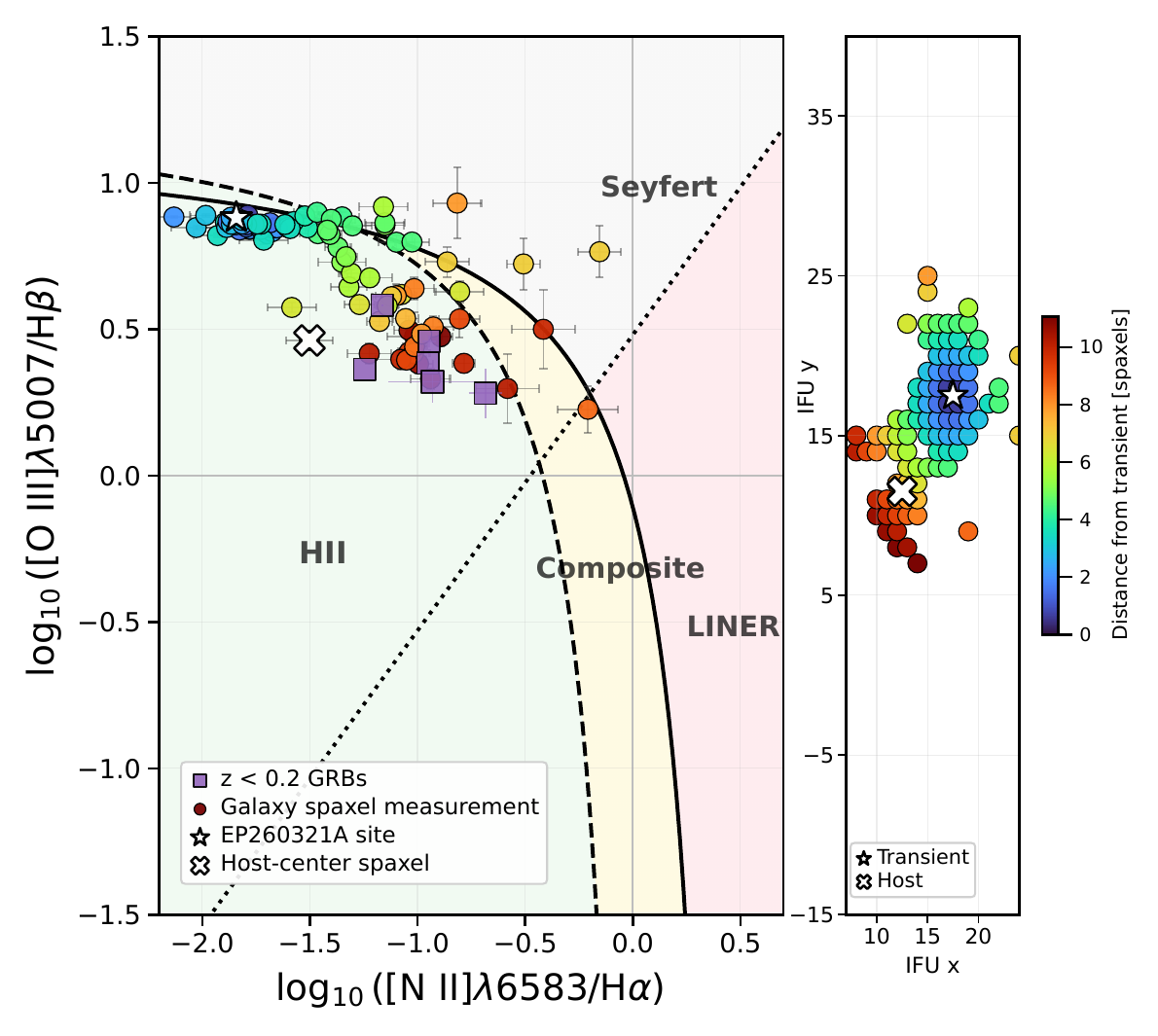}
    \caption{Spatially resolved [N \textsc{ii}]-BPT diagram from HET IFU spectroscopy. Each point represents an individual galaxy spaxel, with emission-line ratios following the BPT diagnostics of \cite{baldwin1981}. Points are color-coded by projected distance from the transient position, as shown in the IFU field map on the right with units of the spaxel ($0.4\arcsec\times 0.4\arcsec$) grid. The dashed, solid and dotted curves indicate the \cite{kauffmann2003} star-forming/composite boundary, the \cite{kewley2001} maximum-starburst boundary, and the \cite{kewley2006} Seyfert/LINER division, respectively. The shaded regions show the corresponding H \textsc{ii}-like region, composite, Seyfert, and LINER-like classifications. The transient site is marked by a white star, while the location of the galaxy center is shown with a white cross. GRB explosion site measurements at $z$\,$<$\,$0.2$ are shown as purple squares \citep{Christensen2008,Han2010,Levesque2011,Levesque2012,Thone2014,Izzo2017,Thone2024}.}
    \label{fig:het_space}
\end{figure}

Additionally, for each spaxel we measured the narrow emission-line fluxes by fitting the H$\beta$, [O \textsc{iii}]$\lambda5007$, H$\alpha$ and [N \textsc{ii}]$\lambda6583$ lines with Gaussian profiles. These fluxes were used to construct a [N \textsc{ii}]-based Baldwin, Phillips \& Terlevich (BPT) diagram \citep{baldwin1981} as described in Figure \ref{fig:het_space}. Uncertainties on the measured fluxes were estimated from the continuum noise in the vicinity of each fitted line. The corresponding line ratio uncertainties were then propagated from the measured flux uncertainties in logarithmic space. To ensure a robust classification, we retained only spaxels for which all relevant diagnostic lines were securely detected, requiring a minimum S/N of at least 3. 

The resulting spatially resolved [N \textsc{ii}] BPT diagram shows that the vast majority of spaxels lie within the H \textsc{ii} region locus. An [O \textsc{i}]-based BPT diagram provides a consistent conclusion. Figure \ref{fig:het_space} indicates that the ionization across the region surrounding the transient, and the host galaxy, is dominated by star formation. We further observe that the transient location differs from the rest of the host galaxy in terms of its location in the BPT diagram. The explosion site of EP260321a/SN 2026gzf has a substantially lower value of the ratio N2$\equiv\log\big([$N\,{\sc ii}$]/$H$\alpha\big)$, suggesting a metallicity gradient from the explosion site to the galaxy center. In particular, at the supernova location we measure $\log\big([$O\,{\sc iii}$]/$H$\beta\big)$\,$=$\,$0.87\pm0.01$ and N2\,$=$\,$-1.84 \pm 0.03$, placing it in the upper left region of the H \textsc{ii} region in Figure \ref{fig:het_space}. 

We additionally derive O3N2$\equiv$  $\log\big(([{\rm O\,III}]\,5007/H\beta)/([{\rm N\,II}]\,6583/H\alpha)\big)$ $=$\,$2.72\pm0.03$. 
We use the N2 and O3N2 ratios to derive the metallicity at the explosion site using the calibrations from \citet{PettiniPagel2004} and \citet{Marino2013}. This yields a consistently low metallicity inferred from the oxygen abundance  12 + log(O/H) $\approx$\,$7.85$\,$-$\,$7.95$. 
This corresponds to an extremely subsolar metallicity of $0.15\,Z_\odot$ to  $0.2\,Z_\odot$ \citep{AllendePrieto2001,Asplund2004,Asplund2009}, and more comparable to the SMC inferred values \citep{Russell1990}. 
We note that the statistical uncertainty ($\pm0.02$) on the oxygen abundance is significantly smaller than the systematic uncertainty ($0.1-0.2$ dex) on these calibrations \citep{PettiniPagel2004,Marino2013}. We note our inferred values for O3N2 and N2 lie outside the range of values used to define these oxygen abundance calibrations, and should be interpreted as generally indicative of an extremely low metallicity environment. As we derive consistent values across multiple diagnostic ratios and multiple calibrations, we consider this a robust interpretation of the data. 

While we observe a clear trend in the BPT diagram (Figure \ref{fig:het_space}) of decreasing N2, and therefore decreasing metallicity, moving from the transient location to the rest of the host galaxy, the metallicity of the galaxy is still extremely sub-solar, lying at a similar location to other GRB explosion sites \citep{Christensen2008,Han2010,Levesque2011,Levesque2012,Thone2014,Izzo2017,Thone2024}. Using the spectra extracted from the spaxel covering the center of the host galaxy, we derive N2\,$=$\,$-1.18\pm0.06$ and O3N2\,$=$\,$1.70\pm0.06$, yielding 12 + log(O/H) $=$\,$8.20\pm0.04$, otherwise denoted as $0.3\,Z_\odot$ \citep{Asplund2009}. The derived value agrees between the different calibrations \citep{PettiniPagel2004,Marino2013}. 

Additionally, by extracting spectra from within a 3 spaxel radius of the center of the transient and host galaxy, we derive H$\alpha$ inferred \citep{Kennicutt1998} star formation rates (SFRs) of $0.20\pm0.01\, M_\odot$ yr$^{-1}$  and $0.07\pm0.01\, M_\odot$ yr$^{-1}$ for the transient explosion site and host galaxy, respectively. In this analysis, we adopted a Chabrier \citep{Chabrier2003} initial mass function (IMF). 

\begin{figure}
    \centering
    \includegraphics[width=\linewidth]{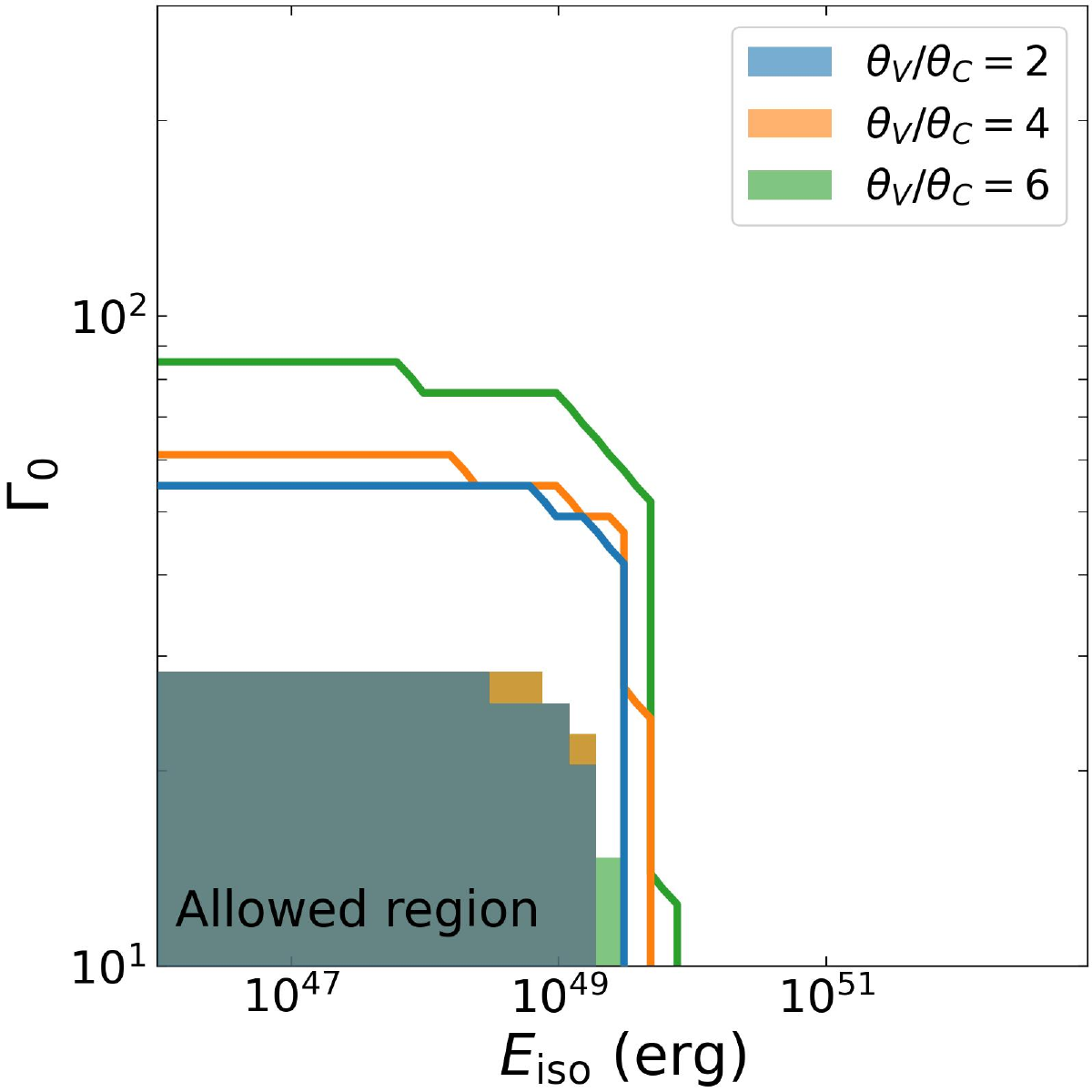}
    \caption{Allowed parameter space (shaded regions) of initial Lorentz factor $\Gamma_0$ at the jet's core and the jet's core kinetic energy $E_\textrm{kin}$ for afterglow non-detection, assuming a Gaussian structured jet at 158 Mpc ($z$\,$=$\,$0.0344$) based on the available X-ray and radio upper limits. In a wind environment, the allowed parameter space is weakly dependent on the viewing angle $\theta_\textrm{v}/\theta_\textrm{c}$. We have fixed $\theta_\textrm{c}$\,$=$\,$0.15$ rad, $A_\ast$\,$=$\,$1$, $\varepsilon_\textrm{e}$\,$=$\,$0.1$, $\varepsilon_\textrm{B}$\,$=$\,$0.01$, and $p$\,$=$\,$2.2$. As an example of the impact of decreasing density, we show the allowed parameter space for $A_\ast$\,$=$\,$0.1$ as solid lines (see also Figure \ref{fig:afterglowconst2} in Appendix \ref{sec:aftappendix}).
}
    \label{fig:afterglowconst1}
\end{figure}

\subsection{Afterglow Constraints}
\label{sec:outflowconstraints}

Here we compute the allowed parameters for non-detection of a synchrotron afterglow \citep{Sari1998,Granot2002} using the \texttt{VegasAfterglow} package \citep{VegasAfterglow}. We model the afterglow with a Gaussian structured jet propagating into a wind environment characterized by $\rho(r)$\,$\propto$\,$r^{-2}$. The physical setup is specified by eight parameters: the isotropic-equivalent kinetic energy at the jet's core $E_\textrm{kin}$, the Lorentz factor at the jet's core $\Gamma_0$, the jet's core half-opening angle $\theta_\textrm{c}$, the observer's viewing angle $\theta_\textrm{v}$, the density $A_*$, the magnetic and electron energy fractions $\varepsilon_\textrm{B}$ and $\varepsilon_\textrm{e}$, and the electron powerlaw index $p$. 

Following \citet{OConnor25ulz}, we generate afterglow models across a broad grid of parameters and compare the flux density at each time and frequency to our X-ray upper limits. We also include multi-frequency radio upper limits from the VLA \citep[This work, \S \ref{sec:vla}; and][]{2026GCN.44229....1L} and require that the early X-ray afterglow is fainter than the initial EP detections \citep{2026GCN.44075....1H}. 

Over a dense scan spanning $p=2.2-2.8$, $A_\ast=10^{-4}-10^{2}$, $E_{\rm kin}=10^{48}-10^{52}$ erg, $\theta_c=0.05-0.25$, $\theta_{\rm v}/\theta_{\rm c}=0-6$, and $\epsilon_B=10^{-4}-10^{-1}$ with fixed $\Gamma_0$\,$=$\,$100$, only $\sim 1\%$ of models remain allowed by the X-ray and radio upper limits. 
More specifically, wind models with $\theta_{\rm v}/\theta_{\rm c} \lesssim 3$ are almost entirely excluded, and even near $\theta_{\rm v}/\theta_{\rm c} \sim 3$ only a very small region at low $A_\ast$ and low $\epsilon_B$ survives. We note that excluding the requirement that the X-ray afterglow is fainter than the initial EP detections increases the allowed models over this grid to 28\%. However, the lack of an X-ray afterglow at early times and lack of a powerlaw X-ray spectrum are confirmed by the early EP/FXT observations reported by \citet{Yuan2026}.

The allowed region shrinks rapidly as either $A_\ast$ or $\epsilon_B$ increases. For example, in the fixed slice with $\theta_{\rm c}=0.15$ and $\epsilon_B=10^{-2}$, the entire explored wind parameter space for $A_\ast=0.1-10$ and $\theta_{\rm v}/\theta_{\rm c}=0,2,4,6$ is excluded. For densities below $A_*$\,$\lesssim$\,$0.1$, the allowed parameter space rapidly expands (see Figure \ref{fig:afterglowconst1} and Figure \ref{fig:afterglowconst2} in Appendix \ref{sec:aftappendix}).  However, given the pre-explosion activity of the transient (Appendix \ref{sec:decamarch}; \citealt{Yuan2026,Chen2026}), which is likely due to instabilities and eruptive mass-loss, we find it unlikely that the surrounding environment is pristine, and instead expect a higher density.

As such, for fixed $A_*$\,$=$\,$1$, $\theta_\textrm{c}$\,$=$\,$0.15$ rad, $\varepsilon_\textrm{e}$\,$=$\,$0.1$, $\varepsilon_\textrm{B}$\,$=$\,$0.01$, and $p$\,$=$\,$2.2$, we compute the allowed Lorentz factor and isotropic-equivalent kinetic energy at the jet's core. The results are shown in Figure \ref{fig:afterglowconst1}.  
In a wind environment, the entire surface of the jet decelerates rapidly at small radii (due to the higher density closer to the star), which reduces the dependence of the flux on viewing angle compared with an uniform density medium. Therefore, these results are very weakly dependent on the viewing angle. 
For these assumed parameters, most specifically the density (see Figure \ref{fig:afterglowconst2} in Appendix \ref{sec:aftappendix}), we find that only low Lorentz factors $\Gamma_0$\,$\lesssim$\,$30$ and low kinetic energies $E_\textrm{kin}$\,$\lesssim$\,$10^{49}$ erg are allowed for $A_*$\,$\gtrsim$\,$1$. If we relax the density to $A_*$\,$=$\,$0.1$ (see Figure \ref{fig:afterglowconst1}), then the requirement decreases to $\Gamma_0$\,$\lesssim$\,$100$ and $E_\textrm{kin}$\,$\lesssim$\,$10^{50}$ erg. This implies that if the stellar explosion drove a successful relativistic jet into the surrounding stellar wind, it would have to be extremely weak and mildly relativistic if the environment was typical of those observed for long GRBs. However, if the surrounding density is lower by $>10\times$ than this assumption (i.e., if $A_*$\,$\lesssim$\,$0.1$), then we cannot rule out an ultrarelativistic jet ($\Gamma_0$\,$>$\,$100$) with energies $E_\textrm{kin}$\,$\lesssim$\,$10^{50}$, significantly reducing the required parameter space, especially for the Lorentz factor. For additional analyses of the presence of a relativistic outflow, see \citet{Yuan2026,MartinCarrillo2026,Rastinejad2026}.

\begin{figure}
    \centering
    \includegraphics[width=\linewidth]{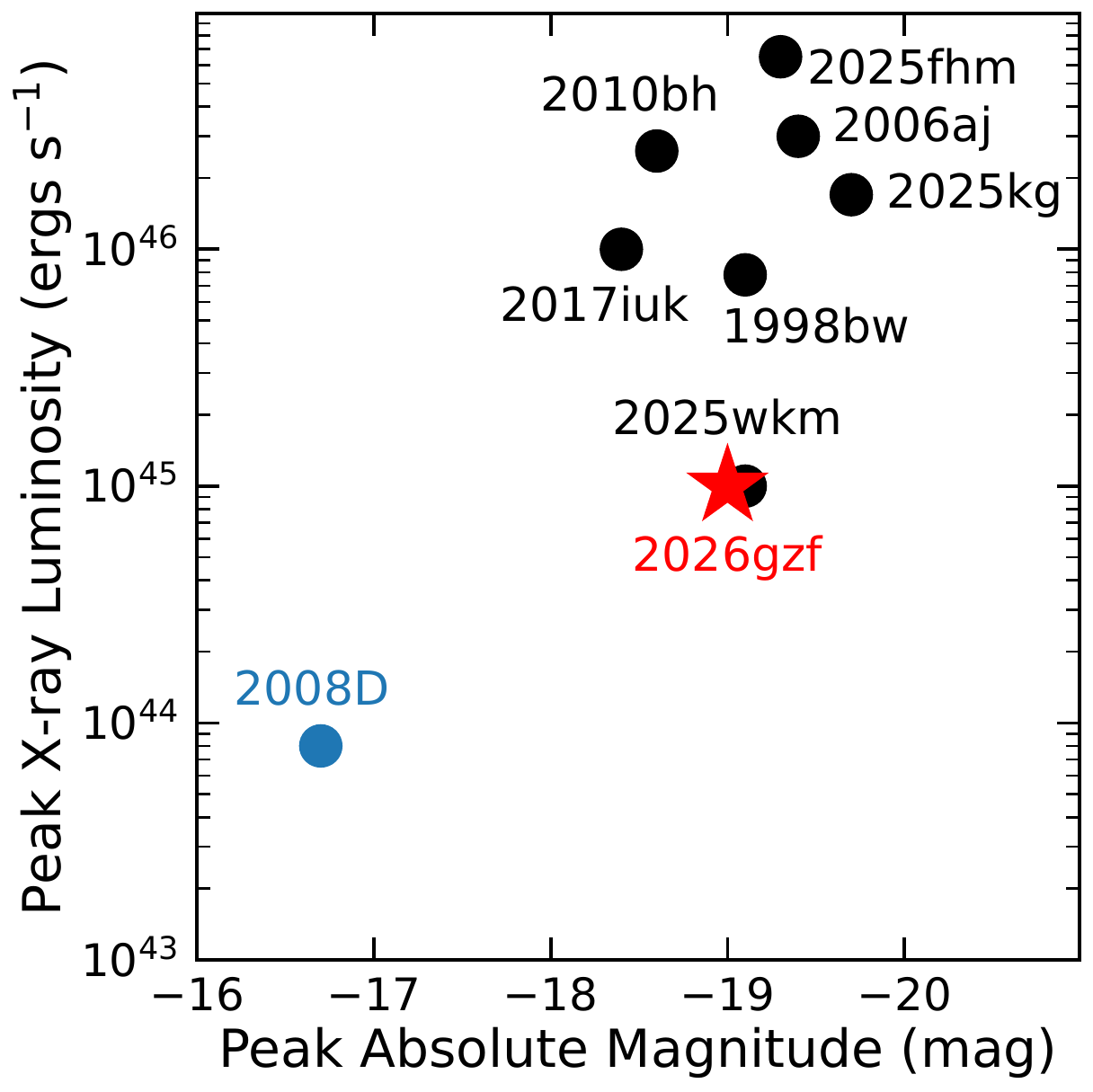}
    \caption{X-ray luminosity versus peak absolute magnitude of supernova shock breakout candidates detected by \textit{Swift} (GRB 980425/SN 1998bw; GRB 060218/SN 2006aj; XRF 080109/SN 2008D; GRB 100316D/SN 2010bh; GRB 171205A/SN 2017iuk) and EP (EP 250108a/SN 2025kg; EP250827b/SN 2025wkm). EP260321a/SN 2026gzf is shown as a red star \citep{Yuan2026}. SN Ic-BL are shown as black circles. SN 2008D, a SN Ib, is shown as a blue circle. Data have been taken from \citet{Galama1998,Soderberg2006grb060218,Soderberg2008,Starling2011,Cano2011,Izzo2019,Li2025,Rastinejad2025EP,Srinivasaragavan2025EP0108a,Srinivasaragavan2026,Cotter2026}. This figure is reproduced from \citet{Rastinejad2025EP} with the addition of newer EP events (SNe 2025wkm and 2026gzf). Here, we have utilized the dust corrected absolute magnitudes of SN 2008D and SN 2010bh. 
    }
    \label{fig:xraylum}
\end{figure}

\begin{figure}
    \centering
    \includegraphics[width=\linewidth]{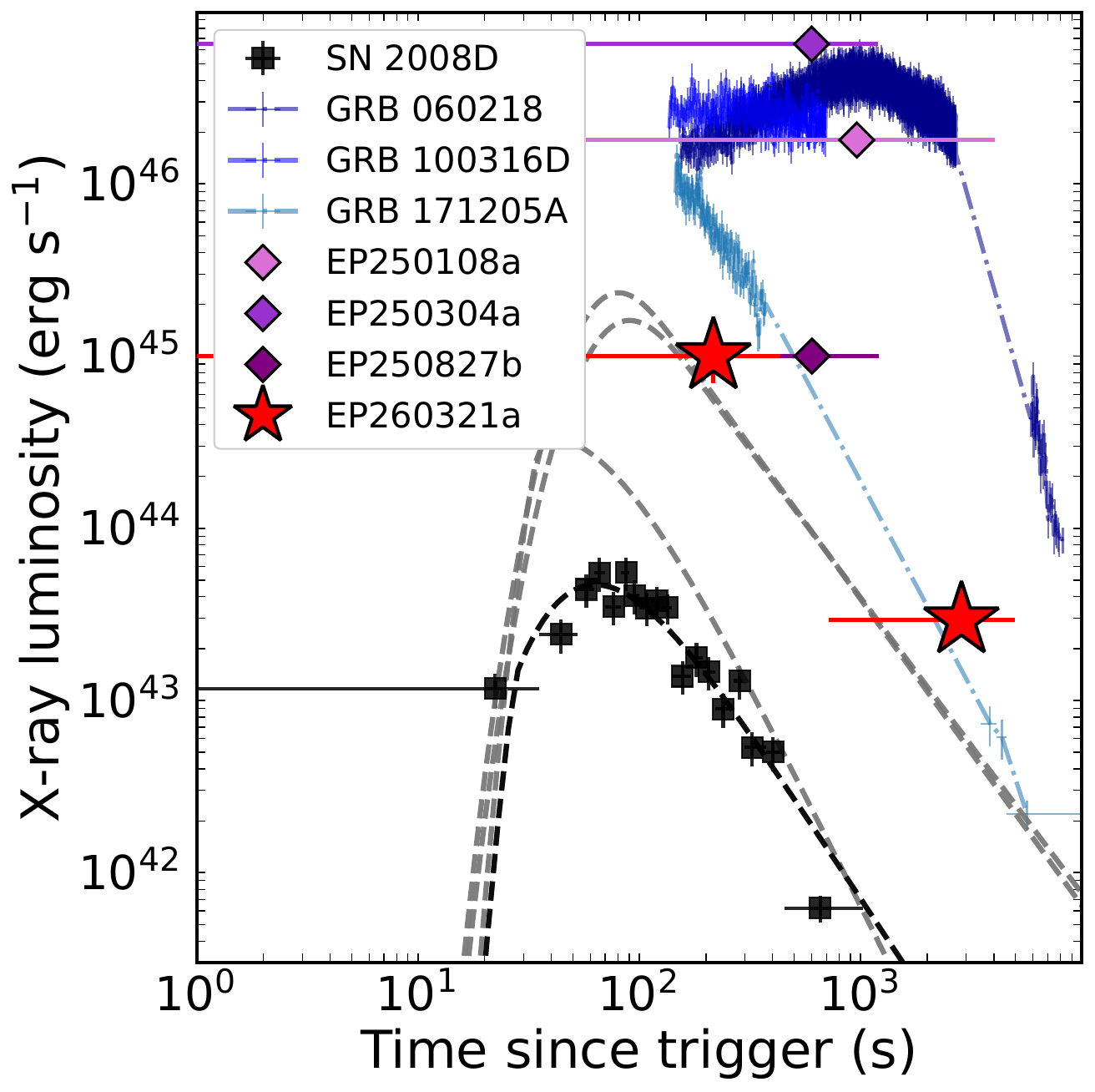}
    \caption{X-ray lightcurves of shock breakout candidates detected by \textit{Swift} and EP. The EP events are based on time-averaged spectra, and the initial datapoint for each event is in the $0.5$\,$-$\,$4$ keV band while all other points (and all \textit{Swift} events) are in the $0.3$\,$-$\,$10$ keV band. However, the EP events display soft spectra and are unlikely to be very different in a broader energy band. Data have been taken from \citet{Soderberg2008,Sun2024,Li2025,Srinivasaragavan2026,Yuan2026}. Dashed lines represent a compilation of Ib and Ic-BL shock breakout models from \citet{Fryer2026}. We note that delay in the start of the GRB lightcurves is due to the slew time of \textit{Swift} following the gamma-ray triggers.
    }
    \label{fig:xraybreakoutlum}
\end{figure}

\begin{figure*}
    \centering
    \includegraphics[width=\linewidth]{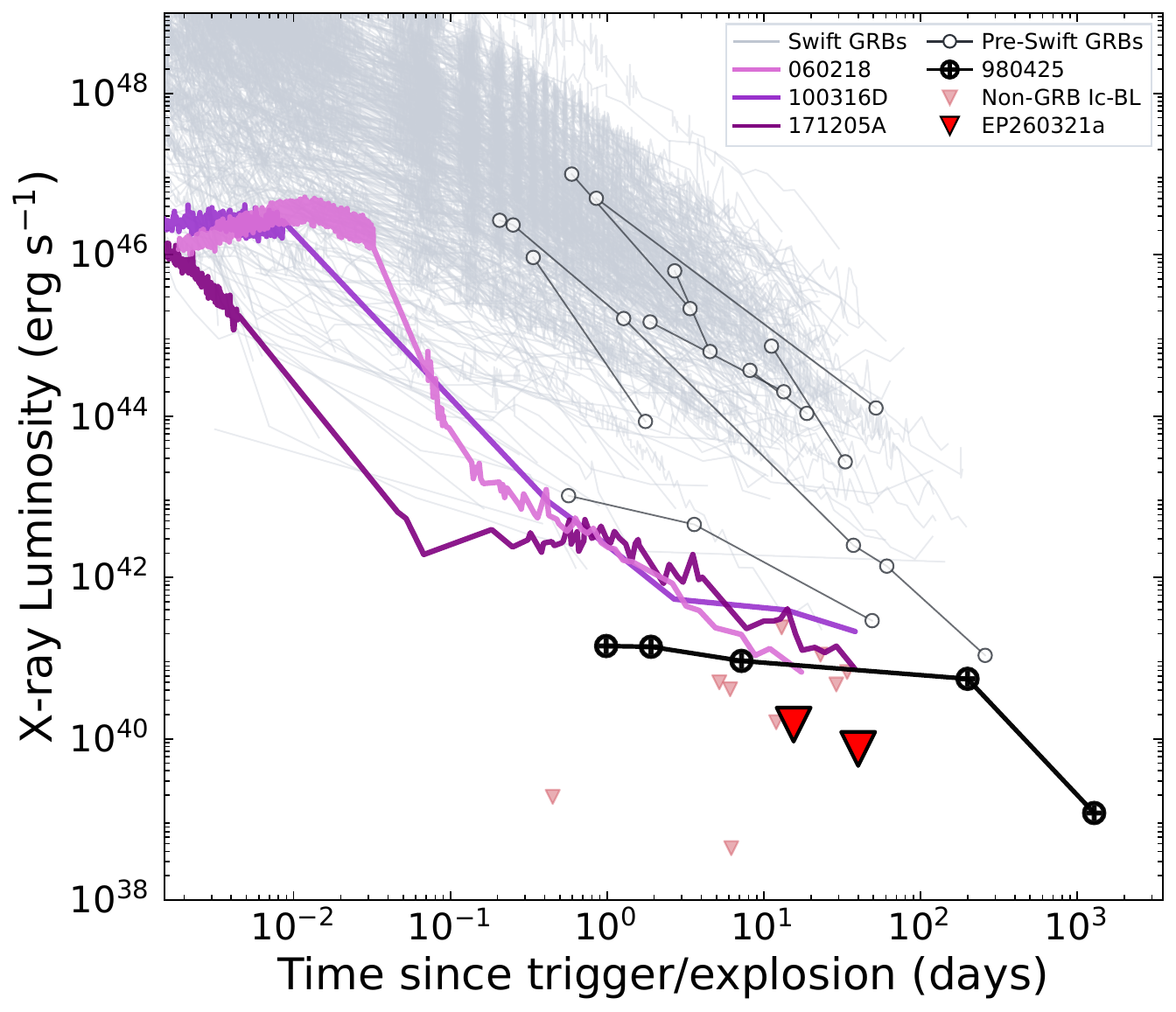}
    \caption{Observer-frame X-ray afterglow ($0.3$\,$-$\,$10$ keV) lightcurves of gamma-ray bursts and fast X-ray transients. The X-ray upper limits from \textit{Chandra} for EP260321a are shown as downward red triangles. For comparison, we show both long and short GRBs (gray) from \textit{Swift} \citep{Evans2007,Evans2009}, and specifically highlight a sample of low-luminosity GRBs. Additional X-ray data for GRBs 060218 and 100316D are compiled from \citet{Soderberg2006grb060218} and \citet{Margutti2013-100316D}. Pre-\textit{Swift} GRB and XRF X-ray afterglows are reproduced from \citet{Pian1999,Pian2000,Pian2004,Kouveliotou2004,Tiengo2004,Watson2004} and references therein. X-ray upper limits for non-GRB associated Ic-BL supernovae (orange downward triangles) were obtained from \citet{ODwyer2025}.}
    \label{fig:xray}
\end{figure*}

\section{Discussion}
\label{sec:discuss}

\subsection{Origin of the X-ray Emission}


Despite their theoretical ubiquity, detections of the onset of supernova shock breakouts are extremely rare and notoriously difficult to catch due to their faint nature, short duration, and largely soft X-ray or ultraviolet emission that lies in a little probed region of parameter space. At high-energy wavelengths, the most convincing example of shock breakout is associated with the Ib SN 2008D \citep{Soderberg2008,Mazzali2008,Malesani2009}. Additional, generally accepted, examples of relativistic shock breakout are associated with low-luminosity GRBs, such as GRBs 060218 ($z$\,$=$\,$0.0331$), 100316D ($z$\,$=$\,$0.0591$), and 171205A ($z$\,$=$\,$0.0368$), see, e.g., \citet{Campana2006,Ghisellini2006,Starling2011,DElia2018,Izzo2019}. However, in those cases there is theoretical disagreement over  the exact origin of the prompt gamma-ray and X-ray emission in terms of relative contributions from the jet and breakout emission \citep[e.g.,][]{Waxman2007,Irwin2016,Irwin2025-1}. Each of these low-luminosity GRB/shock breakout candidates have been associated with relativistic outflows and Ic-BL supernovae, while SN 2008D displayed at most transrelativistic ejecta \citep{Soderberg2008} and was associated with a Ib supernova.

The prompt high-energy emission of EP260321a is unusual in the context of the known population of nearby broad-lined Type Ic supernovae with X-ray or gamma-ray triggers (Figures \ref{fig:xraylum} and \ref{fig:xraybreakoutlum}). The EP/WXT spectrum was well described by a soft thermal component with $kT=132^{+30}_{-21}$ eV and a time-averaged unabsorbed luminosity of $L_{\rm X}=(1.0\pm0.3)\times10^{45}$ erg s$^{-1}$ \citep{Yuan2026}. This luminosity is 10$\times$ fainter than the peak X-ray luminosities of the nearby low-luminosity GRBs GRB 060218/SN 2006aj \citep{Campana060218,Soderberg2006grb060218} and GRB 100316D/SN 2010bh \citep{Chornock2010,Starling2011,Cano2011} and the recent EP Type Ic-BL shock breakout candidates EP250108a/SN 2025kg \citep[$z$\,$=$\,$0.176$;][]{Li2025,Srinivasaragavan2025EP0108a,Eyles-Ferris2025EP,Rastinejad2025EP} and EP250304a/SN 2025fhm \citep[$z$\,$=$\,$0.2$;][]{Cotter2026}, see Figure \ref{fig:xraylum}. The peak luminosity is the same as reported for the sub-threshold EP transient EP250827b/SN 2025wkm \citep[$z$\,$=$\,$0.119$;][]{Srinivasaragavan2026}. At the same time, the associated supernova resides in the typical range of peak\footnote{The peak absolute magnitude refers to the brightness at the SN lightcurve peak at $\sim$\,$10$\,$-$\,$20$ d post-explosion and not the early shock cooling emission seen in low-luminosity GRBs \citep[e.g.,][]{Campana060218}.} optical brightness for Ic-BL supernovae associated with both GRBs and FXTs (Figure \ref{fig:xraylum}; see also \S \ref{sec:grbsncompare}). Thus, EP260321a occupies a region of parameter space in which the supernova robustly resembles the energetic Ic-BL explosions associated with GRBs and FXTs, extending the low-luminosity end of the shock breakout population and more comparable to SN 2008D, despite the higher photospheric velocities and absolute magnitude of SN 2026gzf (Figures \ref{fig:compare} and \ref{fig:velcompare}).

This combination conclusively demonstrates that the luminosity of the high-energy breakout emission is not influenced by the optical luminosity of the supernova (Figure \ref{fig:xraylum}). Instead, the X-ray output depends sensitively on the structure of the fastest ejecta, the Lorentz factor of the shock, the breakout radius, and the immediate circumstellar environment \citep[see, e.g.,][]{Fryer2026}. Thus, EP260321a represents a lower luminosity extension of the same family of high-energy shock breakouts previously identified through \textit{Swift} and EP. This interpretation is also consistent with recent models predicting that the majority of detectable stellar shock breakouts should be substantially fainter than the classical low-luminosity GRBs and therefore most readily discovered by wide-field soft X-ray monitors such as EP \citep{Sun2022,Bayless2022,Fryer2026}. The location of EP260321a in Figure \ref{fig:xraylum} therefore argues against a direct mapping between Ic-BL supernova luminosity and prompt X-ray luminosity, and instead points to a broad luminosity function for high-energy breakout, requiring a more extended distribution of physical parameters at explosion to bridge the gap between SN 2008D and GRB 060218. A broad luminosity function, dominated by fainter breakouts, also provides a more natural explanation for the mismatch between theoretical predictions for the rate of shock breakouts that EP should detect and the true rate of detected events \citep[see, e.g.,][]{Sun2022,Fryer2026}.

Interestingly, while the duration of the X-ray emission in EP260321a is shorter than in low-luminosity GRBs (Figure \ref{fig:xraylum}), it is longer than in SN 2008D and is much longer than the light-crossing time of a Wolf-Rayet star, which is of order 10 s.  This may require an asymmetric explosion or a breakout which occurs in dense material outside the star (see \citealt{Irwin2021} for further discussion, and, e.g., \citealt{Svirski2014-1,Svirski2014-2}).  The presence of such extended material is consistent with the progenitor’s apparent pre-explosion activity \citep{Chen2026}, see also Appendix \ref{sec:decamarch}.

There are multiple proposed explanations for low-luminosity GRBs and the multi-component broadband emission observed from GRB 060218-like events, see, e.g., \citet{Irwin2016,Irwin2025-2} for an in depth summary. Leading models include that of mildly relativistic or subrelativistic shock breakout \citep{Soderberg2006grb060218,Campana2006,Waxman2007,Irwin2025-2}, a choked jet \citep{Nakar2015}, and models that decouple the gamma-ray emission from the rest of the components, such that the X-rays and gamma-rays are produced by a weak jet and the optical emission is produced by the cooling post-breakout ejecta \citep[e.g.,][]{Irwin2016}. In these models, the multiple components are generally considered to be isotropic, with the exception of jetted gamma-ray emission. As EP260321a does not display gamma-ray emission or low-luminosity GRB-like X-ray or radio emission, and it is noticeably weaker in its overall X-ray luminosity (and spectral shape) than other events of this class \citep[see also][]{Yuan2026}, it may favor models that decouple the presence of gamma-rays from the shock breakout. 


However, it should be noted that GRBs 060218, 100316D, and 171205A were initially detected in gamma-rays as roughly constant flux image triggers \citep[see, e.g.,][]{Lien2014} by \textit{Swift}'s Burst Alert Telescope (BAT; \citealt{Gehrels2004,Barthelmy2005}), as opposed to a shorter duration events like typical GRBs. Using BAT trigger simulations \citep{Lien2014,Moss2022,Moss2026}, these events are not detectable by BAT beyond $400$ Mpc, which is why these events were all identified at much closer distances of $\sim$\,$150$\,$-$\,$260$ Mpc and why their observed event rate has not substantially increased over the decades. Therefore, the lack of prompt gamma-ray detection for EP260321a can be due to different causes, such as the current suspended operation status of \textit{Swift}, and the location of the source on the sky with respect to GRB detectors (e.g., \textit{Fermi} was in the South Atlantic Anomaly). 

Despite these caveats, a main difference between EP260321a and these other events is the lack of an additional powerlaw component in the X-ray spectrum \citep{Yuan2026}, which was observed in both the low-luminosity GRBs that were likely produced by relativistic shock breakout \citep{Campana060218,Starling2011,DElia2018,Izzo2019} as well as SN 2008D \citep{Soderberg2008,Modjaz2009}. A powerlaw component, as opposed to the observed low $kT$ thermal emission \citep{Yuan2026}, would more naturally suggest that gamma-ray emission was possible. 
Therefore, EP260321a/SN 2026gzf demonstrates that prompt X-ray emission can be produced in the absence of simultaneous gamma-rays, more strongly suggesting these components may have had separate origins in nearby low-luminosity GRBs. Alternatively, this may also be simply due to higher Lorentz factor outflows launched by these other events \citep[e.g.,][]{Fryer2026}. 
We also note that powerlaw X-ray emission does not necessarily imply the presence of a jet, as it may also be produced by shock breakout under certain conditions \citep[see, e.g,][]{Irwin2025-1,Irwin2025-2, Irwin2025-3}. The lack of a powerlaw component significantly constrains the  properties leading to the observed shock breakout emission, and in particular may require shock velocities which do not significantly exceed $0.1c$ \citep[see, e.g,][]{Irwin2025-1,Irwin2025-2, Irwin2025-3}.

\subsection{Comparison to GRB-SNe and FXT-SNe}
\label{sec:grbsncompare}

Following the ground-breaking discovery of GRB 980425/SN 1998bw \citep{Galama1998}, over the last 28 years there have been $\sim$\,$70$ identified GRB-SNe of which about half are spectroscopically classified \citep[see, e.g.,][and references therein]{Woosley2006,Hjorth2012sn,Cano2017,GRBSNtool}, though only about a dozen have high signal-to-noise, multi-epoch spectral sequences. EP260321a/SN 2026gzf is the 16th high-energy (X-ray or gamma-ray) transient associated with a Ic-BL supernova within $z$\,$<$\,$0.2$, only the 11th within $z$\,$<$\,$0.1$, and the 5th closest ever, including being the closest X-ray triggered Ic-BL\footnote{The previous closest is EP250827b at $z$\,$=$\,$0.119$ \citep{Srinivasaragavan2026} followed by EP250108a at $z$\,$=$\,$0.176$ \citep{Srinivasaragavan2025EP0108a,Li2025,Eyles-Ferris2025EP,Rastinejad2025EP}.}. Thus, the association of SN 2026gzf with EP260321a places it among a small but rapidly growing sample of nearby Ic-BL supernovae discovered through prompt high-energy emission.

We have presented a comprehensive investigation placing SN 2026gzf in the context of both GRB-SNe and FXT-SNe, as well as optically selected Ic-BL SNe \citep{Taddia2018,Taddia2019,Srinivasaragavan2024}. 
We find an overall good agreement to the general population in terms of lightcurve shape (Figures \ref{fig:lightcurve}, \ref{fig:lightcurveshapecompare}, and \ref{fig:compare}), absolute magnitude (Figure \ref{fig:compare}), spectral evolution (Figures \ref{fig:specsequence} and \ref{fig:speccompare}), and ejecta velocities (Figures \ref{fig:vel} and \ref{fig:velcompare}). In Figure \ref{fig:compare}, we compare the peak magnitude, peak time, and post-peak decline rate of SN 2026gzf to literature samples of GRB-SNe, FXT-SNe, and optically selected Ic-BL SNe. SN 2026gzf peaks at $M_r$\,$\approx$\,$-19$ mag on a rest-frame timescale of $\approx$\,$15$ d in $r$-band, with a decline rate of $\Delta m_{15,r}$\,$\approx$\,$0.6$ mag, falling in a region of this multi-dimensional parameter space that is closest to previously known GRB-SNe and FXT-SNe. This is important because the X-ray emission of EP260321a is substantially fainter than that of the low-luminosity GRBs and other EP shock breakout candidates (Figure \ref{fig:xraylum}), yet the optical supernova itself is not correspondingly faint.

The spectral evolution leads to the same conclusion. The DESI spectrum obtained near optical maximum at $T_0+17.7$ d shows broad absorption features characteristic of Type Ic-BL SNe and closely resembles spectra of the GRB-SNe SN 2006aj, SN 2013dx, and SN 2017iuk at similar phases (Figure \ref{fig:speccompare}). The overall spectral morphology and its evolution (Figure \ref{fig:specsequence}) is clearly that of a high velocity stripped-envelope explosion. The match to GRB-SNe is particularly notable because EP260321a was not accompanied by a detected GRB and because its prompt X-ray spectrum was characterized by soft, thermal emission \citep{2026GCN.44075....1H}. Thus, the optical spectra identify SN 2026gzf as a member of the energetic Ic-BL class, independent of the unusual high-energy properties of EP260321a (Figures \ref{fig:xraylum},  \ref{fig:xraybreakoutlum}, and \ref{fig:xray}).

The velocity evolution further supports this interpretation. In Figure \ref{fig:velcompare}, we compare the Fe\,{\sc ii} $\lambda5169$ and Si\,{\sc ii} $\lambda6355$ velocities of SN 2026gzf to the Ic-BL and GRB-SN sample of \citet{Finneran2025}. At the earliest phases, SN 2026gzf shows very high velocities, with both Fe\,{\sc ii} and Si\,{\sc ii} indicating material moving at $\sim$\,$30,000$ km s$^{-1}$. These velocities decline rapidly over the first two weeks and then evolve more slowly, reaching values near the median of the GRB-SN/Ic-BL comparison sample around peak light. The early high velocities are most similar to the upper envelope of the GRB-SN population, while the later velocities remain consistent with the broader Ic-BL distribution, and are substantially higher than SN 2008D \citep{Mazzali2008,Malesani2009}. 


SN 2026gzf therefore adds to the emerging evidence that stripped-envelope supernovae with associated prompt X-ray and gamma-ray emission span a continuum in explosion properties. At one end are classical and low-luminosity GRBs, where a relativistic outflow produces gamma-rays and long-lived X-ray emission. At the other end are events like SN 2008D \citep{Soderberg2008}, where the high-energy emission is best understood as shock breakout from a non-GRB Type Ib explosion. EP260321a/SN 2026gzf lies between these regimes, as while its supernova properties agree well with known GRB-SNe its high-energy emission is dominated by a low-luminosity, thermal X-ray transient that more closely resembles the X-ray emission from SN 2008D (Figures \ref{fig:xraylum} and \ref{fig:xraybreakoutlum}). This makes EP260321a one of the clearest examples that the wide-field soft X-ray survey conducted by EP is uncovering events at the boundary between ordinary stripped-envelope supernovae, relativistic shock breakouts, and failed or weak jet explosions.

\subsection{Nature of the Supernova Power Source}

In the collapsar model for GRB-SNe and energetic Ic-BL supernovae \citep{Woosley1993,MacFadyen1999,Woosley2006}, the formation and longevity of an accretion disk around the remnant black hole provides a physical connection between the relativistic jet and the optical supernova. 
If the SN lightcurve is powered by the decay of $^{56}$Ni (primarily produced in the disk wind), the brightness of the supernova depends primarily on the properties of the disk~\citep{2020MNRAS.499.4097Z,2024MNRAS.529..178C,2024PhRvD.109h3010D}.  
The jet power, and hence Lorentz factor of the breakout material should also depend on the properties of the disk and we would expect a correlation between the breakout and supernova emission.  SN 2026gzf presents a challenge to this expectation. 

Recent semi-analytic shock breakout calculations show that the X-ray luminosity and duration are sensitive to the energy distribution in the fastest ejecta $E(\Gamma\beta)=E_0\,\Gamma^{-p}$, the maximum Lorentz factor $\Gamma_{\rm max}$, and the effective emitting area \citep{Fryer2026}. The low X-ray luminosity of EP260321a (Figure \ref{fig:xraybreakoutlum}) relative to relativistic shock breakout events such as GRB 060218/SN 2006aj therefore suggests a lower $\Gamma_{\rm max}$ and less kinetic energy in material with $\Gamma$\,$\gtrsim$\,$2$. \citet{Fryer2026} assumed a maximum Lorentz factor from a weak jet ($\Gamma_{\rm max} = 10$) to a strong jet ($\Gamma_{\rm max} = 100$).  Only a subset of the $\Gamma_{\rm max} = 10$ runs match these faint X-ray fluxes (see Figure \ref{fig:xraybreakoutlum}), arguing that the bulk of the ejecta in EP260321a has Lorentz factors below 10 (models with $\Gamma_{\rm max} = 2-5$ from \citet{Fryer2026} can produce better matches to the observed luminosity). The reduced requirements are below the ``standard’’ parameter range in \citet{Fryer2026} for disk/jet models but on the high end of the standard parameter suite they used for convective engine explosions, requiring a larger exploration over a broader parameter range which we leave to future work.  Thus, based on the model gird of \citet{Fryer2026}, EP260321a lies on the edge between disk/jet and neutrino-driven engines.

We note that shock breakout and cooling ejecta only dominates the emission at very early times \citep{Chen2026,MartinCarrillo2026}  In the shock breakout models \citep{Fryer2026}, the required mass of high Lorentz factor material ($\Gamma > 2$) is very small, less than $0.01-0.1\,M_\odot$. The evolution of the lightcurve after the initial cooling phase probes the nature of the non-relativistic ejecta (the bulk of the ejecta mass) and requires an additional power source: e.g., $^{56}$Ni decay, central engine (fallback accretion or a magnetar) or additional shock interactions. 
Our simple Arnett model \citep{Arnett1982} fit to the bolometric lightcurve (Figure \ref{fig:bolometric}) implies a large $^{56}$Ni mass of $M_{\rm Ni}=0.45\pm0.02\,M_\odot$, which is consistent with the typical values inferred for GRB-SNe associated to relativistic outflows \citep{Cano2017}.

Such high $^{56}$Ni ejecta masses are possible from models of disk winds and jet outflows, but well beyond the typical  $^{56}$Ni masses ejected through the neutrino-driven engine (0.05-0.2\,M$_\odot$; \citealt{Andrews2020}).
The tension between the maximum Lorentz factor and inferred Nickel mass suggests either that the jet and disk wind were not tightly coupled in EP260321a, or that the peak optical luminosity does not directly trace the true $^{56}$Ni yield. In the first case, the disk may have produced a strong nickel rich wind while the jet was choked before escaping the progenitor. In the second case, additional energy sources, such as shock heating or circumstellar interaction \citep[see, e.g.,][]{Chen2026,MartinCarrillo2026}, may have contributed to the bolometric lightcurve, causing a purely radioactive Arnett model to overestimate the nickel mass \citep{Niblett2025}. Under this interpretation, our inferred $M_{\rm Ni}$ should be regarded as an upper limit to the true $^{56}$Ni mass \citep[see, e.g.,][]{Meza2020A&A...641A.177M,Afsariardchi2021ApJ...918...89A,Rodriguez2023ApJ...955...71R}. Despite this caveat, the comparison of our inference of the $^{56}$Ni mass with other Ic-BL SNe is still valuable as nickel masses for many Ic-BL and GRB associated supernovae are derived with similar one-zone radioactive models that also neglect possible shock heating \citep{Cano2017}. EP260321a therefore shows either that there is a weaker correlation between the jet-disk system and the disk wind produced $M_{\rm Ni}$, or that part of the luminosity commonly attributed to $^{56}$Ni in such events may instead arise from a non-radioactive power source \citep[see, e.g.,][]{Niblett2025,Srinivasaragavan2026}.
The interpretation can be tested in the future with better models designed to understand the $^{56}$Ni production in disks and the disk's connection to the observed shock breakout emission.

\begin{figure}
    \centering
    \includegraphics[width=\linewidth]{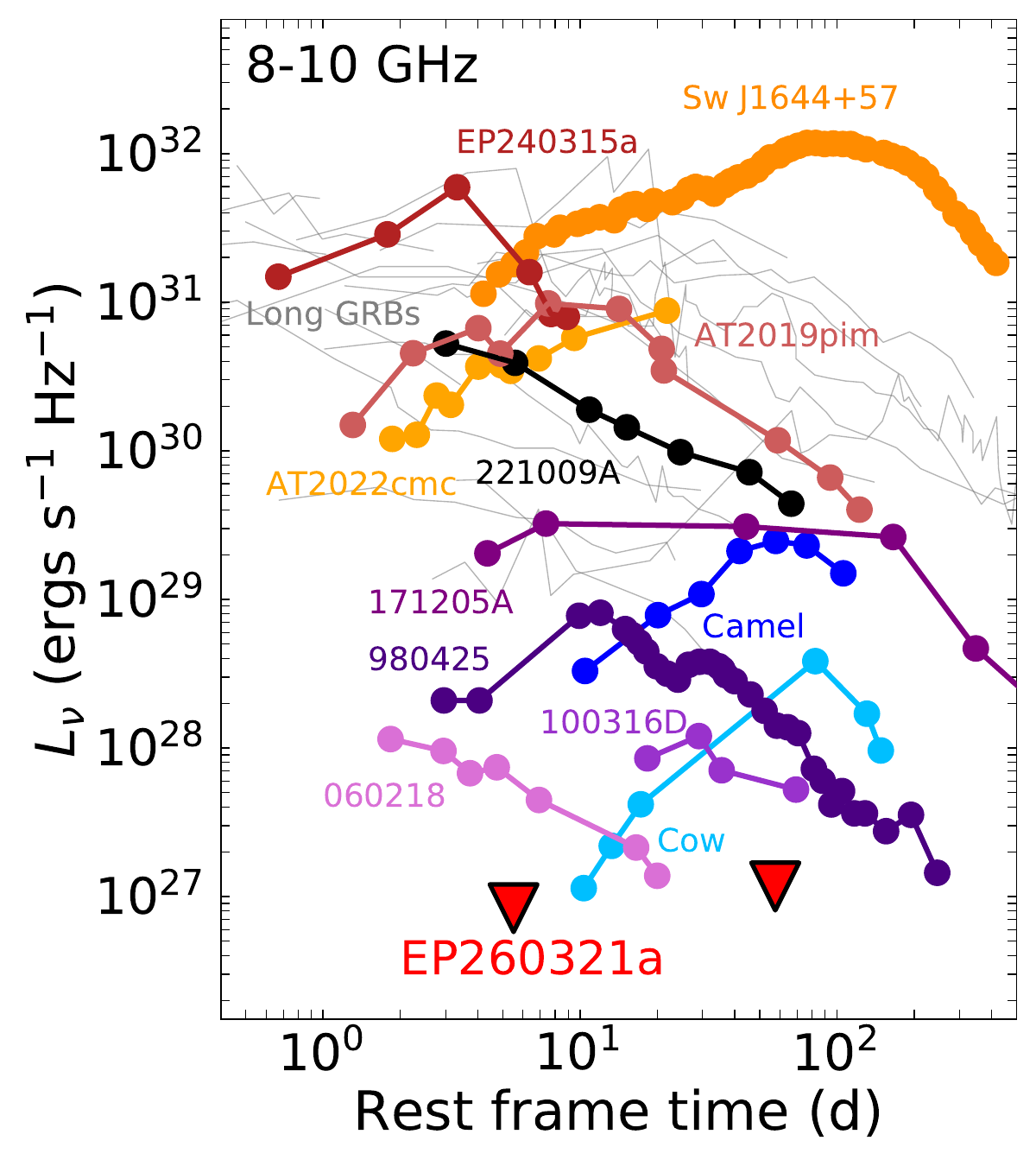}
    \caption{Comparison between the radio upper limits for EP260321a (downward red triangles) versus the radio luminosity of multiple classes of energetic transients, including GRBs \citep{Soderberg2006grb060218,ChandraFrail2012,Laskar2022,Perley2024,Ricci2025}, jetted tidal disruption events \citep{Zauderer2011,Andreoni2022}, and fast blue optical transients \citep{Ho2019cow,Margutti2019cow,HoKoala,Coppejans2020cow}. The figure is reproduced from \citet{OConnor2025}. }
    \label{fig:radio}
\end{figure}

\subsection{Constraints on a Relativistic Outflow}

In contrast to low-luminosity GRBs, EP260321 only displayed prompt X-ray emission and was subsequently undetected by \textit{Chandra} at $T_0+15.4$ d and $T_0+39.0$ d. These limits are far below the X-ray luminosities of the known low-luminosity GRBs (see Figure \ref{fig:xray}) at comparable phases. We note that the low-luminosity GRBs 060218, 100316D, and 171205A, shown in Figure \ref{fig:xray}, are the faintest \textit{Swift} afterglows and each has X-ray data at the phase of our \textit{Chandra} observations. The exception being the pre-\textit{Swift} GRB 980425 which requires a slight extrapolation, though given its lack of evolution to $>100$ d we consider this to be reasonable. Although the physical origin of the late-time X-ray emission in events such as GRB 060218 and GRB 100316D remains debated \citep[e.g.,][]{Soderberg2006grb060218,Margutti2013}, our \textit{Chandra} limits show that EP260321a did not produce a comparable long-lived X-ray component. This marks a genuine observational difference between EP260321a and previously known GRB-SN shock breakout events. More generally, these \textit{Chandra} limits are capable of excluding all previously known X-ray afterglows of both short and long GRBs, XRFs, and EP FXTs \citep[e.g.,][]{Cotter2026} that have a known redshift, with the single exception of GRB 170817A \citep[e.g.,][]{Troja2017,Margutti2017}. The existing radio limits produce similarly stringent constraints (Figure \ref{fig:radio}). 

The absence of a luminous late-time X-ray or radio counterpart strongly disfavors the presence of a relativistic outflow, as seen in standard GRBs. As discussed in \S \ref{sec:outflowconstraints}, our \textit{Chandra} limits, together with available radio upper limits \citep{2026GCN.44229....1L}, exclude a broad region of parameter space for a relativistic jet propagating into a wind environment, with a very weak dependence on viewing angle. For typical densities $A_*$\,$=$\,$1$ (Figure \ref{fig:afterglowconst1}), we require both a slower moving, low energy jet with isotropic-equivalent $E_\textrm{kin}$\,$<$\,$10^{49}$ erg and Lorentz factor $\Gamma_0$\,$<$\,$30$ at the jet's core. As demonstrated in Figure \ref{fig:afterglowconst1}, these results have only a very weak dependence on viewing angle, but a strong dependence on the assumed density (see also Figure \ref{fig:afterglowconst2} in Appendix \ref{sec:aftappendix}) such that for densities below $A_*$\,$\lesssim$\,$0.1$ the allowed parameter space greatly relaxes and we would no longer be capable of excluding ultrarelativistic outflows or more energetic jets. 
Our conclusions are similar to and consistent with the independent analyses of \citet{Yuan2026,Rastinejad2026,MartinCarrillo2026}.

We conclude that both a weak jet with an extremely low kinetic energy propagating into a very low density environment or a choked jet whose energy is deposited into a mildly relativistic cocoon could sufficiently suppress the observed gamma-ray and late-time afterglow emission. However, regardless of the interpretation, our deep \textit{Chandra} limits require that any such component be substantially fainter than the afterglows of known nearby GRB-SNe and FXT-SNe, suggesting the absence of a luminous, relativistic jet. Therefore, not all luminous Ic-BL supernovae that display high-energy emission are necessarily associated with relativistic ejecta. Similar conclusions have been drawn for optically selected Ic-BL supernovae (see Figure \ref{fig:xray}; \citealt{Corsi2023,Schroeder2025,ODwyer2025}), though each of those lacked any high-energy association.

\begin{figure}
    \centering
    \includegraphics[width=\linewidth]{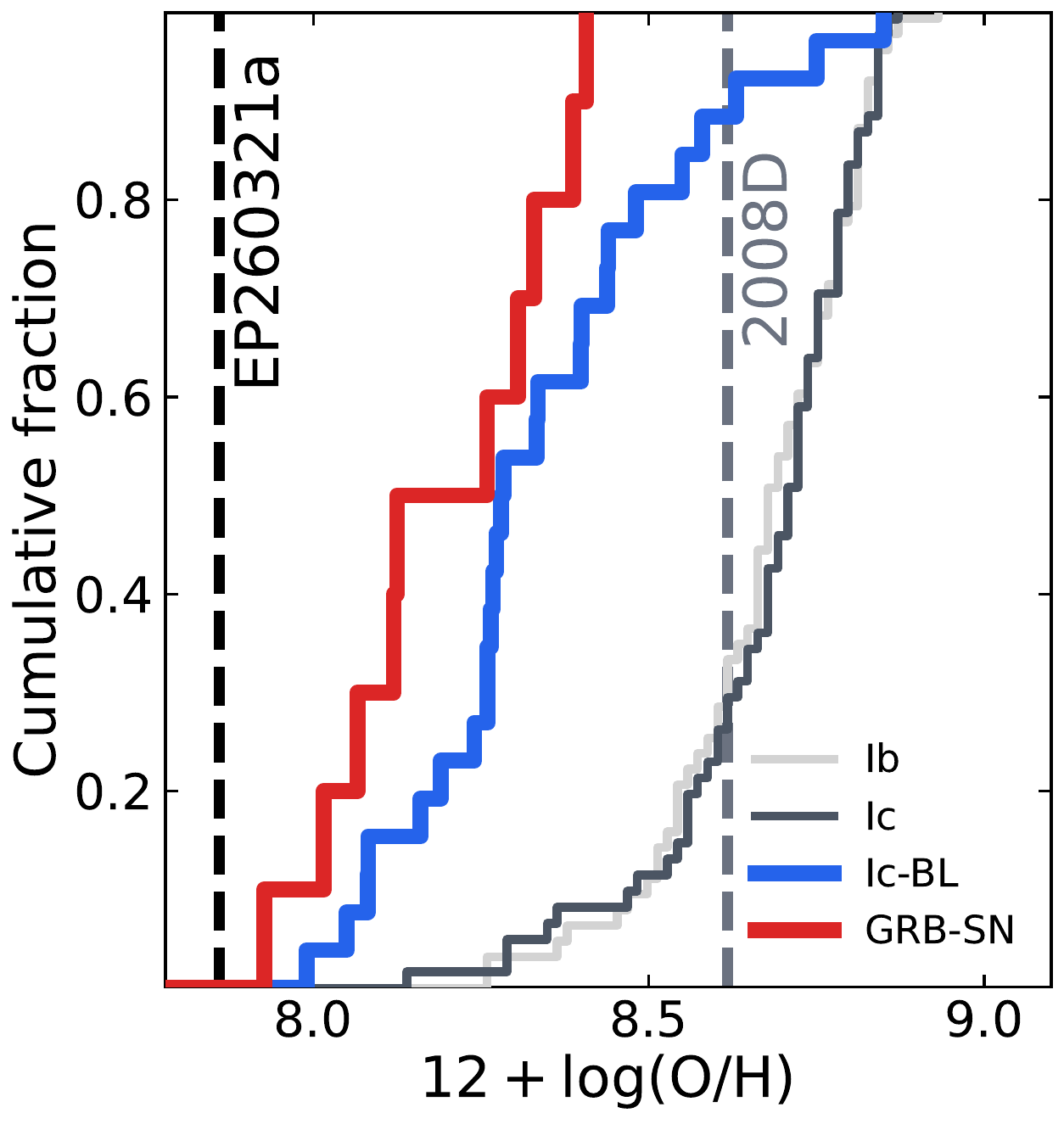}
    \caption{Cumulative distribution of oxygen abundances for low redshift ($z$\,$<$\,$0.2$) broad-lined Type Ic supernovae without detected GRBs (blue) and GRB-SNe/Ic-BL events (red), compiled from the PP04 O3N2 \citep{PettiniPagel2004} measurements from \citet{Japelj2018} and \citet{Modjaz2020}, and local ($z$\,$<$\,$0.025$) SNe Ib and Ic (gray) from \citet{Modjaz2011,Gans2022,Gans2025} converted to the PP04 O3N2 scale. The metallicity at the explosion sites of EP260321a and SN 2008D are shown as a dashed vertical line.}
    \label{fig:metallicity}
\end{figure}

\subsection{Local Environment and Metallicity}


The host galaxy of SN 2026gzf is a blue, dwarf galaxy with stellar mass $\sim$\,$10^9$ $M_\odot$ \citep{Chen2026,Rastinejad2026} that is actively star forming (${\rm SFR}_{\rm H\alpha}$\,$\approx$\,$0.07\,M_\odot\,{\rm yr}^{-1}$) and has a subsolar metallicity $12+\log({\rm O/H})\approx8.2$. This is already a prime environment for the production of a Ic-BL SN progenitor. However, the transient is located 2.5 kpc from the center of the host (Figure \ref{fig:FC}), and the local environment is most relevant to the progenitor properties (see \S \ref{sec:hetenv}). The explosion site of EP260321a/SN2026gzf is actively star forming at a higher rate than the rest of the host galaxy with ${\rm SFR}_{\rm H\alpha}$\,$\approx$\,$0.2\,M_\odot\,{\rm yr}^{-1}$. Additionally, the location of the transient has a lower metallicity than the rest of the host, with $12+\log({\rm O/H})\approx7.9$, see Figure \ref{fig:het_space}. This trend is commonly observed in low-$z$ GRBs, such that the local environment is less metal enriched compared to the rest of the galaxy \citep{Levesque2011,Thone2014,Kruhler2015,Kruhler2017,Izzo2017,Cano2016jca,Tanga2018,Michalowski2018,Melandri2019,Thone2021,Thone2024}.

The lack of metal enrichment in the local environment is consistent with observations of GRB associated Ic-BL SNe \citep[e.g.,][]{Levesque2010-1,Levesque2010-2,Kruhler2015} and non-GRB associated Ic-BL discovered through untargeted optical surveys \citep{Japelj2018,Modjaz2020}. In Figure \ref{fig:metallicity}, we compare the metallicity at the explosion site of EP260321a/SN 2026gzf to these two populations of low redshift $z$\,$<$\,$0.2$ Ic-BL supernovae with and without associated GRBs using the O3N2 calibration of \citet{PettiniPagel2004}. The local environment of EP260321a is clearly extremely metal poor, and one of the lowest metallicity environments observed for a Ic-BL supernova both in general and, in particular, for a Ic-BL without an associated GRB. The metallicity also differs significantly from the environments of Ib and Ic supernovae (see Figure \ref{fig:metallicity}; \citealt{Gans2022,Gans2025}), such as SN 2008D \citep{Thone2009,Modjaz2011}. As EP260321a both lacked a GRB association and occurred in such a low metallicity environment, it strongly suggests that metallicity cannot be the only factor in producing a progenitor star population of ``collapsars'' \citep{Woosley1993,MacFadyen1999} capable of launching relativistic outflows \citep[see also][and references therein]{Wolf2007,Schady2015,Disberg2025,Briel2025}. 

We note that our results are in agreement with the analysis of \citet{MartinCarrillo2026} and \citet{Chen2026} who likewise find this to be an extremely low metallicity environment. We must note however, as stated in \S \ref{sec:hetenv}, that the derived values of O3N2 and N2 that lead to this metallicity lie outside of the range of values used to define the oxygen abundance calibrations and additionally that these calibrations have systematic uncertainties of $0.1$\,$-$\,$0.2$ dex \citep[see][]{PettiniPagel2004,Marino2013}. This is of course always the case when extending to such low metallicity, and while the exact value of $12+\log({\rm O/H})$ may therefore be uncertain, the derived O3N2 and N2 strongly point to an extremely metal poor local environment of SN 2026gzf that is different from the bulk of the Ic-BL and Ibc populations (Figure \ref{fig:metallicity}).

\section{Conclusions}
\label{sec:conclusions}

We have presented a comprehensive photometric ($ugrizJK_s$) and spectroscopic observing campaign of EP260321a/SN 2026gzf over the first 55 days of its evolution using FTW, DECam, DESI, SALT, and HET. Our photometric observations with FTW begin 10 hours after the detection of the X-ray shock breakout emission. We subsequently observe the rising emission from the supernova and our spectral observations starting at 3.3 d after explosion capture high velocity $\sim$\,$30,000$ km s$^{-1}$ ejecta that we observe to decelerate and produce numerous broad absorption features typical of broad-lined Ic supernovae. We find that SN 2026gzf is a typical Ic-BL supernova in all respects, from its lightcurve and spectral evolution to its overall bolometric luminosity.

We additionally obtained deep X-ray imaging with the \textit{Chandra X-ray Observatory} at 15 and 39 d after explosion and radio imaging with the Very Large Array at 60 d. Combined with the unprecedented sensitivity of \textit{Chandra} and the nearby distance of EP260321a at 158 Mpc, our observations can exclude the afterglow of all previously known gamma-ray bursts (that have a redshift), with the singular exception of the exceptional off-axis binary neutron star merger GW170817 \citep{Troja2017,Margutti2018}, down to an X-ray luminosity of $10^{40}$ erg s$^{-1}$ between $15-60$ d after explosion. We apply these constraints to exclude a collimated relativistic outflow with both high kinetic energies and high Lorentz factors, requiring instead $E_\textrm{kin}$\,$<$\,$10^{49}$ erg and $\Gamma_0$\,$<$\,$30$ for a typical stellar wind density of $A_*$\,$\gtrsim$\,$1$. This requirement can be strongly relaxed for lower assumed densities, such that for $A_*$\,$=$\,$0.1$ the constraints relax to $E_\textrm{kin}$\,$<$\,$10^{50}$ erg and $\Gamma_0$\,$<$\,$100$. Therefore, while we cannot fully exclude all possible afterglow parameters, we can conclude that observationally any afterglow arising from this event differs from past observations of low-luminosity GRBs.

We interpret EP260321a as a low-luminosity, thermally dominated shock breakout event associated with an otherwise typical energetic Ic-BL supernova. The event bridges SN 2008D-like X-ray shock breakout and the Ic-BL supernovae associated with low-luminosity GRBs. Unlike SN 2008D, however, the associated supernova is a broad-lined Type Ic with higher ejecta velocities, and unlike the classical nearby low-luminosity GRBs, EP260321a lacks both a gamma-ray detection and a late-time X-ray or radio afterglow. This combination is naturally explained if EP260321a was produced by a mildly relativistic shock breakout caused by a weak jet/cocoon in which the fastest ejecta produced the observed soft X-rays but did not generate either a bright X-ray or radio afterglow or substantial gamma-ray emitting material along the line of sight.

While observations of shock breakout events have been limited over the last three decades due to the requirement of serendipitous timing of the explosion with observations from narrow-field instruments \citep[e.g.,][]{Soderberg2008,Alp2020}, the launch of the Einstein Probe \citep{EP2015,EP2022,Yuan2025} in 2024 January has enabled a drastic increase in their rate of discovery over the last few years. The launch of the \textit{BlackCAT} mission \citep{Colosimo2024ApJ...969..138C,Blackcat} and upcoming wide-field ultraviolet missions, \textit{ULTRASAT} \citep{Ultrasat-1,Ultrasat}
and \textit{UVEX} \citep{uvex}, will aid in driving a shock breakout revolution and expand our understanding of the deaths of massive stars and their pre-explosion mass loss and eruptions.

\begin{acknowledgments}

The authors acknowledge the anonymous referee for their careful reading of the manuscript and for their constructive comments that led to a clearer presentation of our results.

BO and XJH thank Moses Mogotsi, Lee Townsend, and Danièl Groenewald for approving a SALT DDT request and for assistance in obtaining the observations. BO acknowledges the staff of the Chandra X-ray Observatory, including Pat Slane, Dan Schwartz, Jack Steiner, and Vinay Kashyap, for approving and rapidly scheduling the Chandra observations. 

BO is supported by the McWilliams Postdoctoral Fellowship in the McWilliams Center for Cosmology and Astrophysics at Carnegie Mellon University. JH acknowledges
support from NASA under award number 80GSFC24M0006. CMI is supported by the JST FOREST Program (JPMJFR2136) and the JSPS Grant-in-Aid
for Scientific Research (20H05639, 20H00158, 23H01169, 23H04900, 26K07149). Support was provided by Schmidt Sciences, LLC. for K. Malanchev. AF was funded by the European Union ERC-2022-STG - BOOTES - 101076343. Views and opinions expressed are however those of the author(s) only and do not necessarily reflect those of the European Union or the European Research Council Executive Agency. Neither the European Union nor the granting authority can be held responsible for them.

This work used resources on the Vera Cluster at the Pittsburgh Supercomputing Center (PSC). Vera is a dedicated cluster for the McWilliams Center for Cosmology and Astrophysics at Carnegie Mellon University. We thank the PSC staff for their support of the Vera Cluster.

Support for this work was provided by NASA through Chandra Award No. GO6-27037X, issued by the Chandra X-ray Center, which is operated by the Smithsonian Astrophysical Observatory for and on behalf of NASA under contract NAS8-03060. 

The scientific results reported in this article are based on observations made by the Chandra X-ray Observatory (CXO). This research has made use of data obtained from the Chandra Data Archive provided by the Chandra X-ray Center (CXC). This paper employs a list of Chandra datasets, obtained by the Chandra X-ray Observatory, contained in the Chandra Data Collection \dataset[doi:10.25574/cdc.617]{https://doi.org/10.25574/cdc.617}. This research has made use of software provided by the Chandra X-ray Center (CXC) in the application package \texttt{CIAO}. 

This paper contains data obtained at the Wendelstein Observatory of the Ludwig-Maximilians University Munich. Funded by the Deutsche Forschungsgemeinschaft (DFG, German Research Foundation) under Germany's Excellence Strategy – EXC-2094/2 – 390783311.

Some of the observations reported in this paper were obtained with the Southern African Large Telescope (SALT).

This material is based upon work supported by the U.S. Department of Energy (DOE), Office of Science, Office of High-Energy Physics, under Contract No. DE–AC02–05CH11231, and by the National Energy Research Scientific Computing Center, a DOE Office of Science User Facility under the same contract. Additional support for DESI was provided by the U.S. National Science Foundation (NSF), Division of Astronomical Sciences under Contract No. AST-0950945 to the NSF’s National Optical-Infrared Astronomy Research Laboratory; the Science and Technology Facilities Council of the United Kingdom; the Gordon and Betty Moore Foundation; the Heising-Simons Foundation; the French Alternative Energies and Atomic Energy Commission (CEA); the National Council of Humanities, Science and Technology of Mexico (CONAHCYT); the Ministry of Science, Innovation and Universities of Spain (MICIU/AEI/10.13039/501100011033), and by the DESI Member Institutions: \url{https://www.desi.lbl.gov/collaborating-institutions}. The authors are honored to be permitted to conduct scientific research on I'oligam Du'ag (Kitt Peak), a mountain with particular significance to the Tohono O’odham Nation. Any opinions, findings, and conclusions or recommendations expressed in this material are those of the author(s) and do not necessarily reflect the views of the U. S. National Science Foundation, the U. S. Department of Energy, or any of the listed funding agencies.

This material is based upon work supported in part by the National Science Foundation through Cooperative Agreements AST-1258333 and AST-2241526 and Cooperative Support Agreements AST-1202910 and 2211468 managed by the Association of Universities for Research in Astronomy (AURA), and the Department of Energy under Contract No. DE-AC02-76SF00515 with the SLAC National Accelerator Laboratory managed by Stanford University. Additional Rubin Observatory funding comes from private donations, grants to universities, and in-kind support from LSST-DA Institutional Members.

Based on observations obtained with the Samuel Oschin Telescope 48-inch and the 60-inch Telescope at the Palomar Observatory as part of the Zwicky Transient Facility project. ZTF is supported by the National Science Foundation under Grants No. AST-1440341, AST-2034437, and currently Award AST-2407588. ZTF receives additional funding from the ZTF partnership. Current members include Caltech, USA; Caltech/IPAC, USA; University of Maryland, USA; University of California, Berkeley, USA; University of Wisconsin at Milwaukee, USA; Cornell University, USA; Drexel University, USA; University of North Carolina at Chapel Hill, USA; Institute of Science and Technology, Austria; National Central University, Taiwan, and OKC, University of Stockholm, Sweden. Operations are conducted by Caltech’s Optical Observatory (COO), Caltech/IPAC, and the University of Washington at Seattle, USA.

The Babamul alerts broker and BOOM software infrastructure \citep{Babamul} is co-developed by the California Institute of Technology and the University of Minnesota. This work acknowledges support from the National Science Foundation through AST Award No. 2432476 (PI Kasliwal; co-PI Coughlin) and leverages experience from the Zwicky Transient Facility (co-PIs Graham and Kasliwal).

This research is based on data obtained from the Astro Data Archive at NSF NOIRLab. NOIRLab is managed by the Association of Universities for Research in Astronomy (AURA) under a cooperative agreement with the U.S. National Science Foundation. This project used data obtained with the Dark Energy Camera (DECam), which was constructed by the Dark Energy Survey (DES) collaboration. Funding for the DES Projects has been provided by the U.S. Department of Energy, the U.S. National Science Foundation, the Ministry of Science and Education of Spain, the Science and Technology Facilities Council of the United Kingdom, the Higher Education Funding Council for England, the National Center for Supercomputing Applications at the University of Illinois at Urbana-Champaign, the Kavli Institute of Cosmological Physics at the University of Chicago, Center for Cosmology and Astro-Particle Physics at the Ohio State University, the Mitchell Institute for Fundamental Physics and Astronomy at Texas A\&M University, Financiadora de Estudos e Projetos, Fundacao Carlos Chagas Filho de Amparo, Financiadora de Estudos e Projetos, Fundacao Carlos Chagas Filho de Amparo a Pesquisa do Estado do Rio de Janeiro, Conselho Nacional de Desenvolvimento Cientifico e Tecnologico and the Ministerio da Ciencia, Tecnologia e Inovacao, the Deutsche Forschungsgemeinschaft and the Collaborating Institutions in the Dark Energy Survey. The Collaborating Institutions are Argonne National Laboratory, the University of California at Santa Cruz, the University of Cambridge, Centro de Investigaciones Energeticas, Medioambientales y Tecnologicas-Madrid, the University of Chicago, University College London, the DES-Brazil Consortium, the University of Edinburgh, the Eidgenossische Technische Hochschule (ETH) Zurich, Fermi National Accelerator Laboratory, the University of Illinois at Urbana-Champaign, the Institut de Ciencies de l’Espai (IEEC/CSIC), the Institut de Fisica d’Altes Energies, Lawrence Berkeley National Laboratory, the Ludwig Maximilians Universitat Munchen and the associated Excellence Cluster Universe, the University of Michigan, NSF NOIRLab, the University of Nottingham, the Ohio State University, the University of Pennsylvania, the University of Portsmouth, SLAC National Accelerator Laboratory, Stanford University, the University of Sussex, and Texas A\&M University.

This paper contains data from observations obtained with the Hobby-Eberly Telescope (HET), which is a joint project of the University of Texas at Austin, the Pennsylvania State University, Ludwig-Maximilians-Universität München, and Georg-August Universität Göttingen. The HET is named in honor of its principal benefactors, William P. Hobby and Robert E. Eberly. We acknowledge the Texas Advanced Computing Center (TACC) at The University of Texas at Austin for providing high-performance computing, visualization, and storage resources that have contributed to the results reported within this paper. The Low Resolution Spectrograph 2 (LRS2) was developed and funded by the University of Texas at Austin, McDonald Observatory, Department of Astronomy, and Pennsylvania State University. We thank the Leibniz-Institut für Astrophysik Potsdam (AIP) and the Institut für Astrophysik Göttingen (IAG) for their contributions to the construction of the integral field units.

The National Radio Astronomy Observatory is a facility of the National Science Foundation operated under cooperative agreement by Associated Universities, Inc.

This work made use of data supplied by the UK \textit{Swift} Science Data Centre at the University of Leicester. 
This research has made use of the XRT Data Analysis Software (XRTDAS) developed under the responsibility of the ASI Science Data Center (ASDC), Italy. 
This research has made use of data and/or software provided by the High Energy Astrophysics Science Archive Research Center (HEASARC), which is a service of the Astrophysics Science Division at NASA/GSFC. This research has made use of the Astrophysics Data System, funded by NASA under Cooperative Agreement 80NSSC21M00561.

\end{acknowledgments}





%
\facilities{WO:2m, Mayall, SALT, HET, Blanco, ZTF, Rubin, CXO
}

\software{\texttt{Astropy} \citep{2018AJ....156..123A,2022ApJ...935..167A}, \texttt{SFFT} \citep{Hu2022}, \texttt{AstrOmatic} \citep{Bertin1996,Bertin2006,Bertin2010,2002ASPC..281..228B}, \texttt{XSPEC} \citep{Arnaud1996}, \texttt{CIAO} \citep{Ciao}, \texttt{HEASoft} \citep{2014ascl.soft08004N}, \texttt{CASA} \citep{CASA2022}, \texttt{HAFFET} \citep{HAFFET}, \texttt{VegasAfterglow} \citep{VegasAfterglow}
}


\appendix

\section{Log of Observations}

In Tables \ref{tab:photometry}, \ref{tab:rubin-ztf-phot}, \ref{tab:spectra}, we present a log of all photometric and spectroscopic observations obtained and analyzed in this work.

\section{Afterglow Constraints}
\label{sec:aftappendix}

\begin{figure*}
    \centering
    \includegraphics[width=\linewidth]{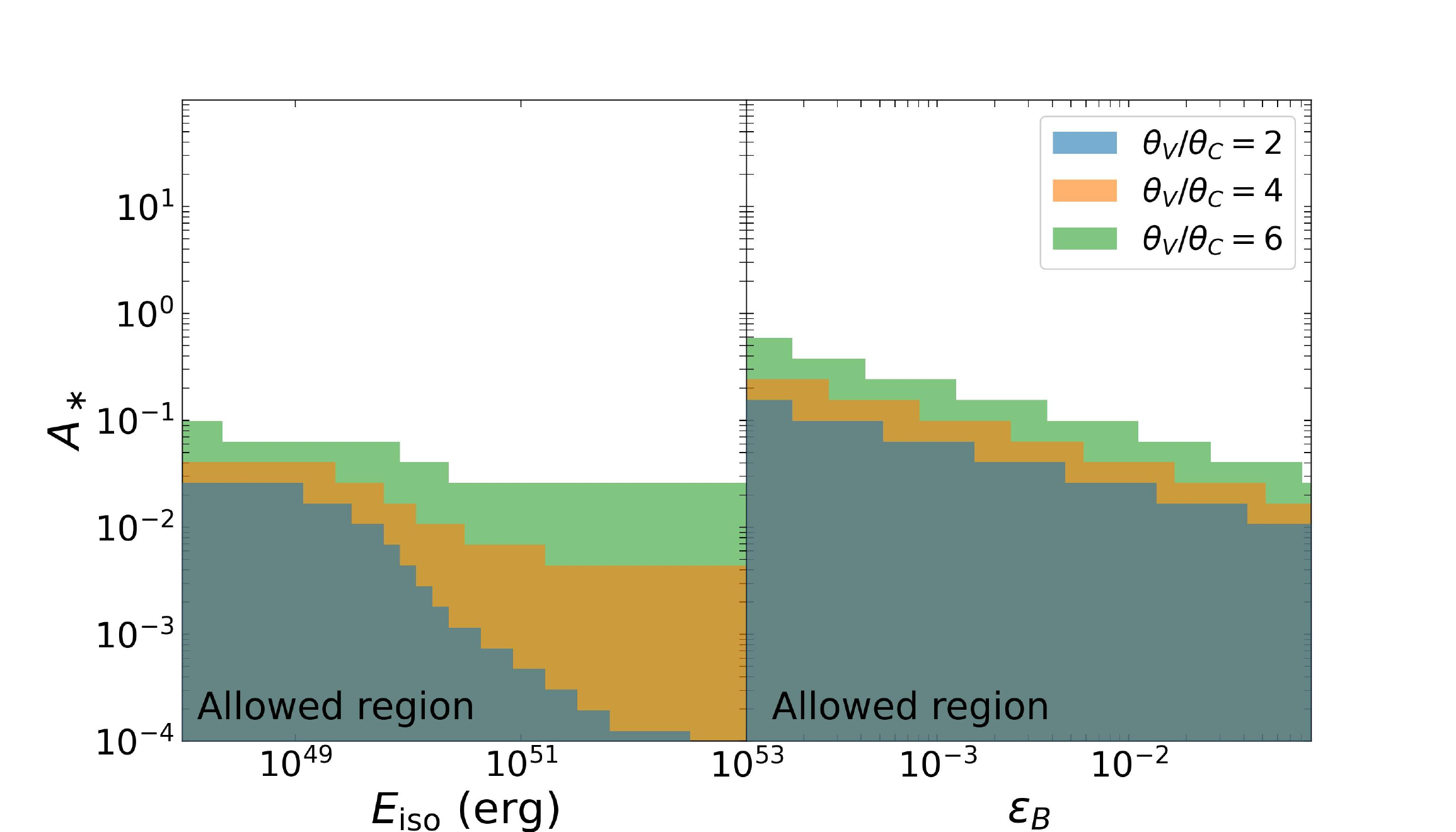}
    \caption{Allowed parameter space (shaded regions) for afterglow non-detection, assuming a Gaussian structured jet, at 158 Mpc ($z$\,$=$\,$0.0344$). \textbf{Left:} Allowed values of the density $A_*$ versus the isotropic-equivalent kinetic energy at the jet's core $E_\textrm{kin}$ for different viewing angles $\theta_\textrm{v}/\theta_\textrm{c}$. We have fixed $\theta_\textrm{c}$\,$=$\,$0.15$ rad, $\varepsilon_\textrm{e}$\,$=$\,$0.1$, $\varepsilon_\textrm{B}$\,$=$\,$0.01$, $p$\,$=$\,$2.2$, and $\Gamma_0$\,$=$\,$100$. \textbf{Right:} Similar as the left panel but instead as a function of $\varepsilon_\textrm{B}$ for fixed $E_{\rm kin}$\,$=$\,$10^{48}$ erg.}
    \label{fig:afterglowconst2}
\end{figure*}

In Figure \ref{fig:afterglowconst2}, we show additional afterglow model constraints generated using \texttt{VegasAfterglow}, see \S \ref{sec:outflowconstraints} for further discussion.

\section{Archival DECam Time Series}
\label{sec:decamarch}

\begin{figure}
    \centering
    \includegraphics[width=\linewidth]{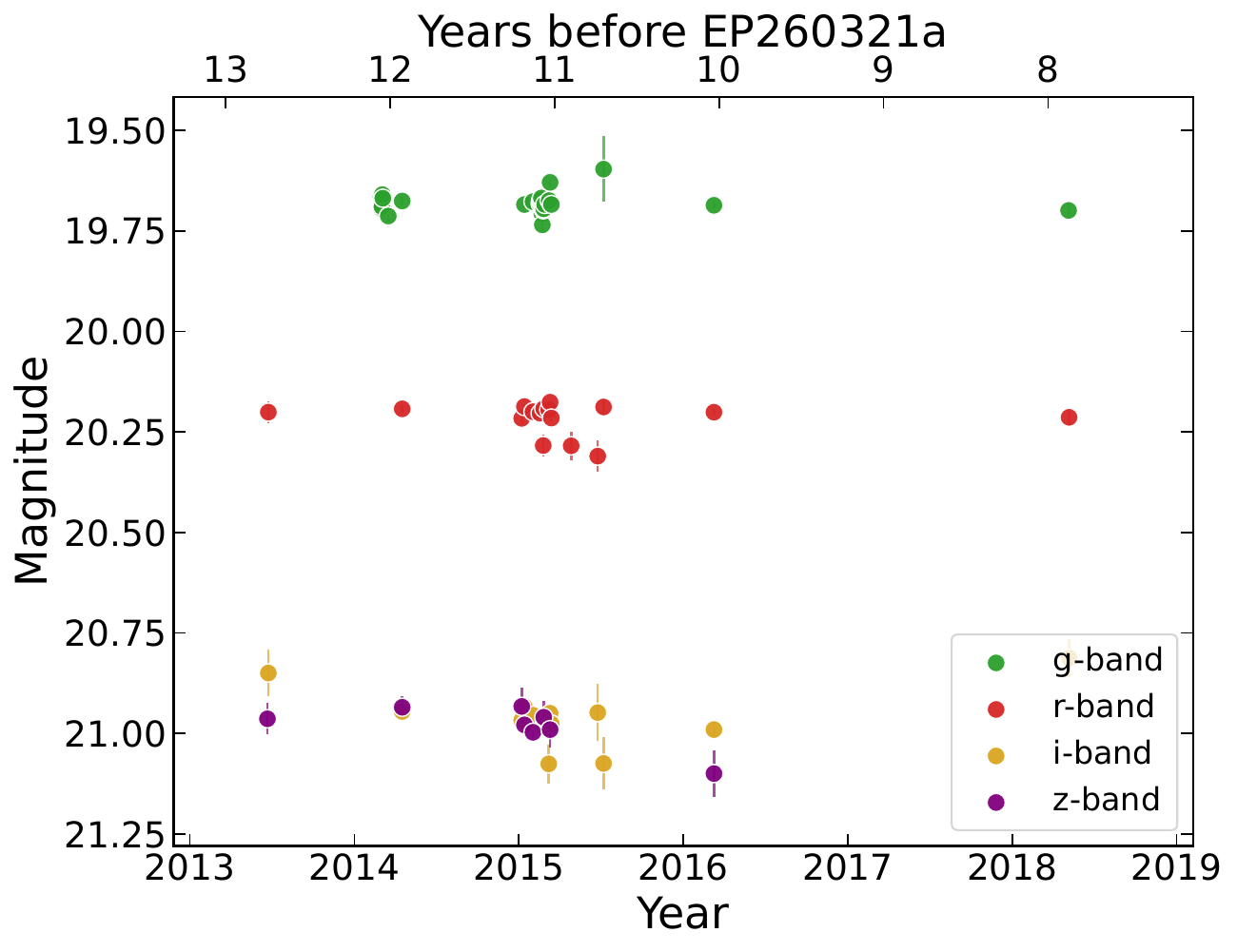}
    \caption{Archival DECam lightcurve of the pre-explosion source (Figure \ref{fig:FC}) at the location of EP260321a/SN 2026gzf in the $griz$ bands.}
    \label{fig:decamarch}
\end{figure}

We built an archival lightcurve of the source prior to its explosion using DECam data spanning 12.8 to 7.9 years prior to explosion, see Figure \ref{fig:decamarch}. The source is visible in all previous images of this field (Figure \ref{fig:FC}), including in other surveys such as PS1\footnote{\url{https://www.wis-tns.org/object/2018mtl}} \citep[e.g., AT 2018mtl;][]{2026GCN.44089....1S}, SDSS, or DSS. This is potentially due to pre-explosion activity akin to that seen in luminous blue variables \citep[LBVs;][]{Weis2020,Smith2026}, as first suggested by \citet{2026GCN.44082....1T} and shown more clearly by \citet{Chen2026} based on archival PS1 observations. We do not observe significant variability in the DECam data (which is consistent with \citet{Chen2026} who also analyze this data, see their Figure 7). In the DECam data, the source is persistently blue in color ($g-i$\,$\approx$\,$-1.3$ mag). The absolute magnitude is $M_r$\,$\approx$\,$-15.7$ mag at a redshift of $z$\,$=$\,$0.0344$. This brightness is at the very extreme end of previous LBV activity \citep[e.g.,][]{Smith2010,Smith2011,Kochanek2011,Smith2014,Smith2020,Smith2022} and may suggest a dominant contribution from an existing HII region or compact dwarf galaxy as suggested by \citet{Chen2026,Rastinejad2026,MartinCarrillo2026,Yuan2026} though with varying inferences on the stellar mass of the compact source ranging from $10^7$ to $10^8$\,$M_\odot$. This is due to different choices regarding the aperture size across different surveys/filters. We note that given this source is visible in all images of this field, it is unclear what contribution is persistent versus transient behavior as difference imaging can only be performed using the oldest images, which may still contain transient activity. 
Regardless, the PS1 variability observed by \citet{Chen2026}, which is at the level of  $M_r$\,$\approx$\,$-14$ mag (see their Figure 3), points to an extreme behavior of the progenitor star leading up to terminal collapse. The exact nature of this pre-explosion source, and the contribution of the possible LBV brightness, will become clear as the supernova fades away.

\clearpage

\begin{longtable}{llllll}
\caption{Log of photometry of EP260321a/SN 2026gzf obtained with FTW and DECam in this work. Times $\Delta T$ are given relative to the EP trigger. The photometry has not been corrected for Galactic reddening $E(B-V)=0.02$ mag \citep{Schlafly2011}.} \label{tab:photometry}\\
\hline
\textbf{Start Time (UT)} & \textbf{$\Delta T$ (d)} & \textbf{Exposure (s)} & \textbf{Telescope} & \textbf{Filter} & \textbf{AB magnitude} \\
\hline
\endfirsthead

\hline
\textbf{Start Time (UT)} & \textbf{$\delta T$ (d)} & \textbf{Exposure (s)} & \textbf{Telescope} & \textbf{Filter} & \textbf{AB magnitude} \\
\hline
\endhead

\hline
\endfoot

\hline
\endlastfoot
2026-03-21T22:29:50 & 0.45 & 5400 & FTW & i & $20.93 \pm 0.08$ \\
2026-03-21T22:29:50 & 0.45 & 5400 & FTW & r & $20.41 \pm 0.01$ \\
2026-03-21T22:30:08 & 0.45 & 5092 & FTW & J & $20.69 \pm 0.26$ \\
2026-03-23T20:13:21 & 2.34 & 2340 & FTW & i & $18.67 \pm 0.01$ \\
2026-03-23T20:13:21 & 2.34 & 2340 & FTW & r & $18.38 \pm 0.01$ \\
2026-03-23T20:13:40 & 2.34 & 2207 & FTW & J & $19.10 \pm 0.06$ \\
2026-03-24T19:12:49 & 3.29 & 1697 & FTW & J & $18.69 \pm 0.05$ \\
2026-03-24T19:49:19 & 3.32 & 1464 & FTW & Ks & $19.40 \pm 0.27$ \\
2026-03-25T02:32:44 & 3.59 & 70 & DECam & g & $17.80 \pm 0.01$ \\
2026-03-25T02:36:21 & 3.59 & 110 & DECam & i & $18.11 \pm 0.01$ \\
2026-03-25T02:38:40 & 3.59 & 90 & DECam & z & $18.41 \pm 0.01$ \\
2026-03-28T01:47:50 & 6.55 & 70 & DECam & g & $17.24 \pm 0.01$ \\
2026-03-28T01:49:29 & 6.56 & 90 & DECam & r & $17.40 \pm 0.01$ \\
2026-03-28T01:51:28 & 6.56 & 110 & DECam & i & $17.53 \pm 0.01$ \\
2026-03-28T01:53:47 & 6.56 & 90 & DECam & z & $17.74 \pm 0.01$ \\
2026-03-30T03:28:12 & 8.62 & 90 & DECam & r & $17.21 \pm 0.01$ \\
2026-03-30T03:30:10 & 8.63 & 110 & DECam & i & $17.36 \pm 0.01$ \\
2026-03-30T03:32:29 & 8.63 & 90 & DECam & z & $17.48 \pm 0.01$ \\
2026-03-31T01:58:20 & 9.56 & 70 & DECam & g & $17.05 \pm 0.01$ \\
2026-03-31T02:01:57 & 9.56 & 110 & DECam & i & $17.30 \pm 0.01$ \\
2026-03-31T02:04:17 & 9.57 & 90 & DECam & z & $17.39 \pm 0.01$ \\
2026-04-03T02:08:51 & 12.57 & 70 & DECam & g & $17.10 \pm 0.01$ \\
2026-04-03T02:12:28 & 12.57 & 110 & DECam & i & $17.22 \pm 0.01$ \\
2026-04-03T02:14:45 & 12.57 & 90 & DECam & z & $17.20 \pm 0.01$ \\
2026-04-04T23:54:58 & 14.48 & 360 & FTW & g & $17.38 \pm 0.04$ \\
2026-04-04T23:54:58 & 14.48 & 1080 & FTW & z & $17.11 \pm 0.02$ \\
2026-04-06T01:33:58 & 15.54 & 70 & DECam & g & $17.31 \pm 0.01$ \\
2026-04-06T01:35:37 & 15.55 & 90 & DECam & r & $17.08 \pm 0.01$ \\
2026-04-06T01:37:36 & 15.55 & 110 & DECam & i & $17.21 \pm 0.01$ \\
2026-04-06T01:39:54 & 15.55 & 90 & DECam & z & $17.06 \pm 0.01$ \\
2026-04-07T20:47:42 & 17.38 & 1620 & FTW & i & $17.20 \pm 0.01$ \\
2026-04-07T20:47:42 & 17.38 & 1620 & FTW & r & $16.99 \pm 0.01$ \\
2026-04-07T20:58:58 & 17.39 & 1260 & FTW & g & $17.52 \pm 0.01$ \\
2026-04-07T20:58:58 & 17.39 & 3240 & FTW & z & $17.03 \pm 0.01$ \\
2026-04-07T21:47:36 & 17.40 & 900 & FTW & u & $19.42 \pm 0.11$ \\
2026-04-08T01:20:08 & 17.54 & 70 & DECam & g & $17.52 \pm 0.01$ \\
2026-04-08T01:23:47 & 17.54 & 110 & DECam & i & $17.22 \pm 0.01$ \\
2026-04-08T01:26:07 & 17.54 & 90 & DECam & z & $17.02 \pm 0.01$ \\
2026-04-08T19:00:20 & 18.31 & 1620 & FTW & i & $17.17 \pm 0.01$ \\
2026-04-08T19:00:20 & 18.31 & 1440 & FTW & r & $17.05 \pm 0.01$ \\
2026-04-08T19:11:18 & 18.31 & 1440 & FTW & g & $17.63 \pm 0.01$ \\
2026-04-08T19:11:18 & 18.31 & 3240 & FTW & z & $17.02 \pm 0.01$ \\
2026-04-08T19:28:36 & 18.32 & 1440 & FTW & u & $19.57 \pm 0.10$ \\
2026-04-11T18:53:28 & 21.27 & 1080 & FTW & z & $17.08 \pm 0.01$ \\
2026-04-11T19:04:02 & 21.28 & 360 & FTW & u & $19.90 \pm 0.16$ \\
2026-04-11T19:15:13 & 21.29 & 720 & FTW & i & $17.23 \pm 0.01$ \\
2026-04-11T19:15:13 & 21.28 & 360 & FTW & r & $17.09 \pm 0.01$ \\
2026-04-11T19:24:08 & 21.29 & 360 & FTW & g & $17.81 \pm 0.01$ \\
2026-04-15T01:12:14 & 24.53 & 70 & DECam & g & $18.24 \pm 0.01$ \\
2026-04-15T01:15:50 & 24.53 & 110 & DECam & i & $17.35 \pm 0.01$ \\
2026-04-15T01:18:08 & 24.53 & 90 & DECam & z & $17.09 \pm 0.01$ \\
2026-04-15T20:11:36 & 25.33 & 720 & FTW & z & $17.17 \pm 0.01$ \\
2026-04-15T20:35:53 & 25.35 & 960 & FTW & i & $17.36 \pm 0.01$ \\
2026-04-15T20:35:53 & 25.34 & 300 & FTW & r & $17.29 \pm 0.02$ \\
2026-04-15T20:50:56 & 25.36 & 660 & FTW & g & $18.33 \pm 0.01$ \\
2026-04-16T19:12:39 & 26.28 & 1080 & FTW & z & $17.28 \pm 0.01$ \\
2026-04-16T19:35:59 & 26.30 & 720 & FTW & i & $17.36 \pm 0.01$ \\
2026-04-16T19:45:44 & 26.31 & 360 & FTW & g & $18.37 \pm 0.02$ \\
2026-04-17T20:37:54 & 27.35 & 900 & FTW & u & $20.63 \pm 0.32$ \\
2026-04-17T20:37:54 & 27.36 & 2160 & FTW & z & $17.22 \pm 0.01$ \\
2026-04-17T21:11:17 & 27.38 & 1440 & FTW & i & $17.44 \pm 0.01$ \\
2026-04-17T21:11:17 & 27.38 & 720 & FTW & r & $17.40 \pm 0.01$ \\
2026-04-17T21:20:12 & 27.39 & 720 & FTW & g & $18.45 \pm 0.01$ \\
2026-04-18T00:26:12 & 27.50 & 70 & DECam & g & $18.46 \pm 0.01$ \\
2026-04-18T00:29:47 & 27.50 & 110 & DECam & i & $17.49 \pm 0.01$ \\
2026-04-18T00:32:06 & 27.50 & 90 & DECam & z & $17.20 \pm 0.01$ \\
2026-04-18T20:02:05 & 28.32 & 1260 & FTW & u & $20.65 \pm 0.33$ \\
2026-04-18T20:09:08 & 28.33 & 900 & FTW & z & $17.33 \pm 0.01$ \\
2026-04-18T20:23:47 & 28.33 & 240 & FTW & r & $17.39 \pm 0.01$ \\
2026-04-18T20:44:38 & 28.34 & 240 & FTW & i & $17.47 \pm 0.01$ \\
2026-04-18T20:48:12 & 28.35 & 240 & FTW & g & $18.55 \pm 0.01$ \\
2026-04-22T19:09:35 & 32.33 & 5400 & FTW & z & $17.47 \pm 0.01$ \\
2026-04-22T19:09:51 & 32.33 & 5092 & FTW & J & $17.40 \pm 0.01$ \\
2026-04-22T19:48:31 & 32.32 & 1440 & FTW & u & $20.82 \pm 0.31$ \\
2026-04-22T20:03:43 & 32.35 & 1440 & FTW & i & $17.66 \pm 0.01$ \\
2026-04-22T20:03:43 & 32.34 & 720 & FTW & r & $17.63 \pm 0.01$ \\
2026-04-22T20:03:58 & 32.35 & 1273 & FTW & Ks & $18.31 \pm 0.08$ \\
2026-04-22T20:14:39 & 32.35 & 720 & FTW & g & $18.77 \pm 0.01$ \\
2026-04-24T20:04:36 & 34.32 & 371 & FTW & z & $17.56 \pm 0.01$ \\
2026-04-24T20:11:08 & 34.33 & 371 & FTW & Ks & $18.46 \pm 0.11$ \\
2026-04-24T20:41:45 & 34.34 & 371 & FTW & r & $17.77 \pm 0.01$ \\
2026-04-24T20:47:59 & 34.35 & 371 & FTW & i & $17.71 \pm 0.01$ \\
2026-04-26T21:19:40 & 36.37 & 849 & FTW & J & $17.57 \pm 0.02$ \\
2026-04-26T21:38:28 & 36.39 & 637 & FTW & Ks & $18.44 \pm 0.14$ \\
2026-05-05T20:01:32 & 45.32 & 720 & FTW & i & $18.24 \pm 0.01$ \\
2026-05-05T20:01:32 & 45.32 & 720 & FTW & r & $18.14 \pm 0.01$ \\
2026-05-05T20:01:45 & 45.32 & 679 & FTW & J & $18.00 \pm 0.05$ \\
2026-05-09T19:45:12 & 49.31 & 900 & FTW & i & $18.34 \pm 0.01$ \\
2026-05-09T19:45:12 & 49.31 & 900 & FTW & r & $18.41 \pm 0.01$ \\
2026-05-09T19:45:17 & 49.31 & 903 & FTW & J & $18.14 \pm 0.05$ \\
2026-05-09T20:03:58 & 49.32 & 720 & FTW & g & $19.36 \pm 0.02$ \\
2026-05-09T20:03:58 & 49.32 & 720 & FTW & z & $18.02 \pm 0.02$ \\
2026-05-09T20:04:02 & 49.32 & 723 & FTW & Ks & $>18.74$ \\
2026-05-17T20:21:32 & 57.33 & 482 & FTW & Ks & $>19.14$ \\

\end{longtable}

\begin{longtable}{lllll}
\caption{Log of public photometry of EP260321a/SN 2026gzf obtained by Rubin/LSST \citep{Ivezic2019} and ZTF \citep{bellm_zwicky_2018}. The photometry was queried through the Babamul alert broker \citep{Babamul} using Object ID 314003014107006318. Times $\Delta T$ are given relative to the EP trigger. The photometry has not been corrected for Galactic reddening $E(B-V)=0.02$ mag \citep{Schlafly2011}.}\label{tab:rubin-ztf-phot}\\
\hline
\textbf{Start Time (UT)} & \textbf{$\Delta T$ (d)} & \textbf{Telescope} & \textbf{Filter} & \textbf{AB magnitude} \\
\hline
\endfirsthead

\hline
\textbf{Start Time (UT)} & \textbf{$\delta T$ (d)} & \textbf{Telescope} & \textbf{Filter} & \textbf{AB magnitude} \\
\hline
\endhead

\hline
\endfoot

\hline
\endlastfoot
2026-03-22T07:12:13 & 0.78 & ZTF & r & $19.90 \pm 0.18$ \\
2026-03-22T07:35:25 & 0.80 & ZTF & g & $19.70 \pm 0.18$ \\
2026-03-26T05:04:13 & 4.69 & ZTF & r & $17.61 \pm 0.05$ \\
2026-03-26T06:27:25 & 4.75 & ZTF & i & $17.84 \pm 0.09$ \\
2026-03-26T07:43:20 & 4.80 & ZTF & g & $17.62 \pm 0.06$ \\
2026-03-27T03:53:42 & 5.64 & ZTF & r & $17.45 \pm 0.05$ \\
2026-03-27T04:40:42 & 5.67 & ZTF & g & $17.38 \pm 0.05$ \\
2026-03-27T05:52:30 & 5.72 & ZTF & i & $17.63 \pm 0.07$ \\
2026-04-02T04:46:20 & 11.68 & ZTF & i & $17.20 \pm 0.04$ \\
2026-04-02T05:36:23 & 11.71 & ZTF & r & $17.02 \pm 0.06$ \\
2026-04-05T04:02:34 & 14.65 & ZTF & r & $16.98 \pm 0.04$ \\
2026-04-06T00:29:23 & 15.50 & Rubin & r & $17.01 \pm 0.01$ \\
2026-04-06T00:33:38 & 15.50 & Rubin & z & $17.07 \pm 0.01$ \\
2026-04-06T03:43:42 & 15.63 & Rubin & i & $17.20 \pm 0.01$ \\
2026-04-07T03:27:01 & 16.62 & Rubin & i & $17.19 \pm 0.01$ \\
2026-04-07T03:54:24 & 16.64 & ZTF & i & $17.18 \pm 0.08$ \\
2026-04-07T04:05:27 & 16.65 & ZTF & r & $16.99 \pm 0.05$ \\
2026-04-09T01:42:28 & 18.55 & Rubin & g & $17.67 \pm 0.01$ \\
2026-04-09T01:45:33 & 18.55 & Rubin & i & $17.23 \pm 0.01$ \\
2026-04-09T04:05:02 & 18.65 & ZTF & r & $16.98 \pm 0.06$ \\
2026-04-09T05:05:00 & 18.69 & ZTF & g & $17.66 \pm 0.07$ \\
2026-04-10T04:57:21 & 19.69 & ZTF & i & $17.20 \pm 0.05$ \\
2026-04-10T05:04:20 & 19.69 & ZTF & g & $17.75 \pm 0.06$ \\
2026-04-11T01:48:16 & 20.55 & Rubin & i & $17.23 \pm 0.01$ \\
2026-04-11T02:59:46 & 20.60 & Rubin & g & $17.86 \pm 0.01$ \\
2026-04-11T03:00:27 & 20.60 & Rubin & g & $17.87 \pm 0.01$ \\
2026-04-11T04:08:01 & 20.65 & Rubin & i & $17.22 \pm 0.01$ \\
2026-04-12T01:33:59 & 21.54 & Rubin & u & $19.68 \pm 0.02$ \\
2026-04-12T01:37:30 & 21.55 & Rubin & r & $17.11 \pm 0.01$ \\
2026-04-12T01:41:01 & 21.55 & Rubin & z & $17.12 \pm 0.01$ \\
2026-04-13T00:04:50 & 22.48 & Rubin & u & $19.84 \pm 0.02$ \\
2026-04-13T01:21:55 & 22.54 & Rubin & i & $17.27 \pm 0.01$ \\
2026-04-13T01:25:25 & 22.54 & Rubin & u & $19.84 \pm 0.02$ \\
2026-04-14T01:26:31 & 23.54 & Rubin & r & $17.14 \pm 0.01$ \\
2026-04-14T01:30:03 & 23.54 & Rubin & u & $19.95 \pm 0.02$ \\
2026-04-15T03:01:11 & 24.60 & Rubin & i & $17.33 \pm 0.01$ \\
2026-04-15T03:01:51 & 24.61 & Rubin & i & $17.32 \pm 0.01$ \\
2026-04-16T04:24:29 & 25.66 & ZTF & i & $17.37 \pm 0.06$ \\
2026-04-16T04:33:35 & 25.67 & ZTF & r & $17.27 \pm 0.06$ \\
2026-04-16T06:45:25 & 25.76 & ZTF & g & $18.47 \pm 0.10$ \\
2026-04-17T06:44:31 & 26.76 & ZTF & r & $17.35 \pm 0.06$ \\
2026-04-20T03:42:01 & 29.63 & ZTF & g & $18.63 \pm 0.12$ \\
2026-04-20T04:04:45 & 29.65 & ZTF & r & $17.46 \pm 0.04$ \\
2026-04-21T00:39:56 & 30.51 & Rubin & i & $17.60 \pm 0.01$ \\
2026-04-21T00:40:37 & 30.51 & Rubin & i & $17.59 \pm 0.01$ \\
2026-04-21T00:44:13 & 30.51 & Rubin & u & $20.62 \pm 0.02$ \\
2026-04-21T00:45:01 & 30.51 & Rubin & u & $20.57 \pm 0.02$ \\
2026-04-21T06:56:32 & 30.77 & ZTF & g & $18.66 \pm 0.13$ \\
2026-04-22T01:59:46 & 31.56 & Rubin & r & $17.63 \pm 0.01$ \\
2026-04-24T03:52:48 & 33.64 & ZTF & i & $17.68 \pm 0.08$ \\
2026-04-25T04:26:18 & 34.66 & ZTF & g & $18.96 \pm 0.20$ \\
2026-04-25T05:07:58 & 34.69 & ZTF & r & $17.91 \pm 0.11$ \\
2026-04-29T04:24:56 & 38.66 & ZTF & i & $17.97 \pm 0.09$ \\
2026-04-30T00:03:48 & 39.48 & Rubin & z & $17.79 \pm 0.01$ \\
2026-04-30T00:11:28 & 39.49 & Rubin & r & $18.04 \pm 0.01$ \\
2026-04-30T04:05:09 & 39.65 & ZTF & i & $17.97 \pm 0.09$ \\
2026-04-30T06:37:46 & 39.76 & ZTF & r & $18.04 \pm 0.13$ \\
2026-05-01T02:13:55 & 40.57 & Rubin & g & $19.16 \pm 0.01$ \\
2026-05-01T02:14:36 & 40.57 & Rubin & g & $19.14 \pm 0.01$ \\
2026-05-02T01:32:03 & 41.54 & Rubin & r & $18.09 \pm 0.01$ \\
2026-05-02T01:34:43 & 41.54 & Rubin & z & $17.89 \pm 0.01$ \\
2026-05-02T01:37:23 & 41.55 & Rubin & r & $18.10 \pm 0.01$ \\
2026-05-02T01:40:05 & 41.55 & Rubin & z & $17.89 \pm 0.01$ \\
2026-05-02T01:54:00 & 41.56 & Rubin & z & $17.89 \pm 0.01$ \\
2026-05-02T04:09:19 & 41.65 & ZTF & g & $19.22 \pm 0.21$ \\
2026-05-02T06:30:10 & 41.75 & ZTF & r & $18.08 \pm 0.12$ \\
2026-05-03T04:02:26 & 42.65 & ZTF & g & $19.19 \pm 0.22$ \\
2026-05-07T03:56:16 & 46.64 & ZTF & i & $18.32 \pm 0.09$ \\
2026-05-07T04:55:06 & 46.68 & ZTF & r & $18.21 \pm 0.08$ \\
2026-05-07T06:07:56 & 46.73 & ZTF & g & $19.35 \pm 0.16$ \\
2026-05-08T04:07:28 & 47.65 & ZTF & r & $18.28 \pm 0.09$ \\
2026-05-08T04:39:09 & 47.67 & ZTF & g & $19.44 \pm 0.16$ \\
2026-05-10T04:26:36 & 49.66 & ZTF & r & $18.37 \pm 0.07$ \\
2026-05-10T05:00:37 & 49.69 & ZTF & g & $19.45 \pm 0.08$ \\
2026-05-12T00:03:22 & 51.48 & Rubin & r & $18.43 \pm 0.01$ \\
2026-05-12T23:35:53 & 52.46 & Rubin & u & $21.36 \pm 0.05$ \\
2026-05-12T23:37:33 & 52.46 & Rubin & u & $21.36 \pm 0.05$ \\
2026-05-16T23:40:26 & 56.47 & Rubin & g & $19.54 \pm 0.01$ \\
2026-05-16T23:44:26 & 56.47 & Rubin & u & $21.42 \pm 0.05$ \\
2026-05-17T23:56:03 & 57.48 & Rubin & u & $21.58 \pm 0.05$ \\
2026-05-17T23:56:53 & 57.48 & Rubin & u & $21.51 \pm 0.05$ \\
2026-05-17T23:57:58 & 57.48 & Rubin & u & $21.53 \pm 0.06$ \\
2026-05-18T00:02:29 & 57.48 & Rubin & r & $18.57 \pm 0.01$ \\
2026-05-18T23:35:12 & 58.46 & Rubin & g & $19.58 \pm 0.01$ \\
2026-05-18T23:35:53 & 58.46 & Rubin & g & $19.59 \pm 0.01$ \\
2026-05-18T23:38:53 & 58.46 & Rubin & u & $21.39 \pm 0.05$ \\
2026-05-19T23:50:44 & 59.47 & Rubin & u & $21.62 \pm 0.06$ \\
2026-05-19T23:51:32 & 59.47 & Rubin & u & $21.46 \pm 0.06$ \\
2026-05-25T23:29:47 & 65.46 & Rubin & z & $18.52 \pm 0.01$ \\
2026-05-26T23:28:32 & 66.46 & Rubin & g & $19.66 \pm 0.01$ \\
2026-05-29T00:47:35 & 68.51 & Rubin & g & $19.63 \pm 0.01$ \\
2026-05-29T00:51:43 & 68.51 & Rubin & i & $18.83 \pm 0.01$ \\
2026-05-30T00:57:00 & 69.52 & Rubin & z & $18.65 \pm 0.01$ \\
2026-05-30T23:25:19 & 70.45 & Rubin & g & $19.71 \pm 0.01$ \\
2026-05-30T23:26:00 & 70.46 & Rubin & g & $19.71 \pm 0.01$ \\
2026-06-04T00:51:41 & 74.51 & Rubin & i & $18.94 \pm 0.01$ \\
\end{longtable}

\begin{table*}
\centering
\caption{Log of spectroscopic observations of EP260321a/SN 2026gzf.}
\label{tab:spectra}
\begin{tabular}{lcccccc}
\hline
Start Date (UTC) & $\delta t$ (days) & Telescope & Instrument & Grating & Angle (deg) & Exposure (s) \\
\hline
2026-03-24 19:11:27 & 3.28  & SALT           & RSS           & pg0700  &  4.6  & 2100 \\
2026-03-25 03:35:06 & 3.63  & HET           & LRS2           & LRS2-B  &  4.6   & 1800 \\
2026-03-27 18:40:03 & 6.26  & SALT           & RSS           & pg0700   & 4.6  & 2100 \\
 2026-04-01 19:02:08   &  11.27  & SALT           & RSS           & pg0700   &  4.6 & 2100 \\
2026-04-05 17:50:18    & 15.22  & SALT           & RSS           & pg0700   & 4.6  & 2100 \\
2026-04-08 04:33:00    & 17.67  & Mayall        & DESI             & \nodata     &   \nodata   &  1200 \\
2026-04-13 18:36:10    &  23.25 & SALT           & RSS           & pg0700  & 4.6    & 2100 \\
2026-04-23 03:38:37   & 32.63  & Mayall        & DESI             & \nodata      &   \nodata  &  1200 \\
 2026-05-01 17:25:57   &  41.21 & SALT           & RSS           & pg0700   &  7.6 & 2100 \\
2026-05-09 17:00:30    & 49.19  & SALT           & RSS           & pg0700   &  7.6 & 2100 \\
 2026-05-14 03:40:38   &  53.63 & Mayall        & DESI             & \nodata     &  \nodata    &  1200 \\
 2026-05-14 17:17:51   & 54.20  & SALT           & RSS           & pg0700   &  4.6 & 2100 \\
\hline
\end{tabular}
\end{table*}

\bibliography{bib}{}
\bibliographystyle{aasjournal}



\end{document}